\documentclass[12pt]{article}
\pdfoutput=1
\usepackage[top=30truemm,bottom=30truemm,left=25truemm,right=25truemm]{geometry}
\usepackage{authblk}
\usepackage{amsthm}
\usepackage{amsmath,amssymb,bm}
\usepackage{cite}
\usepackage[]{hyperref}
\hypersetup{colorlinks,linkcolor={blue},citecolor={blue},urlcolor={blue}}  
\usepackage[titletoc]{appendix}
\usepackage{url}
\usepackage{graphicx,color}
\usepackage{bbold}
\usepackage{braket}
\usepackage{qcircuit}
\usepackage{here}
\usepackage{mathrsfs}

\newtheorem{assumption}{Assumption}

\newtheorem{innercustomthm}{Theorem}
\newenvironment{customthm}[1]
  {\renewcommand\theinnercustomthm{#1}\innercustomthm}
  {\endinnercustomthm}
  
\newtheorem{innercustomstat}{Statement}
\newenvironment{customstat}[1]
  {\renewcommand\theinnercustomstat{#1}\innercustomstat}
  {\endinnercustomstat}

\newcommand{\imineq}[2]{\vcenter{\hbox{\includegraphics[height=#2ex]{#1}}}}

\def\sinh{{\mathrm{sinh}}}
\def\cosh{{\mathrm{cosh}}}
\def\tanh{{\mathrm{tanh}}}

\def\CT{{\mathcal{T}}}
\def\CI{{\mathcal{I}}}

\def\m{{\mu}}
 
 \def\n{{\nu}}

 \def\frac#1#2{{#1\over #2}}

 \def\CC{{\cal C}}
 \def\CD{{\cal D}}
 \def\CE{{\cal E}}
 \def\CI{{\cal I}}
 \def\CO{{\cal O}}

 \def\CN{{\cal N}}
 \def\CM{{\cal M}}
 \def\CL{{\cal L}}

 \def\p{\partial}

\def\be{\begin{equation}}
\def\ee{\end{equation}}
\def\ba{\begin{eqnarray}}
\def\ea{\end{eqnarray}}

\numberwithin{equation}{section}

\AtBeginDocument{%
  {\footnotesize\global\footnotesep=0.8\baselineskip}%
}

\begin{document}

\begin{titlepage}
\thispagestyle{empty}

\begin{flushright}
YITP-21-51 \\ 
~ \\
\end{flushright}

\bigskip

\begin{center}
\noindent{\bf \LARGE Causal Structures and Nonlocality in}\\
\vspace{0.5cm}
\noindent{\bf \LARGE Double Holography}\\
\vspace{1.6cm}

{\bf \normalsize Hidetoshi Omiya$^a$ and Zixia Wei$^b$}
\vspace{1cm}\\

{\it $^a$Department of Physics, Kyoto University, \\
Kitashirakawa Oiwakecho, Sakyo-ku, Kyoto 606-8502, Japan}\\

{\it $^b$Center for Gravitational Physics,\\
Yukawa Institute for Theoretical Physics,
Kyoto University, \\
Kitashirakawa Oiwakecho, Sakyo-ku, Kyoto 606-8502, Japan}
\vspace{0.5cm}\\

\end{center}

\begin{abstract}
Double holography plays a crucial role in recent studies of Hawking radiation and information paradox by relating an intermediate picture, in which a dynamical gravity living on an end-of-the-world brane is coupled to a non-gravitational heat bath, to a much better-understood BCFT picture as well as a bulk picture. In this paper, causal structures in generic double holographic setups are studied. We find that the causal structure in the bulk picture is compatible with causality in the BCFT picture, thanks to a generalization of the Gao-Wald theorem. On the other hand, consistency with the bulk causal structure requires the effective theory in the intermediate picture to contain a special type of superluminal and nonlocal effect which is significant at long range or IR. These are confirmed by both geometrical analysis and commutators of microscopic fields. Subregion correspondences in double holography are discussed with the knowledge of this nonlocality. Possible fundamental origins of this nonlocality and its difference with other types of nonlocality will also be discussed. 
\end{abstract}

\end{titlepage}

\newpage
\setcounter{page}{1}
\tableofcontents

\newpage

\section{Introduction}

The information loss problem in black hole evaporation \cite{Hawking74,Page93,AHMST20} has puzzled physicists for decades and recently gets a great development \cite{Penington19,AEMM19}. Starting from a pure initial state and tracking the time evolution from the formation to the evaporation of a black hole, the entanglement entropy between its interior and exterior is expected, from unitarity, to start increasing from zero and finally return back to zero again. This behavior is known as a Page curve \cite{Page93}. Hawking's original computation \cite{Hawking74} was performed with a local quantum field theory living on a classical spacetime with a black hole. By simply factorizing the Hilbert space into interior and exterior on the classical spacetime, Hawking's computation suggests that the entanglement entropy monotonically grows and leads to a breakdown of unitarity, i.e. loss of information. 

Recent studies have resolved this problem by coupling a gravitational region containing a black hole to a non-gravitational region working as a heat bath, and studying the entanglement entropy between them. It is found that another saddle point which Hawking did not count dominates at late time and reproduces the expected Page curve \cite{Penington19,AEMM19}. The existence and dominance of this saddle point is justified both by a class of doubly holographic models\cite{AMMZ19} and by gravitational path integral \cite{AHMST19,PSSY19}. 

\begin{figure}[H]
    \centering
    \includegraphics[width=14cm]{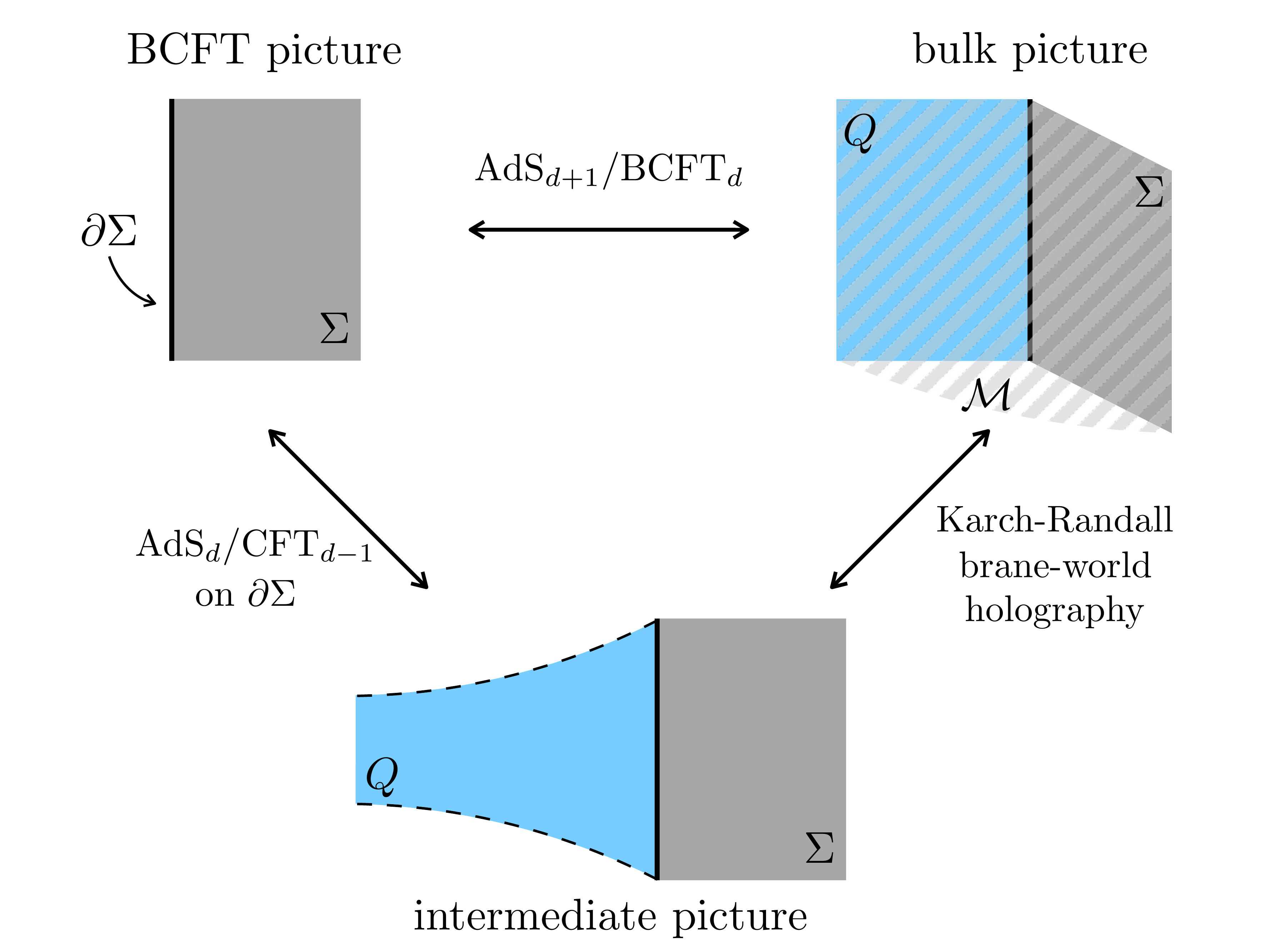}
    \caption{The three equivalent pictures of a doubly holographic model. The asymptotic boundary $\Sigma$ and the end-of-the-world brane $Q$ are shown in grey and blue, respectively. The bulk region $\CM$ is shaded.}
    \label{fig:DHtriangle}
\end{figure}

Consider a $d$-dimensional AdS gravity living on $Q$ interacting with a $d$-dimensional CFT living on $\Sigma$ through a $(d-1)$-dimensional interface $\partial Q = \partial \Sigma$. Double holography relates the current setup to two different but equivalent theories. One is a boundary CFT (BCFT) on $\Sigma$, which can be obtained by applying the AdS/CFT correspondence to $Q$ and regarding it as a $(d-1)$-dimensional CFT living on $\partial\Sigma$. The other one is an AdS$_{d+1}$ gravity with an end-of-the-world brane floating in it. The asymptotic boundary and the end-of-the-world brane are identified with $\Sigma$ and $Q$, respectively. In this paper, we call the latter two equivalent pictures the BCFT picture and the bulk picture respectively. At the same time, we call the original setup the intermediate picture, in the sense that it can be regarded as an intermediate process when jumping between the BCFT picture and the bulk picture. See figure \ref{fig:DHtriangle} for a sketch. 

Although the terminology ``double holography" is relatively new, the correspondence between the three pictures has been known for a long time since \cite{KR00,KR01}. In particular, the duality between the intermediate picture and the bulk picture is often called the Karch-Randall type brane-world holography. On the other hand, the duality between the bulk picture and the BCFT picture is further explored in \cite{Takayanagi11,FTT11} and called the AdS/BCFT correspondence. With double holography, the second saddle point in Hawking radiation can be thought to come from boundary OPE in the BCFT picture and minimal surfaces ending on the end-of-the-world brane in the bulk picture \cite{RSJvRWW19,CMNRS20,CMNRS20-2}. On the other hand, from a gravitational path integral point of view, this saddle point comes from spacetime configurations with higher topology \cite{AHMST19,PSSY19}. 

While dynamics in double holography plays a crucial role in recent studies of Hawking radiation, discussions in the Lorentzian signature\cite{AKSTW20,AKSTW21,MW21} are limited and mostly focusing on specific spacetime configurations. One of the most important ingredients in a Lorentzian theory is the causal structure. In the following of this paper, we study the causal structures in double holography for generic spacetime dimensions and configurations. 

The causal structure in holography was firstly discussed by Gao and Wald \cite{GW00}. They studied the AdS/CFT correspondence \cite{Maldacena97,GKP98,Witten98} and proved a theorem which implies that the causal structure in AdS is compatible with causality (the property that one cannot send a signal outside of the light cone) in CFT. As for double holography, we will firstly show that the causal structure in the bulk picture is compatible with causality in the BCFT picture, as a generalization of the Gao-Wald theorem. On the other hand, we will see that, the intermediate picture is expected to have a unique causality violation, for it to be compatible with the bulk causal structure. More specifically, a signal propagating within the gravitational region $Q$ or within the non-gravitational region $\Sigma$ cannot travel faster than light, while superluminal information propagation is allowed when sending the signal from $Q$ to $\Sigma$ and vice versa. These features will also be confirmed by computing commutators of microscopic fields. See \cite{Ishihara00} for a previous study of causality within the end-of-the-world brane in the context of Randall-Sundrum type brane-world holography \cite{RS99-2,RS99}.

This causal structure implies that the effective theory in the intermediate picture is a nonlocal one. To our knowledge, this property has not been explicitly examined in previous works. The nonlocality prompts us to reconsider the notion of domain of dependence which is usually discussed in local theories. Based on this, the subregion correspondence in double holography will be discussed. In ordinary AdS/CFT or AdS/BCFT correspondence, a subregion in the (B)CFT picture is equivalent to its entanglement wedge in the bulk picture \cite{HHLM14,ADH14,DHW16}. We will introduce an analogy of the entanglement wedge for intermediate subregions instead of BCFT subregions and call it the tentative entanglement wedge. It will be shown that the tentative entanglement wedge is not equivalent to the corresponding intermediate subregion, in contrast to what one may expect at first glance. 

The causality violation and nonlocality in the intermediate picture becomes significant when zooming out to IR and neglectable when zooming in to UV. Going beyond double holography, features in the intermediate picture suggest a possibility that effective theories of quantum gravity show similar nonlocality in general. We will argue that the effect from spacetime configuration with higher topology is a possible origin of such nonlocality and discuss its difference from nonlocality known in early studies of quanutm gravity \cite{LPSTU95,Giddings06}. 

This paper is organized as follows. In section \ref{sec:preandsum}, we present a careful review of a doubly holographic model starting from the AdS/BCFT correspondence. After that, we review compatibility of causality in holographic dualities and summarize the main technical results about causal structures in double holography found in this paper. In section \ref{sec:geoex}, we focus on the vacuum configuration of double holography as a simple concrete example and verify our results by explicitly writing down the geodesics. In section \ref{sec:causalitySigma}, we show that causality in the BCFT picture is compatible with the bulk causal structure for generic configurations with reasonable assumptions, and discuss its physical consequences in AdS/BCFT. In section \ref{sec:causalityQ}, we study the causal structure in the intermediate picture and discuss its relation with other holographic models. In section \ref{sec:commutator}, we compute the commutator of a light primary operator in the intermediate picture using a holographic computation and confirm that it is consistent with the results shown in the previous sections. In section \ref{sec:EWRC}, we discuss physical consequences coming from the nonlocality in the intermediate picture. After clarifying confusing notions of subregions and reduced states associated to them, and examining the correspondence between subregions in the three pictures, we discuss more fundamental features of nonlocality in quantum gravity. In section \ref{sec:conclu}, we summarize our results and discuss future directions. In appendix \ref{app}, basics of (globally) AdS spacetime are reviewed. 

We would like to finish the introduction with a reading guide. Section \ref{sec:causalitySigma}, \ref{sec:causalityQ}, \ref{sec:commutator} and \ref{sec:EWRC} are relatively independent from each other. Therefore, readers can go straight to each of them after checking the basic results in section \ref{sec:preandsum} and \ref{sec:geoex}. 

{\bf Note added:} While we were writing this paper, we got aware of an interesting work \cite{Neuenfeld21-2}, where subregion correspondence in double holography is independently considered in a setup similar to ours discussed in section \ref{sec:tenwedge}, but from a different perspective.

\section{Preliminaries and Summary of Technical Results}
\label{sec:preandsum}
In this section, we firstly review the setup of double holography and causality compatibility in the AdS/CFT correspondence. Then we will summarize our results about the causal structures in double holography. 

\subsection{Review of Double Holography and Related Topics}
\label{sec:doublehol}

We present a review of double holography in this subsection. Readers who are familiar with this topic may skip to the next subsection after catching a glimpse of table \ref{tab:notations} and figure \ref{fig:DHtriangle} for the notations using in this paper. 

The notion of double holography\footnote{The correspondence between the three pictures itself has been known since \cite{KR00,KR01}.}, which is proposed in \cite{AMMZ19} and further explored in \cite{RSJvRWW19,CMNRS20,CMNRS20-2}, arises naturally when considering a CFT defined on a $d$-dimensional manifold $(\Sigma, \gamma_{ij})$ with a time-like boundary $\partial \Sigma$. When the boundary maximally preserves the conformal symmetry, the theory is called a boundary conformal field theory (BCFT). The gravity dual of a holographic BCFT can be constructed in a bottom-up way called AdS/BCFT correspondence \cite{Takayanagi11,FTT11}. We will consider the Lorentzian signature throughout this paper. 

According to AdS/BCFT, the gravity dual of a holographic BCFT is given by a portion of a $(d+1)$-dimensional asymptotically AdS spacetime (AAdS) $(\CM, g_{\mu\nu})$. The boundary of $\CM$ is given by $\partial \CM = \Sigma \cup Q$ with $\partial \Sigma = \partial Q$. Here, $\Sigma$ is the ordinary asymptotic boundary of $\CM$ on which Dirichlet boundary conditions are imposed to the bulk fields. On the other hand, $(Q, h_{ab})$ is an end-of-the-world brane extended to the bulk from $\partial \Sigma$. In contrast with the asymptotic boundary $\Sigma$, Neumann boundary conditions are imposed to the bulk fields on the end-of-the-world brane $Q$. To be more concrete, the bulk action is given by 
\begin{align}
    I_{\rm bulk} = \frac{1}{16\pi G_N} \int_{\CM} \sqrt{-g} (R-2\Lambda) &+ \frac{1}{8\pi G_N} \int_{\Sigma} \sqrt{-\gamma} B  \nonumber \\
    &+ \frac{1}{8\pi G_N} \int_{Q} \sqrt{-h} (K-T) + I_{\rm matter}, 
    \label{eq:bulkaction}
\end{align}
where the four terms are the Einstein-Hilbert term in the bulk $\CM$, the Gibbons-Hawking term on the asymptotic boundary $\Sigma$, the Gibbons-Hawking term on the brane $Q$, and the action for the bulk matter fields. The metric of $\CM$, $\Sigma$ and $Q$ are denoted by $g_{\mu\nu}$, $\gamma_{ij}$ and $h_{ab}$, respectively. $B_{ij}$ ($K_{ab}$) is the extrinsic curvature of $\Sigma$ ($Q$), and $T$ is the tension of $Q$.\footnote{Tension $T$ should be restricted to $-(d-1)/L < T < (d-1)/L$ for the end-of-the-world brane $Q$ to have a time-like intersection with the asymptotic boundary $\Sigma$ \cite{Takayanagi11,FTT11,NTU12,AKTW20}. On the other hand, if one takes $|T| > (d-1)/L$, $Q$ will have a space-like intersection with $\Sigma$ and become asymptotically dS \cite{AKTW20}. } Variation of the gravity sector at the vicinity of the asymptotic boundary $\Sigma$ is given by 
\begin{align}
    \delta I_{\rm bulk} = \frac{1}{16\pi G_N} \int_{\Sigma} \sqrt{-\gamma} (B_{ij}-B\gamma_{ij}) \delta\gamma^{ij} .
\end{align}
Dirichlet boundary condition $\delta \gamma^{ij} = 0$ is imposed on $\Sigma$. In contrast with this, although the variation at the vicinity of the end-of-the-world brane $Q$ is similarly given by
\begin{align}
    \delta I_{\rm bulk} = \frac{1}{16\pi G_N} \int_{Q} \sqrt{-h} (K_{ab}-K h_{ab}+T h_{ab}) \delta h^{ab} ,
\end{align}
the boundary condition imposed on it is chosen to be the Neumann type
\begin{align}\label{eq:Neumann}
    K_{ab}-K h_{ab} - T h_{ab} = 0 .
\end{align}
At leading order of $G_N^{-1}$, the gravity dual of a holographic BCFT defined on $\Sigma$ is an on-shell configuration of (\ref{eq:bulkaction}). 

Here, we note that there is no matter field localized on the brane $Q$ in (\ref{eq:bulkaction}). We would like to take this as default in this paper, while consequences caused by such a matter field will be discussed in section \ref{sec:vsothers}. 

Now, we have two equivalent pictures for a holographic BCFT. We will call them the BCFT picture and the bulk picture, respectively. We use $\CT^{\rm BCFT}_\Sigma$ to denote the theory in the BCFT picture, and $\CT^{\rm bulk}_\CM$ to denote the theory in the bulk picture. 
The novelty of double holography is that there exists another intermediate picture by simply applying holography only to the $(d-1)$-dimensional boundary $\partial \Sigma$ of the BCFT but not to the ambient region. After doing this, naively, we will get a theory composed of two parts. One is a gravitational theory living in an AAdS$_{d}$ whose asymptotic boundary is identified with $\partial \Sigma$.\footnote{It is highly nontrivial whether the gravity dual of the $d-1$-dimensional field theory on $\partial \Sigma$ is really a $d$-dimensional AAdS after the standard definition given in, for example, \cite{Skenderis02}. However, this does not matter in the discussions in this paper.} Another part is the original holographic CFT$_d$ living in the ambient region without gravity. This intermediate picture plays an important role in recent progress of information paradox, since it couples a gravitational region to a non-gravitational heat bath to which Hawking radiation can escape, and allows us to use better-understood BCFT picture and bulk picture to study the information flow. Let us just call it the intermediate picture. 

Although we have intuitively explained how the intermediate picture arises from the BCFT picture, it can also be derived from the bulk picture, long known as Karch-Randall brane-world holography\cite{KR00,KR01,BR02}. In that context, it is known that the bulk theory (\ref{eq:bulkaction}) in $\CM$ is equivalent to a theory on its boundary $\partial \CM = \Sigma \cup Q$ where $Q$ has a dynamical gravity\footnote{One key feature of this gravity sector is that it acquires a mass at $d\geq 2$. See \cite{GK20} for an excellent review on this point.} due to the Neumann boundary condition (\ref{eq:Neumann}) while $\Sigma$ does not due to the Dirichlet boundary condition. Besides, there is a common holographic CFT living on both $Q$ and $\Sigma$ as a matter theory, and the boundary condition for this CFT on $\partial Q = \partial \Sigma$ is transparent \cite{ADFK03}. These together give the intermediate picture.\footnote{Note that the CFT here arises as a matter field on $Q$ in the intermediate picture, while in the bulk picture, no matter field is put on $Q$ by default, as shown in (\ref{eq:bulkaction}). }
See \cite{Porrati01,Neuenfeld21} for an effective action on the end-of-the-world brane (also called the Karch-Randall brane in this context) $Q$. We use $\CT^{\rm int}_{\Sigma\cup Q}$ to denote the theory in the intermediate picture.

So far, we have introduced a bunch of notations associated to the manifolds considered in this paper. We summarize them in table \ref{tab:notations} so that the readers can refer to it easily. 

In short, the doubly holographic model introduced here has three equivalent pictures: the BCFT picture, the bulk picture and the intermediate picture.\footnote{The BCFT picture and the intermediate picture here is often called ``boundary perspective'' and ``brane perspective'' respectively in related literature \cite{CMNRS20,CMNRS20-2,Neuenfeld21,Neuenfeld21-2}. Here, we use ``BCFT picture" instead of ``boundary perspective" to avoid confusions since we are going to deal with many different notions of ``boundary". On the other hand, we use ``intermediate picture" instead of ``brane perspective" to stress that not only the end-of-the-world brane $Q$ but also the asymptotic boundary $\Sigma$ is an important part of it.} The relationship between these three pictures are shown in figure \ref{fig:DHtriangle}.

\begin{table}[htb]
  \begin{tabular}{|c|c|c|c|c|} \hline
    ~ & bulk $\CM$& \begin{tabular}{c}
         asymptotic  \vspace{-2mm}\\ 
         boundary $\Sigma$
    \end{tabular} & \begin{tabular}{c}
         end-of-world  \vspace{-2mm}\\
         brane $Q$
    \end{tabular}  & \begin{tabular}{c}
         embedding  \vspace{-2mm}\\
         spacetime $\mathbb{R}^{2,d}$
    \end{tabular} \\\hline 
    indices & $\m,\n,\rho,\dots$ & $i,j,k,\dots$ & $a,b,c,\dots$ & $A,B,C,\dots$ \\ \hline 
    metric & $g_{\m\n}$ & $\gamma_{ij}$ & $h_{ab}$ & $(-,-,+,\dots,+)$ \\ \hline
    extrinsic curvature & - & $B_{ij}$ & $K_{ab}$ & - \\ \hline
    total dimension & $d+1$ & $d$ & $d$ & $d+2$ \\ \hline 
    time-like dimension & $1$ & $1$ & $1$ & $2$ \\ \hline
  \end{tabular}
  \caption{Notations for the manifolds considered in this paper.}
  \label{tab:notations}
\end{table}

\subsection{Compatibility of Causality in the AdS/CFT Correspondence}\label{sec:compadscft}
If two theories are equivalent via a duality or correspondence, a physical process in one theory can be translated into another equivalent physical process in the other theory, at least in principle. As a result, if a physical process is not achievable in one theory, so should not be its dual process in the dual theory. This fact, while being a matter of course, is in general non-trivial and can give powerful constraints when studying correspondences which are not completely understood. 

\begin{figure}[H]
    \centering
    \includegraphics[width=11cm]{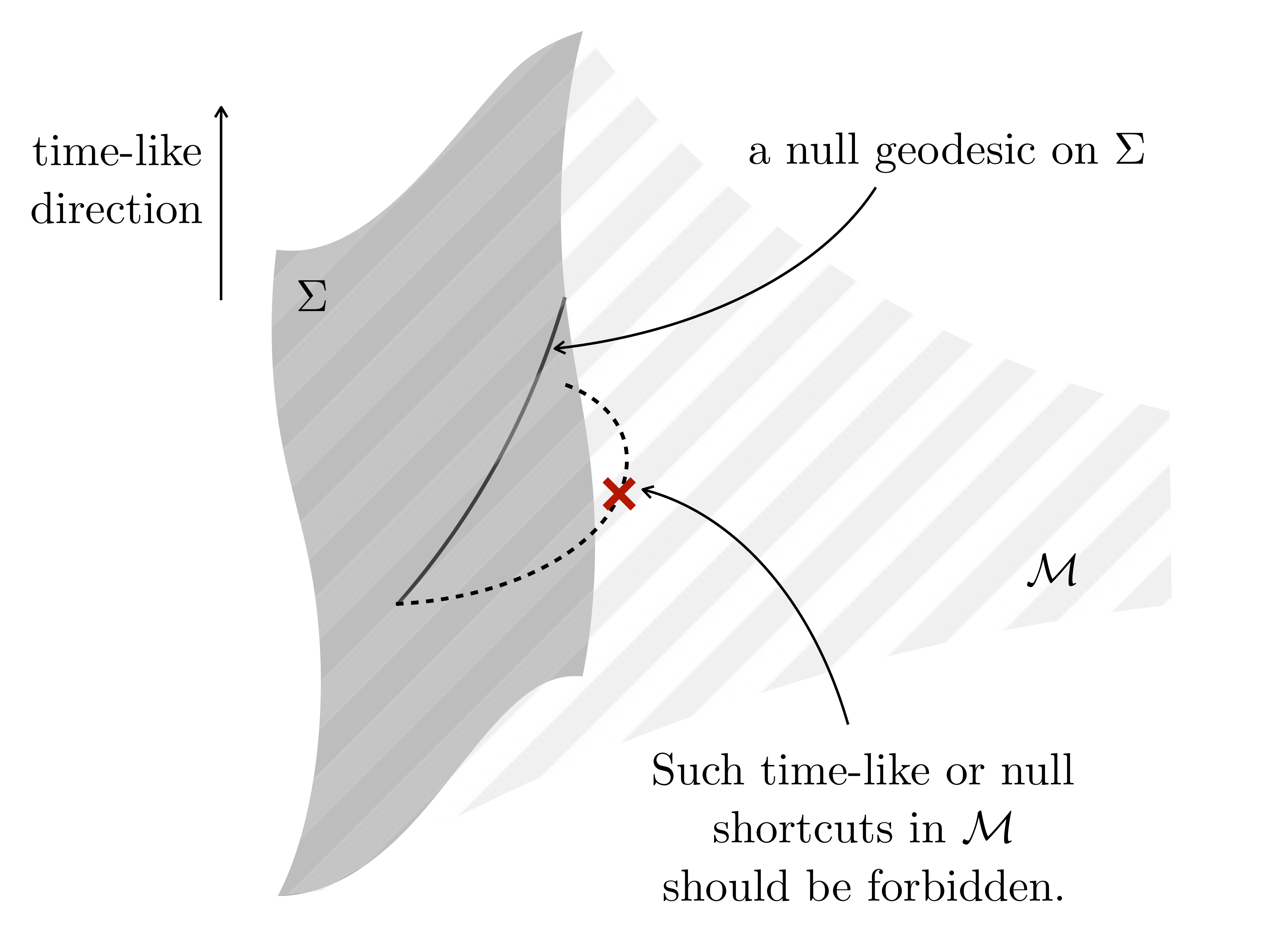}
    \caption{No shortcut should be allowed in the bulk $\CM$ (shaded) when considering signal propagations on the asymptotic boundary $\Sigma$ (shown in grey). The solid line shows a null geodesic on $\Sigma$. The dashed line shows a time-like/null geodesic in $\CM$ connecting two space-like separated (with respect to $\Sigma$) points on $\Sigma$, which should not have existed.}
    \label{fig:gao_wald}
\end{figure}

Let us take the compatibility of causality in the AdS/CFT correspondence as an example. In standard AdS/CFT correspondence, a holographic CFT defined on a $d$-dimensional Lorentzian manifold $\Sigma$ is equivalent to a gravitational theory in an AAdS$_{d+1}$ manifold whose asymptotic boundary is given by $\Sigma$. Here, we consider the case where $\Sigma$ has no boundary. On the CFT side, one cannot send a signal from $p \in \Sigma$ to $q \in \Sigma$ if $p$ and $q$ are space-like separated on $\Sigma$, assuming the CFT is local and unitary. The corresponding process in the bulk theory is to send a signal from $p \in \Sigma$ to $q \in \Sigma$ through the bulk $\CM$, after the GKP-W relation \cite{GKP98,Witten98}. Therefore, we expect the latter process in the bulk $\CM$ should also be impossible, i.e. $p$ and $q$ should also be space-like separated in the bulk $\CM$, for the bulk causal structure to be compatible with causality in CFT. In other words, no shortcut should be allowed in $\CM$ when considering signal propagations on $\Sigma$. The situation is shown in figure \ref{fig:gao_wald}. Let us summarize this in a more clear and standard way. 

\begin{customstat}{A}[Compatibility of Causality in AdS/CFT]~\par
\label{stat:adscft}
In the AdS/CFT correspondence, for any two points $p,q\in\Sigma = \partial \CM$, if $p$ and $q$ are not causally connected on the asymptotic boundary $\Sigma$, then they are not causally connected in the bulk $\CM$ either. 
\end{customstat}

One can check that the statement \ref{stat:adscft} holds in pure AdS by explicitly writing down the geodesics. This point will be reviewed in section \ref{sec:geoexglobal}. 

For more general configurations, the Gao-Wald theorem \cite{GW00} guarantees that the statement \ref{stat:adscft} holds under several reasonable assumptions.\footnote{On the other hand, the no-shortcut statement \ref{stat:adscft} can also be taken as an input to give restrictions on other physical conditions in the AdS/CFT correspondence \cite{EF16,IIM20}.} The Gao-Wald theorem is summarized as follows:

\begin{customthm}{A.0}[Gao-Wald theorem]\label{thm:GW}~\par
    Let $(\mathcal{M},g_{\mu\nu})$ be a spacetime with a time-like boundary $\partial \mathcal{M}=\Sigma$ at asymptotic infinity. Suppose $(\CM,g_{\mu\nu})$ satisfies the following four conditions:
    \begin{enumerate}
        \item $\CM$ is a solution of the Einstein equation in which the matter sector satisfies averaged null energy condition (ANEC),
        \item Null generic condition, 
        \item $\bar{\CM} \equiv \CM\cup\Sigma$ is strongly causal (strong causality condition), 
        \item $J_{\bar{\CM}}^+(p) \cap J_{\bar{\CM}}^{-}( q)$ is compact for any $p,q \in \mathcal{M}$.
    \end{enumerate}
    Let $A_{\Sigma}(p,\CM)$ be
    \begin{align}   
    \begin{aligned}
        A_{\Sigma}(p,\CM) \equiv \{r\in \Sigma|&\text{there exists a future directed causal curve $\lambda$}\\
        &\text{which starts from $p$ and ends at $r$ satisfying $\lambda - (p\cup r) \subset \mathcal{M}$ }\}~,
    \end{aligned}
    \end{align}
    then for $p \in \Sigma$ and $\forall q \in  \partial A_{\Sigma}(p,\CM)$, $q$ satisfies $q \in J_{\bar{\CM}}^{+}(p) \backslash I_{\bar{\CM}}^{+}(p)$. Moreover, any causal curve connecting $p$ and $q$ lies entirely in $\Sigma$, and hence it is also a null geodesic on $(\Sigma,\gamma_{ij})$, i.e. $q \in J_{{\Sigma}}^{+}(p) \backslash I_{{\Sigma}}^{+}(p)$. 
\end{customthm}

Here, $J^+_{\CN}(p)$ ($I^+_{\CN}(p)$) is the causal (chronological) future of $p$ on the manifold $\CN$. 
We briefly comment on the four conditions above. The first one, ANEC, states that the energy-stress tensor $T_{\mu\nu}$ satisfies
        \begin{align}
            \int_{l} T_{\mu\nu} k^\mu k^\nu \ge 0~,
        \end{align}
for any null curve $l$. Here, $\int_l$ stands for the integration along $l$, and $k^{\mu}$ is its tangent vector. The second condition, null generic condition, states that every null geodesic in $\CM$ must contain a point at which
        \begin{align}
            k^\mu k^\nu k_{[\rho} R_{\sigma] \mu\nu [\alpha}k_{\beta]} \neq 0~,
        \end{align}
where $k$ is the tangent vector of the null geodesic \footnote{Note that pure AdS does not satisfy this condition. Therefore, the Gao-Wald theorem does not cover pure AdS.}. This condition together with the ANEC ensures the existence of the conjugate points for any null geodesic \cite{HE73, Witten19}, which is one of the most essential points in the proof of the Gao-Wald theorem. The third condition, strong causality condition, roughly means that no causal curve in $\bar{\CM}$ can be almost closed. More rigorously\footnote{See section 6.4 of \cite{HE73}, for example.}, for any point $p \in \bar{\mathcal{M}}$, every neighborhood of $p$ contains another neighborhood of $p$ which no causal curve can intersect more than once. This condition and the fourth condition guarantee that bulk spacetime under consideration has a sensible causal structure. 

Let us then explain how the Gao-Wald theorem guarantees that statement \ref{stat:adscft} holds. First of all, for points on the asymptotic boundary $\Sigma$, $\partial A_{\Sigma}(p,\CM)$ bounds the region which can receive a signal through the bulk $\CM$, and $J_{{\Sigma}}^{+}(p) \backslash I_{{\Sigma}}^{+}(p)$ bounds the region which can receive a signal through the asymptotic boundary $\Sigma$. Since one consequence of the Gao-Wald theorem is 
\begin{align}
    \partial A_{\Sigma}(p,\CM) ~\subseteq~ J_{{\Sigma}}^{+}(p) \backslash I_{{\Sigma}}^{+}(p),
\end{align}
statement \ref{stat:adscft} follows straightforwardly from the Gao-Wald theorem, for configurations satisfying the four conditions. 

From now on, we are going to consider the causal structures in the three pictures of the doubly holographic model introduced in section \ref{sec:doublehol}.\footnote{It is worth noting that the notion of quantum tasks \cite{HM12} provides a good framework for using the consistency of more general physical processes to study holography \cite{May19,MPS19,May21,MW21}. The physical process we consider can also be regarded as one of the most simple versions of a quantum task, but we will not go deep into it.} 

\subsection{Summary of Results: Causal Structures in Double Holography}
Consider an on-shell configuration $\CM$ of (\ref{eq:bulkaction}) with Dirichlet boundary condition imposed on the asymptotic boundary $\Sigma$ and Neumann boundary condition (\ref{eq:Neumann}) imposed on the end-of-the-world brane $Q$. $\CM$ has the following properties under some reasonable assumptions such as ANEC.
\begin{customstat}{B}\label{stat:adsbcft}
    Let $\Sigma$ be compatible with a BCFT. For any two points $p,q\in\Sigma$, if $p$ and $q$ are not causally connected on the asymptotic boundary $\Sigma$, then they are not causally connected in the bulk $\CM$ either.
\end{customstat}
\begin{customstat}{C}\label{stat:brane}
    For any two points $p,q\in Q$, if $p$ and $q$ are not causally connected on the end-of-the-world brane $Q$, then they are not causally connected in the bulk $\CM$ either.
\end{customstat}
\begin{customstat}{D}\label{stat:across}
    Two points $p\in\Sigma$ and $q\in Q$ can be causally connected in the bulk $\CM$, even if they are not causally connected on $\partial\CM = \Sigma \cup Q$. 
\end{customstat}

In the following few sections, we are going to firstly give a concrete example of $\CM$ and check that it satisfies the statements above by explicitly writing down the null geodesics in section \ref{sec:geoex}. Then we will show that statement \ref{stat:adsbcft} holds for more general configurations in section \ref{sec:causalitySigma}. After that, we will prove statement \ref{stat:brane} for general configurations in section \ref{sec:causalityQ}. Although the statement \ref{stat:brane} looks similar to statement \ref{stat:adsbcft}, the mathematical mechanism is actually totally different. We will see that Neumann boundary condition (\ref{eq:Neumann}) plays the most important role in this consequence. 

The mathematical facts above lead to many important physical consequences. We can see them by picking up two pictures in double holography and comparing them with each other. 

\paragraph{Compatibility of causality in AdS/BCFT}~\par
Comparing the bulk picture and the BCFT picture, statement \ref{stat:adsbcft} implies that, in the AdS/BCFT correspondence, the causal structure in the bulk picture is compatible with causality in the BCFT picture. This gives a further consistency check to AdS/BCFT. 

\paragraph{Causality Structure in the Intermediate Picture}~\par
Comparing the bulk picture and the intermediate picture, statement \ref{stat:adsbcft}, \ref{stat:brane}, \ref{stat:across} imply that the effective theory $\CT^{\rm int}_{\Sigma \cup Q}$ in the intermediate picture has a special causal structure to be compatible with the causal structure in the bulk picture. First of all, superluminal information propagation is not allowed within $\Sigma$ or $Q$. However, it is allowed when sending a signal from $Q$ to $\Sigma$ or vice versa. This point will also be confirmed by computing the commutators of quantum fields in section \ref{sec:commutator}. 

The fact that superluminal effects can be observed only when one can access both $Q$ and $\Sigma$ implies that the intermediate theory $\CT^{\rm int}_{\Sigma \cup Q}$ has a nonlocality which is significant at long-range scale. Physical consequences coming from this nonlocality will be discussed in section \ref{sec:EWRC}.

\section{Vacuum Configuration as an Example}\label{sec:geoex}
In this section, we focus on the vacuum configuration of (\ref{eq:bulkaction}) as the most simple example of $\CM$. This configuration is a portion of pure AdS spacetime. After giving the configuration, we will write down the null geodesics in the pure AdS and our configuration. Finally, we will check that statement \ref{stat:adsbcft}, \ref{stat:brane}, and \ref{stat:across} hold in this specific case.

\subsection{Vacuum Configuration and Null Geodesics}\label{sec:geoexglobal}
For simplicity, let us focus on AdS$_3$. The discussion can be straightforwardly extended to higher dimensions. 

\paragraph{Vacuum configuration}~\par
\begin{figure}[H]
    \centering
    \includegraphics[width=14cm]{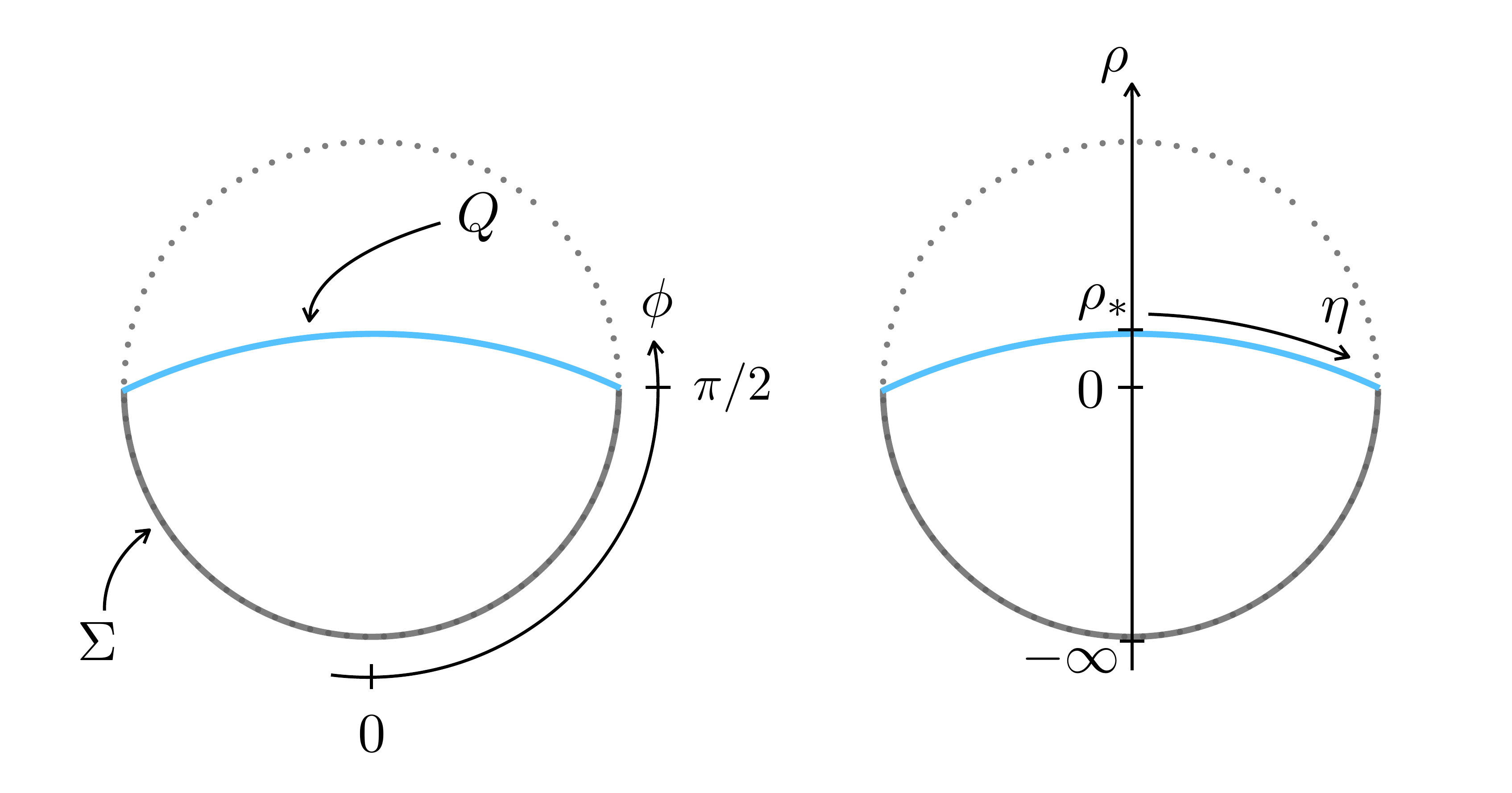}
    \caption{A time slice $\tau = {\rm const.}$ of the vacuum configuration in the $(\tau,r,\phi)$ coordinate (left) and the $(\tau,\eta,\rho)$ coordinate (right) respectively. The asymptotic boundary $\Sigma$ is shown in grey and the end-of-the-world brane $Q$ is shown in blue. The dotted curve shows the asymptotic boundary of global AdS$_3$. The bulk region $\CM$ is surrounded by $\CM = \Sigma \cup Q$. Here, tension $T$ is positive in this figure so that the larger portion of pure AdS$_3$ is identified as the bulk $\CM$.}
    \label{fig:vacuumslice}
\end{figure}

The metric of a pure AdS$_3$ can be written as 
\begin{align}\label{eq:ads3}
    ds^2 = -(r^2+1) d\tau^2 + \frac{dr^2}{r^2+1} + r^2 d\phi^2
\end{align}
in the global coordinate. Here, $-\pi<\phi\leq\pi$. The vacuum configuration of (\ref{eq:bulkaction}) is a portion of pure AdS$_3$ with an end-of-the-world brane $Q$. The location of $Q$ is given by
\begin{align}
    \cos\phi = \frac{T}{\sqrt{1-T^2}}\frac{1}{r}. 
\end{align}
We let $Q$ intersect with the asymptotic boundary at $\phi = \pm \pi/2$. Note that the tension $T$ is restricted to $-1<T<1$. For positive (negative) $T$, the larger (smaller) portion of global AdS$_3$ is identified as the bulk region $\CM$. $Q$ approaches the asymptotic boundary at $|T|\rightarrow1$. See the left panel of figure \ref{fig:vacuumslice}.

Let us introduce another useful coordinate by performing the following transformation. 
\begin{align}
    &r\cos\phi = \sinh \rho, \\
    &1+r^2 = \cosh^2 \rho ~\cosh^2 \eta. 
\end{align}
The metric turns out to be 
\begin{align}
    ds^2 = \cosh^2 \rho \left(-\cosh^2\eta~d\tau^2 + d\eta^2\right) + d\rho^2. 
\end{align}
In this case, the location of the brane is given by 
\begin{align}\label{eq:branelocation}
    \rho = {\rm arctanh}(T) \equiv \rho_{*}. 
\end{align}
Therefore, $Q$ is a pure AdS$_2$. See the right panel of figure \ref{fig:vacuumslice}.

\paragraph{Null geodesics in global AdS$_3$}~\par
\begin{figure}[H]
    \centering
    \includegraphics[width=14cm]{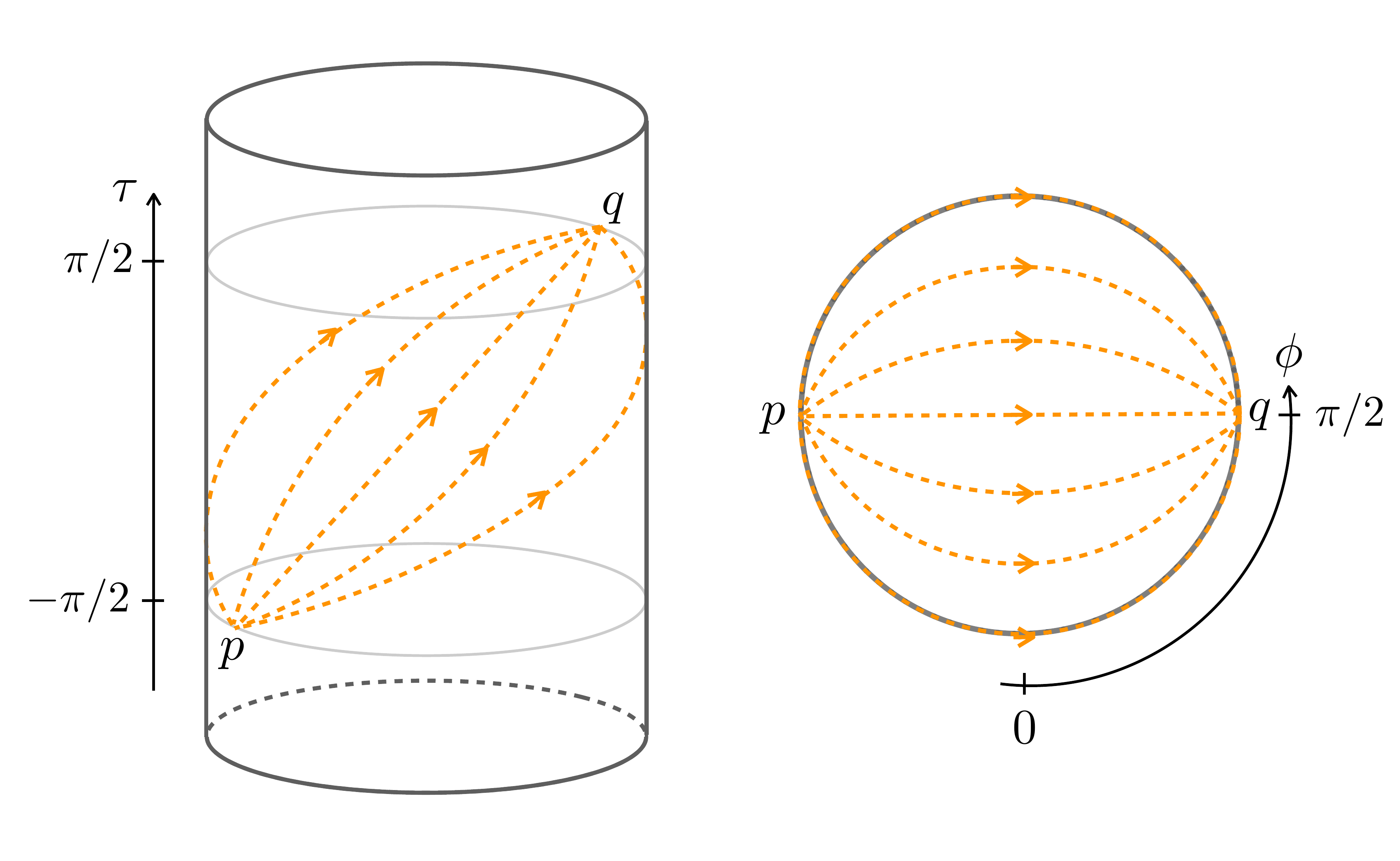}
    \caption{The orange dotted curves in the left panel show some null geodesics launching from $p:(\tau,r,\phi) = (-\pi/2,\infty,-\pi/2)$ in global AdS$_3$. All such geodesics reach $q: (\tau,r,\phi) = (\pi/2,\infty,\pi/2)$. The right panel shows the projection onto a time slice.}
    \label{fig:geoads}
\end{figure}

Let us firstly write down the null geodesics in global AdS$_3$. Since any null geodesic in global AdS$_3$ intersects the asymptotic boundary, it is sufficient to consider those launching from $(\tau,r,\phi) = (-\pi/2,\infty,-\pi/2)$. The location of such a geodesic can be simply expressed in the $(\tau,\eta,\rho)$ coordinate as 
\begin{align}
    &\rho = {\rm const.} \label{eq:nullgeo}~,\\ 
    &\tau = 2\arctan\left(\tanh\frac{\eta}{2}\right). 
\end{align}
It is straightforward to see that all geodesics launching from $(\tau,r,\phi) = (-\pi/2,\infty,-\pi/2)$ reach $(\tau,r,\phi) = (\pi/2,\infty,\pi/2)$. Note that $\rho\rightarrow\pm \infty$ gives the null geodesic on the asymptotic boundary. Therefore, a null geodesic on the asymptotic boundary is also a null geodesic in the bulk in global AdS$_3$. See figure \ref{fig:geoads}. 

As a warm-up, let us check that the no-shortcut statement holds in pure AdS$_3$. Thanks to the symmetries, it is sufficient to check the following statement. 
\begin{customstat}{A.1}\label{stat:adscfttimedelay}
    Consider pure AdS$_3$. Let $p$ be a spacetime point with $p: (\tau,r,\phi) = (\tau_{p},\infty,\phi_{p})$ and $\mathcal{R}$ be an observer localized at $(r,\phi) = (\infty,\phi_{\mathcal{R}})$ on the asymptotic boundary. Consider sending a signal at the speed of light from $p$ to the observer $\mathcal{R}$. Then $\Delta \tau_{\rm asym} = \Delta\tau_{\rm bulk}$, where $\Delta \tau_{\rm asym}$ ($\Delta \tau_{\rm bulk}$) is the required time on the asymptotic boundary (in the bulk) for receiver $\mathcal{R}$ to observe the emitted light ray.
\end{customstat}
To confirm this, we can divide the situations into two cases. If $p$ and $\mathcal{R}$ reside at the antipodal points, i.e. $\phi_{\mathcal{R}} = \phi_p -\pi$, there are infinitely many paths in the bulk which take the same time as the null geodesic on the asymptotic boundary but no shorter path exists. As a result, $\Delta \tau_{\rm asym} = \Delta \tau_{\rm bulk} = \pi$. 
If $\phi_{\mathcal{R}} \neq \phi_p -\pi$, the null geodesic on the asymptotic boundary is the only shortest path. Since this null geodesic is at the same time a bulk null geodesic, $\Delta \tau_{\rm asym} = \Delta \tau_{\rm bulk}$. Therefore, statment \ref{stat:adscfttimedelay} holds in pure AdS$_3$. 

\begin{figure}[H]
    \centering
    \includegraphics[width=16cm]{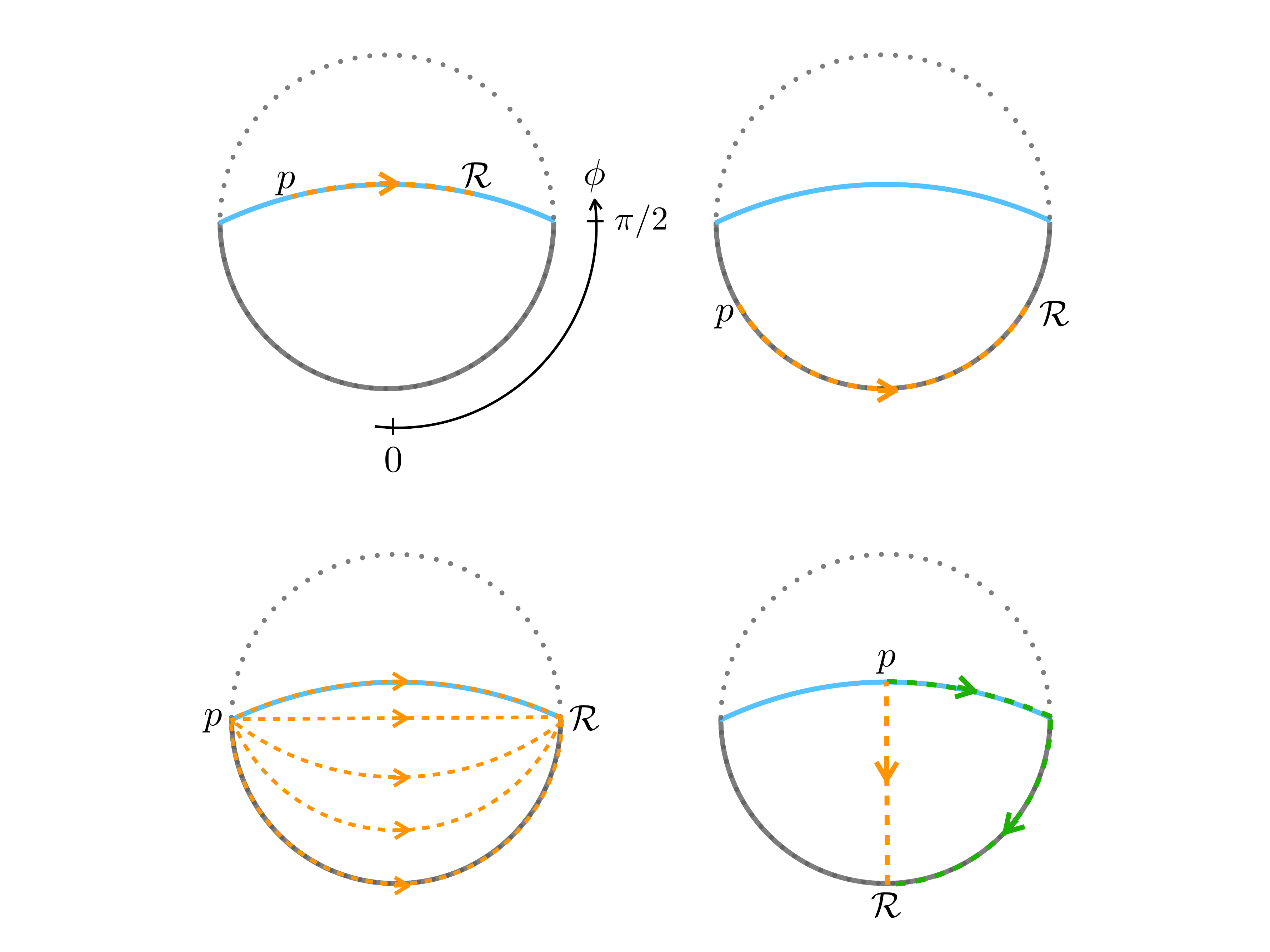}
    \caption{Projection of the null geodesics connecting $p$ and the observer $\mathcal{R}$ on to a time slice for different situations. The orange dotted curves are the the fastest paths in the bulk $\CM$. The green dotted curves are the fastest paths on the boundary $\partial\CM = \Sigma \cup Q$. The shortest path on the boundary $\partial\CM$ is omitted if it coincides that in the bulk $\CM$.}
    \label{fig:shortpath}
\end{figure} 

\paragraph{Null geodesics in the vacuum configuration $\CM$}~\par
Let us then consider the causal structure in the vacuum configuration $\CM$ shown in figure \ref{fig:vacuumslice}. Noticing that the location of a bulk null geodesic (\ref{eq:nullgeo}) and the location of the end-of-the-world brane $Q$ (\ref{eq:branelocation}) have the same form, we find that null geodesics on $Q$ are just given by null geodesics in the bulk $\CM$ with $\rho = \rho_*$. 

Let us then check that statement \ref{stat:adsbcft}, \ref{stat:brane} and \ref{stat:across} hold in this case. Again, thanks to the symmetries, it is sufficient to check the following statements respectively. 
\begin{customstat}{B.1}\label{stat:adsbcfttimedelay}
    Let $p$ be a spacetime point with $p: (\tau,\eta,\rho) = (\tau_p,\eta_{p},-\infty)$ and $\mathcal{R}$ be an observer located at $(\eta,\rho) = (\eta_{\mathcal{R}},-\infty)$ on the asymptotic boundary $\Sigma$. Consider sending a signal at the speed of light from $p$ to $\mathcal{R}$. Then $\Delta \tau_{\Sigma} = \Delta\tau_{\CM}$, where $\Delta \tau_{\Sigma}$ ($\Delta \tau_{\CM}$) is the required time on the asymptotic boundary $\Sigma$ (in the bulk $\CM$) for $\mathcal{R}$ to receive the signal.
\end{customstat}
\begin{customstat}{C.1}\label{stat:branetimedelay}
    Let $p$ be a spacetime point with $p: (\tau,\eta,\rho) = (\tau_p,\eta_{p},\rho_*)$ and $\mathcal{R}$ be an observer located at $(\eta,\rho) = (\eta_{\mathcal{R}},\rho_*)$ on the end-of-the-world brane $Q$. Consider sending a signal at the speed of light from $p$ to $\mathcal{R}$. Then $\Delta \tau_{Q} = \Delta\tau_{\CM}$, where $\Delta \tau_{\Sigma}$ ($\Delta \tau_{\CM}$) is the required time on the brane $Q$ (in the bulk $\CM$).
\end{customstat}
\begin{customstat}{D.1}\label{stat:acrosstimedelay}
    Let $p$ be a spacetime point with $p: (\tau,\eta,\rho) = (\tau_p,\eta_{p},\rho_*)$ on the end-of-the-world brane $Q$     and $\mathcal{R}$ be an observer located at $(\eta,\rho) = (\eta_{\mathcal{R}},-\infty)$ on the asymptotic boundary $\Sigma$. Consider sending a signal at the speed of light from $p$ to $\mathcal{R}$. Then $\Delta \tau_{\Sigma\cup Q} > \Delta\tau_{\CM}$, where $\Delta \tau_{\Sigma\cup Q}$ ($\Delta \tau_{\CM}$) is the required time on $\Sigma\cup Q$ (in the bulk $\CM$).
\end{customstat}

To check these three statements, we consider the following four cases shown in figure \ref{fig:shortpath}. 
\begin{itemize}
    \item (The upper left panel of figure \ref{fig:shortpath}) If $p$ and $\mathcal{R}$ are both on the brane $Q$ but not antipodal points, the null geodesic on the brane $Q$ is the only shortest path, and it is a null geodesic in the bulk $\CM$ at the same time. As a result, $\Delta \tau_{Q} = \Delta\tau_{\CM}$. 
    \item (The upper right panel of figure \ref{fig:shortpath}) $p$ and $\mathcal{R}$ are both on the asymptotic boundary $\Sigma$ but not antipodal points. Similar to the previous case, $\Delta \tau_{\Sigma} = \Delta\tau_{\CM}$. 
    \item (The lower left panel of figure \ref{fig:shortpath}) $p$ and $\mathcal{R}$ are antipodal points. In this case, $p$ and $\mathcal{R}$ are on both $\Sigma$ and $Q$. As we have seen above, there are infinitely many paths in the bulk which take the same time as the null geodesic on the asymptotic boundary but no shorter path exists. As a result, $\Delta \tau_{Q} = \Delta \tau_{\Sigma} = \Delta\tau_{\CM}$.
    \item (The lower right panel of figure \ref{fig:shortpath}) $p$ is on $Q\backslash\partial Q$ and $\mathcal{R}$ is on $\Sigma\backslash\partial\Sigma$. As one can see from the figure, there is a shortcut in the bulk. We can prove the existence of such a shortcut by applying the proposition 4.5.10 of \cite{HE73}, which states that two points joined by a causal curve $\gamma$ which is not a null geodesic, can also be joined by a time-like curve. In this statement, ``a causal curve $\gamma$ which is not a null geodesic" means that either the acceleration vector of $\gamma$ is non-zero and not parallel to its tangent vector on some open interval, or $\gamma$ has some point on which the tangent vector is discontinuous. Consider a null geodesic on $\p\CM$ connecting $p$ and $\mathcal{R}$ (the green dotted curve of lower right panel). Since there is a co-dimension two defect $\partial \Sigma = \partial Q$, the tangent vector in the bulk is discontinuous here. Then proposition 4.5.10 of \cite{HE73} tells us that these two points at $p$ and $\mathcal{R}$ can be joined by a time-like curve in the bulk. Therefore, we can use the null geodesic in the bulk to send information faster than the geodesic on $\p\CM$. As a result, $\Delta \tau_{\Sigma\cup Q} > \Delta\tau_{\CM}$.
\end{itemize}
These four cases cover all possible situations. Therefore, we have confirmed that the statement \ref{stat:adsbcfttimedelay}, \ref{stat:branetimedelay} and \ref{stat:acrosstimedelay} hold. 

Note that in the confirmation of statement \ref{stat:acrosstimedelay}, we did not use any property of the specific configuration, except for the existence of a codimension-2 defect at $\partial\Sigma = \partial Q$. Since double holography naturally introduces such a codimension-2 defect, causality violation in the intermediate picture universally occurs in more general configurations.

\paragraph{Comments on higher dimensions}~\par
The results above can be straightforwardly extended to higher dimensions. The metric of a pure AdS$_{d+1}$ can be written as 
\begin{align}\label{eq:adshigher}
    ds^2 = -(r^2+1) d\tau^2 + \frac{dr^2}{r^2+1} + r^2 d\bm{\Omega}^2
\end{align}
where $d\bm{\Omega}^2$ is the line element of $S^{d-1}$. All the discussions in AdS$_3$ can be similarly repeated in AdS$_{d+1}$ by regarding the $S^1$ parameterized by $\phi$ in (\ref{eq:ads3}) as a equator of $S^{d-1}$.

\subsection{Zooming in to the Poincar\'{e} Patch}

Although we have already checked the statements hold for the vacuum configuration of (\ref{eq:bulkaction}), one deficiency in our discussion so far is that we did not evaluate how faster a signal can be sent in the bulk than on the boundary in statement \ref{stat:acrosstimedelay}. 

\begin{figure}[H]
    \centering
    \includegraphics[width=7cm]{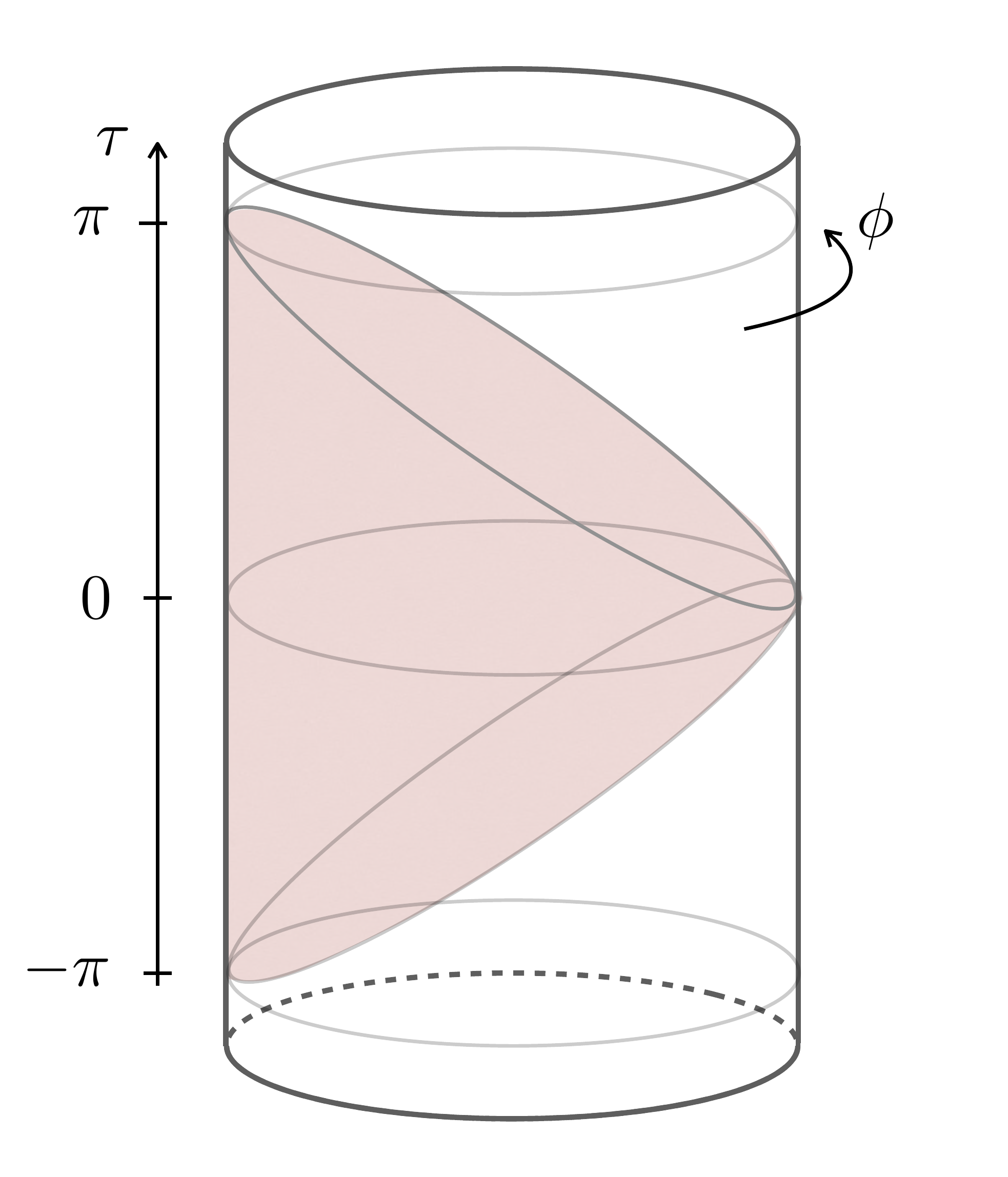}
    \caption{Poincar\'e patch in global AdS$_3$. The shaded region shows the Poincar\'e patch.}
    \label{fig:poincarepatch}
\end{figure}

In this subsection, we restrict our attention to the Poincar\'e patch. See figure \ref{fig:poincarepatch}. The Poincar\'e patch does not cover the whole global AdS spacetime, so it cannot be used to discuss the global causal structure. However, many analytic calculations become extremely simple in it.
We will give the time advance by explicitly writing down the null geodesics in the Poincar\'e patch. 

Performing the following coordinate transformation to global AdS$_{d+1}$ (\ref{eq:adshigher})
\begin{align}
\begin{aligned}
    &\sinh\rho = r \Omega^{d-1}~, & y &= \frac{\sqrt{1 + (r\Omega^{d-1})^2}}{\sqrt{1+r^2}\cos \tau - r \Omega^{d}}~,\\
    &t = \frac{\sqrt{1+r^2} \sin \tau}{\sqrt{1+r^2}\cos \tau - r \Omega^{d}}~, & \xi^{i} &= \frac{r \Omega^{i}}{\sqrt{1+r^2}\cos \tau - r \Omega^{d}}~,
\end{aligned}
\end{align}
we get a new coordinate $(t,y,\rho)$ which describes the Poincar\'e patch. Here, the unit vector $\bm{\Omega} = (\Omega^1,\dots,\Omega^{d})$ denotes a point on the $(d-1)$-dimensional sphere $S^{d-1}$ and $i = 1,\dots,d-2$. The resulting metric is
\begin{align}\label{eq:corrhoty}
	ds^2 = d\rho^2 + \cosh^2\rho \left(\frac{dy^2 - dt^2 + d\bm{\xi}^2}{y^2}\right)~.
\end{align}
The end-of-the-world brane $Q$ locates at 
\begin{align}
    \rho = {\rm arctanh}\left(\frac{T}{d-1}\right) \equiv \rho_{*}. 
\end{align}
and the asymptotic boundary $\Sigma$ locates at $\rho = -\infty$. This is related to the ordinary Poincar\'e metric 
\begin{align}\label{eq:Poincarecor3}
	ds^2 = \frac{dz^2 - dt^2 + d\bm{x}^2}{z^2}
\end{align}
by coordinate transformation 
\begin{align}
\begin{aligned}
	&z = \frac{y}{\cosh \rho}~, & x^{d-1} &= y \tanh \rho~,\\
	&x^{i} = \xi^i ~,& (i &= 1,\dots,d-2)~.
\end{aligned}
\end{align}

\begin{figure}[t]
    \centering
    \includegraphics[width=14cm]{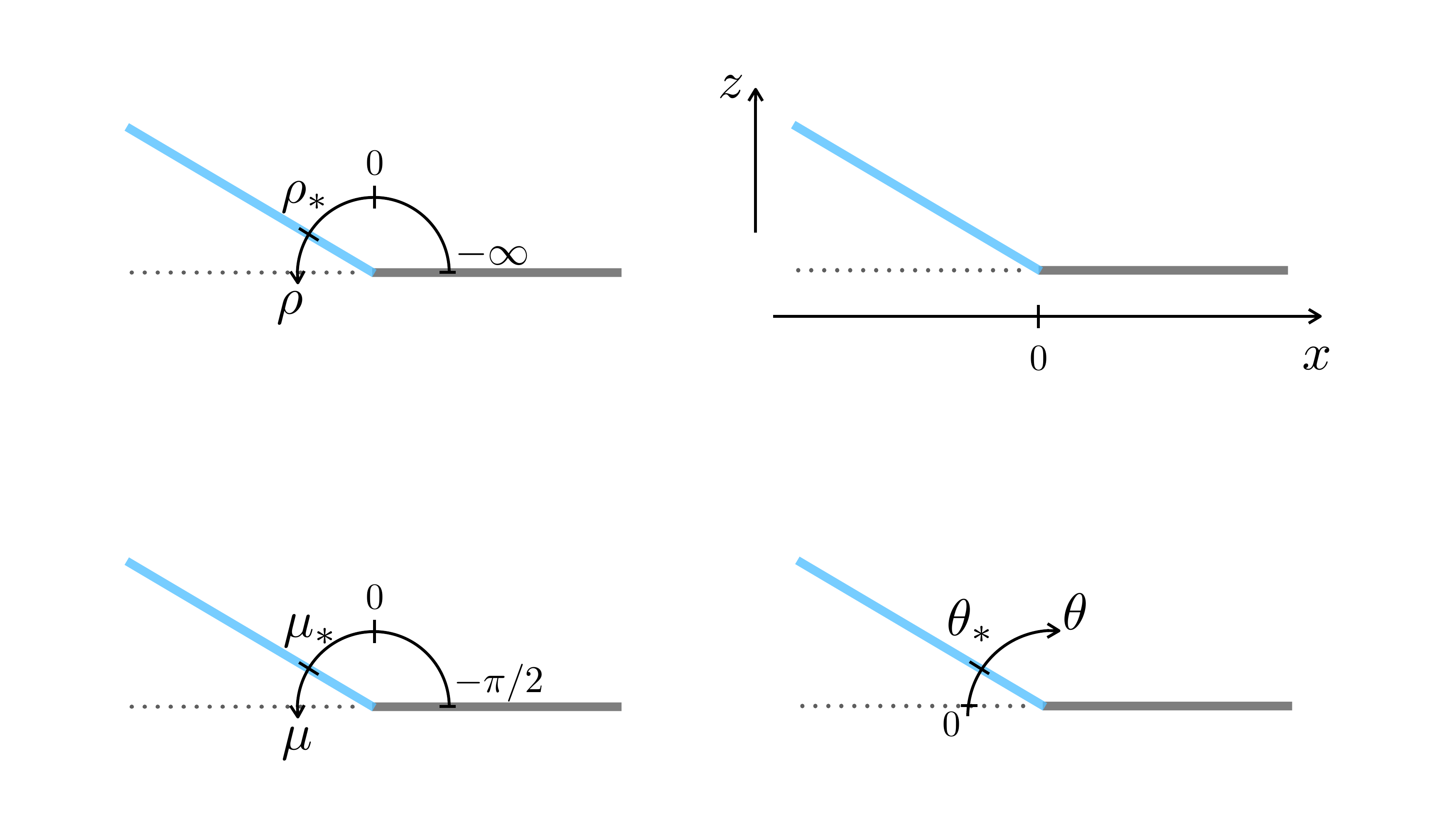}
    \caption{Coordinates describing the Poincar\'e patch shown in a time slice of AdS$_3$. The asymptotic boundary $\Sigma$ is shown in grey and the brane $Q$ is shown in blue.}
    \label{fig:poincoor}
\end{figure}

Let us introduce one more coordinate, which is important when calculating the correlation function on the brane by holography. With a new parameter $\mu$ satisfying 
\begin{align}
    \sinh \rho = \tan \mu, 
\end{align}
the metric becomes
\begin{align}\label{eq:metricFG}
    ds^2 = \frac{d\mu^2}{\cos^2\mu} + \frac{1}{\cos^2\mu}\left(\frac{dy^2 - dt^2 + d\bm{\xi}^2}{y^2}\right)~.
\end{align}
When $\theta \equiv \pi/2 - \mu \ll1$, the metric can be expanded as
\begin{align}\label{eq:metricFG2}
    ds^2 \sim 4\frac{d{\theta}^2}{{\theta}^2} + \frac{4}{{\theta}^2}\left(\frac{dy^2 - dt^2 + d\bm{\xi}^2}{y^2}\right)~.
\end{align}
These coordinates are shown on a time slice of AdS$_3$ in figure \ref{fig:poincoor}. 

Now we calculate the time advance in statement \ref{stat:acrosstimedelay} by considering sending a signal at the speed of light from spacetime point $p \in Q$ to an observer $\mathcal{R}$ residing on $\Sigma$. Let $\Delta t_{\Sigma \cup Q}$ and $\Delta t_{\CM}$ be the time taken by the fastest null geodesic to travel from $p$ to $\mathcal{R}$ through the $\partial \mathcal{M}$ and through $\CM \backslash \partial \CM$, respectively. Here, $\Delta t$ is measured at asymptotic infinity in the $t$ of Poincar\'e coordinate.

We would like to firstly calculate $\Delta t_{\Sigma \cup Q}$. We use $(t,y,\bm{\xi})$ to denote the coordinate on $\Sigma\cup Q$, let the $y>0$ half be $\Sigma$, and let the $y<0$ half be $Q$. 
Consider sending a signal from $p: (t,y,\bm{\xi}) = (0,-y_p,\bm{0})$ on $Q$ to a receiver $\mathcal{R}$ located at $(y,\bm{\xi}) = (y_{\mathcal{R}}, \bm{\xi}_{\mathcal{R}})$ on $\Sigma$. Now the light ray travels through $Q$ and $\Sigma$ with different geometries. Accordingly, the geodesic on $\mathcal{\p \CM}$ can be divided into two pieces: one on $Q$ and one on $\Sigma$. Time needed for the light ray to reach $\mathcal{R}$ is the sum of the time for it to travel through $Q$ from $y = -y_p$ to $y = 0$ and the time for it to travel through $\Sigma$ from $y = 0$ to $y = y_{\mathcal{R}}$:
 \begin{align}\label{eq:tbrane}
 	\Delta t_{\Sigma\cup Q}^2 &= (y_p + y_{\mathcal{R}})^2 + \bm{\xi}_{\mathcal{R}}^2.
 \end{align}
 
Let us then consider sending information through the bulk. Using the bulk Poincar\'e coordinate $(t,z,\bm{\xi},x^{d-1})$, $p$ is given by $(t,z,\bm{\xi},x^{d-1}) = (0, y_p\cos \mu_*,\bm{0},-y_p\sin\mu_*)$ and $\mathcal{R}$ located at $(z,\bm{\xi},x^{d-1}) = (0, \bm{\xi}_{\mathcal{R}}, y_{\mathcal{R}})$. Using the expression for null geodesic (see appendix \ref{sec:AdSgeodesic} for details),
 \begin{align}\label{eq:causalbulk}
 	\Delta t_{\CM}^2 = y_p^2 + y_{\mathcal{R}}^2 + 2 y_p y_{\mathcal{R}}\sin\mu_* + \bm{\xi}_{\mathcal{R}}^2 = (y_p + y_{\mathcal{R}})^2 +\bm{\xi}_{\mathcal{R}}^2 - 2 y_p y_{\mathcal{R}} (1 -  \sin\mu_*)~.
 \end{align}
We can again confirm that
\begin{align}
    \Delta t_{\CM} < \Delta t_{\Sigma\cup Q}~. 
\end{align}
The (squared) time saved by taking a shortcut in the bulk is $\Delta t_{\Sigma\cup Q}^2 - \Delta t_{\CM}^2 = 2y_p y_{\mathcal{R}} (1 - \sin\mu_*)$.

\section{Compatibility of Causality in AdS/BCFT}\label{sec:causalitySigma}
In this section, we show that, under the same assumptions as the Gao-wald theorem \ref{thm:GW} plus one more reasonable assumption on $\Sigma$, any on-shell configuration of the bulk action (\ref{eq:bulkaction}) satisfies statement \ref{stat:adsbcft}. This guarantees that the bulk picture obtained from the AdS/BCFT construction is compatible with causality in the BCFT picture. In the end of this section, we will comment on CFTs defined on a general $\Sigma$ with possibly non-conformal boundaries as well as their gravity duals. 

\subsection{An Assumption on \texorpdfstring{$\Sigma$}{Sigma}}
\label{sec:assumption}
In the discussions below, we make the following assumption. 
\begin{assumption}\label{assum:Sigma}
    $(\Sigma,\gamma_{ij})$ can be mapped to $\mathbb{R}^{1,d-2}\times[0,\infty)$ or $\mathbb{R}^{1,d-2}\times[0,l]$ via a Weyl transformation
    \begin{align}\label{eq:weylSigma}
        \gamma_{ij}(x) \rightarrow \Omega^2(x)\gamma_{ij}(x),
    \end{align}
    where $l$ is a constant. 
\end{assumption}

This assumption restricts the geometry to a very narrow class. Indeed, we will see that this assumption, though sufficient, is not a necessary condition to prove statement \ref{stat:adsbcft} later. However, let us explain why this assumption is made. 

\paragraph{BCFT vs. Locally CFT with Boundaries}~\par
First of all, let us distinguish two different concepts: boundary conformal field theory (BCFT) and locally CFT with boundaries (LCFTB). 

BCFT is used to refer a theory which is not only defined on a manifold with boundaries, but also maximally preserves the conformal symmetry so that CFT techniques including conformal Ward identity can be applied for its analysis \cite{Cardy04}. Therefore, both $\Sigma$ and the boundary condition on $\partial \Sigma$ are highly restricted. 

On the other hand, there exist manifolds which are not compatible with conformal boundary conditions but can still let a locally CFT\footnote{Here, a locally CFT is a theory which has the same Lagrangian density with a CFT.} live on them with no contradiction. We will call a locally CFT living on a manifold with boundaries a locally CFT with boundaries (LCFTB). By definition, BCFTs are LCFTB while the inverse is not true. 

To our knowledge, it is still unsolved to give the most general class of manifolds which are compatible with a BCFT. However, Lorentzian BCFT with time-like boundaries is usually discussed in $\mathbb{R}^{1,d-2}\times[0,\infty)$ or $\mathbb{R}^{1,d-2}\times[0,l]$ and should be straightforwardly generalized to any manifold satisfying assumption \ref{assum:Sigma}. For example, the corresponding boundary condition for a BCFT defined on such a manifold is examined for $d=2$ in \cite{AKSTW21}. Therefore, we would like to just make assumption \ref{assum:Sigma} to restrict our discussions to a class of $\Sigma$ which are compatible with a BCFT. 

\subsection{Showing Statement \ref{stat:adsbcft} for Generic Configurations}
\label{sec:proofadsbcft}
Now we proceed to prove statement \ref{stat:adsbcft}. We have already seen in section \ref{sec:geoex} that this is satisfied when the bulk $\CM$ is a vacuum configuration given by a portion of global AdS. Therefore, in the following, we would like to focus on the case in which the bulk geometry is away from a vacuum configuration by adding matters and gravitational excitations in it. 

We would like to assume that $\CM$ satisfies the four conditions given in the Gao-Wald theorem \ref{thm:GW} as well as assumption \ref{assum:Sigma}. 

Consider the asymptotic boundary $\Sigma$. The region which can receive a signal through the bulk $\CM$ from $p\in\Sigma$ is $A_{\Sigma}(p,\CM)$, and the region which can receive a signal through $\Sigma$ is its causal future $J^{+}_{\Sigma}(p)$. To prove statement \ref{stat:adsbcft}, it is sufficient to show
\begin{align}\label{eq:adsbcftcal}
    {\partial} A_{\Sigma}(p,\CM) ~\subseteq~ J_{{\Sigma}}^{+}(p) . 
\end{align}

Let us firstly see how far we can go without assumption \ref{assum:Sigma}. First of all, we can perform an embedding $\Sigma \subset \Sigma'$, $\CM \subset \CM'$ which satisfy 
\begin{itemize}
    \item ${\rm dim} (\Sigma') = {\rm dim} (\Sigma)$ and ${\rm dim} (\CM') = {\rm dim} (\CM)$.
    \item $\CM'$ is a spacetime with a time-like asymptotic boundary $\CM'=\Sigma'$. 
    \item $\CM'$ satisfies the four conditions given in the Gao-Wald theorem \ref{thm:GW}. 
\end{itemize}
Then the Gao-Wald theorem is applicable to $\CM'$. A consequence is
\begin{align}\label{eq:gaowald}
    A_{\Sigma'}(p,\CM') \cup \partial A_{\Sigma'}(p,\CM') &\subseteq J_{{\Sigma'}}^{+}(p), 
\end{align}
for any point $p\in\Sigma'$. See \cite{GW00} for details.

Let us try to derive \eqref{eq:adsbcftcal} from \eqref{eq:gaowald}.
The way of embedding implies 
\begin{align}
    A_{\Sigma}(p,\CM) \subseteq \left(A_{\Sigma'}(p,\CM') \cap \Sigma \right). 
\end{align}
Then we have
\begin{align}\label{eq:whatGWsay}
    \partial A_{\Sigma}(p,\CM) 
    &\subseteq \left(A_{\Sigma'}(p,\CM') \cap \Sigma \right) ~\cup~ {\partial}\left(A_{\Sigma'}(p,\CM') \cap \Sigma \right) \nonumber\\
    &= \left(A_{\Sigma'}(p,\CM') \cap \Sigma \right) ~\cup~ \left({\partial}A_{\Sigma'}(p,\CM') \cap \Sigma \right) ~\cup~ \left(A_{\Sigma'}(p,\CM') \cap {\partial}\Sigma \right) \nonumber\\
    &= \left[A_{\Sigma'}(p,\CM') \cap (\Sigma \cup \partial\Sigma) \right] ~\cup~ \left({\partial} A_{\Sigma'}(p,\CM') \cap \Sigma \right) \nonumber\\
    &= \left(A_{\Sigma'}(p,\CM') \cap \Sigma \right) ~\cup~ \left({\partial} A_{\Sigma'}(p,\CM') \cap \Sigma \right) \nonumber\\
    &= \left(A_{\Sigma'}(p,\CM') \cup \partial A_{\Sigma'}(p,\CM') \right) ~\cap~ \Sigma \nonumber\\
    &\subseteq \left(J_{{\Sigma'}}^{+}(p)\cap \Sigma \right).
\end{align}
Here, we used the property
\begin{align}
    R_1 \subseteq R_2~~~\Longrightarrow~~~\partial R_1 \subseteq R_2 \cup \partial R_2
\end{align}
in the first line, the identity relation
\begin{align}
    \partial (R_1 \cap R_2) = (\partial R_1 \cap R_2) \cup (R_1 \cap \partial R_2)
\end{align}
in the second line, and 
\begin{align}
    (X\cap Z) \cup (Y\cap Z) = (X\cup Y) \cap Z
\end{align}
in the third line and the fifth line. We also used $\partial \Sigma \subset \Sigma$ in the fourth line and \eqref{eq:gaowald} in the last line.

Since $J_{{\Sigma}}^{+}(p) \subseteq (J_{{\Sigma'}}^{+}(p)\cap \Sigma)$,
in order to derive \eqref{eq:adsbcftcal} from \eqref{eq:whatGWsay}, we need to show $J_{{\Sigma}}^{+}(p) = (J_{{\Sigma'}}^{+}(p)\cap \Sigma)$. However, this does not necessarily hold for general $\Sigma$. This is because two points $p,q\in\Sigma$ which can be causally connected through $\Sigma'$ is not always causally connected through $\Sigma$.

So far, we have not used assumption \ref{assum:Sigma} yet. Let us then see how assumption \ref{assum:Sigma} can rescue the situation.

Consider $p,q \in \partial \Sigma$ connected by a null geodesic $\gamma$ on $\partial \Sigma$, whose tangent vector is given by $u$. The Gauss-Weingarten equation relates the covariant derivative $\tilde{D}_{\tilde{m}}$ on $\partial\Sigma$ and the covariant derivative $D_{i}$ on $\Sigma$ by
    \begin{align}
        u^i D_i u^j = (u^{\tilde{m}} \tilde{D}_{\tilde{m}} u^{\tilde{n}}) {e^j}_{\tilde{n}} - k_{\tilde{m}\tilde{n}}u^{\tilde{m}} u^{\tilde{n}} n^j,
    \end{align}
where $n^{j}$ is the normal vector to $\partial\Sigma$, pointing toward the ambient space. For a geodesic on $\partial\Sigma$
    \begin{align}
        u^{\tilde{m}} \tilde{D}_{\tilde{m}} u^{\tilde{n}} = 0~,
    \end{align}
and hence
    \begin{align}
        u^i D_i u^j = - k_{\tilde{m}\tilde{n}}u^{\tilde{m}} u^{\tilde{n}} n^j.
    \end{align}
This will be a null geodesic in $\Sigma$ if
    \begin{align}\label{eq:extrinsic0}
        k_{\tilde{m}\tilde{n}}u^{\tilde{m}} u^{\tilde{n}} = 0~.
    \end{align}
    
Then we will show that \eqref{eq:extrinsic0} holds under assumption \ref{assum:Sigma}. For $\mathbb{R}^{1,d-2}\times[0,\infty)$ or $\mathbb{R}^{1,d-2}\times[0,l]$, this is trivial. Now consider the Weyl transformation \eqref{eq:weylSigma}. Then the extrinsic curvature $k_{\tilde{m}\tilde{n}}$ transforms as
    \begin{align}
    k_{\tilde{m}\tilde{n}} \to k'_{\tilde{m}\tilde{n}} = \frac{1}{\Omega} k_{\tilde{m}\tilde{n}} + (n^i \partial_i \Omega)h_{\tilde{m}\tilde{n}} = (n^i \partial_i \Omega)h_{\tilde{m}\tilde{n}}~.
    \end{align}
Here, $k'_{\tilde{m}\tilde{n}}$ is the extrinsic curvature defined on the manifold after the weyl transformation and $h_{\tilde{m}\tilde{n}}$ is the induced metric on $\partial \Sigma$. It turns out that the extrinsic curvature is proportional to the induced metric. Therefore, \eqref{eq:extrinsic0} holds, i.e. any null geodesic on $\partial\Sigma$ is also a null geodesic on $\Sigma$. 

Suppose two points $p,q\in\Sigma$ are connected in $\Sigma'$ by a causal curve $\lambda_{p,q}$ which has an overlap with $\Sigma'\backslash\Sigma$, then there exist $p',q'\in\partial\Sigma$ such that 
\begin{align}
    \lambda_{p,q} = \lambda_{p,p'} + \lambda_{p',q'}, + \lambda_{q',q},
\end{align}
and 
\begin{align}
    \lambda_{p,p'}, \lambda_{q,q'} \in \Sigma.
\end{align}
Further more, there exists a causal curve $\lambda'_{p',q'}\in\partial\Sigma$ which connects $p'$ and $q'$, otherwise $q'$ would be outside of the light cone of $p'$ and lead to a contradiction with the fact that any null geodesic on $\partial\Sigma$ is also a null geodesic on $\Sigma$. In one word, any two points $p,q\in\Sigma$ which are causally connected through $\Sigma'$ must be causally connected through $\Sigma$, and hence 
\begin{align}\label{eq:whatassumsay}
    J_{{\Sigma}}^{+}(p) = (J_{{\Sigma'}}^{+}(p)\cap \Sigma).
\end{align}

Combining \eqref{eq:whatassumsay} with \eqref{eq:whatGWsay}, we obtain \eqref{eq:adsbcftcal}. Therefore, statement \ref{stat:adsbcft} is shown. 

\subsection{Comments on More General \texorpdfstring{$\Sigma$}{Sigma}}\label{sec:genericSigma}
We have seen that causality compatibility between the bulk picture and the BCFT picture can be proven for $\Sigma$ satisfying assumption \ref{assum:Sigma}. Let us exclude this assumption here and see what happens. Inspired by the proof given in section \ref{sec:proofadsbcft}, it is convenient to consider the following two cases. 
\begin{itemize}
    \item Case 1: For any null vector $u^{\tilde{m}}$ on $\partial \Sigma$, $k_{\tilde{m}\tilde{n}}u^{\tilde{m}} u^{\tilde{n}} \leq 0$. 
    \item Case 2: There exists a null vector $u^{\tilde{m}}$ on $\partial \Sigma$ such that $k_{\tilde{m}\tilde{n}}u^{\tilde{m}} u^{\tilde{n}} > 0$.
\end{itemize}
As we have already seen in section \ref{sec:proofadsbcft}, $\Sigma$ which satisfies assumption \ref{assum:Sigma} saturates the condition in case 1. 

\begin{figure}[H]
    \centering
    \includegraphics[width=5cm]{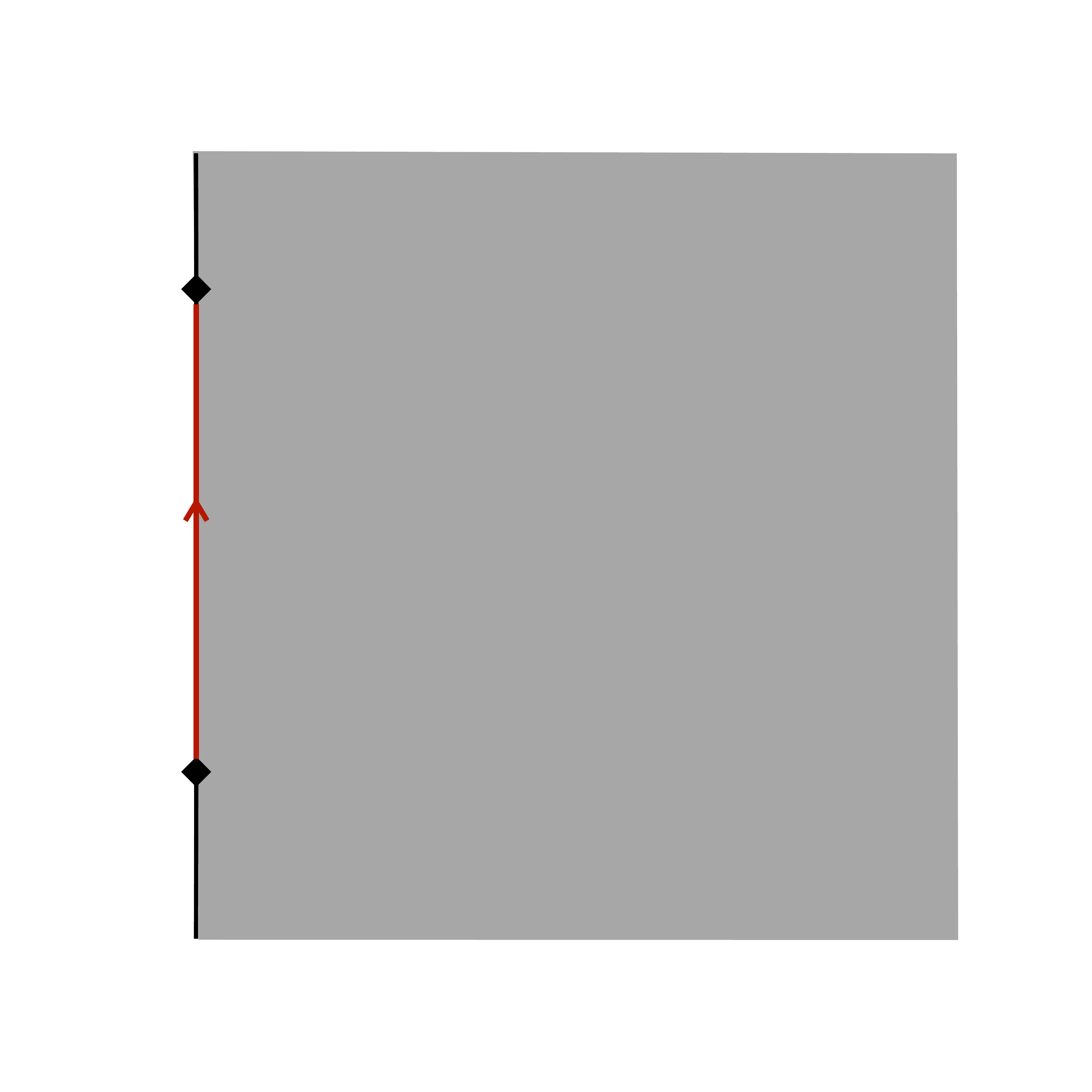}
    \includegraphics[width=5cm]{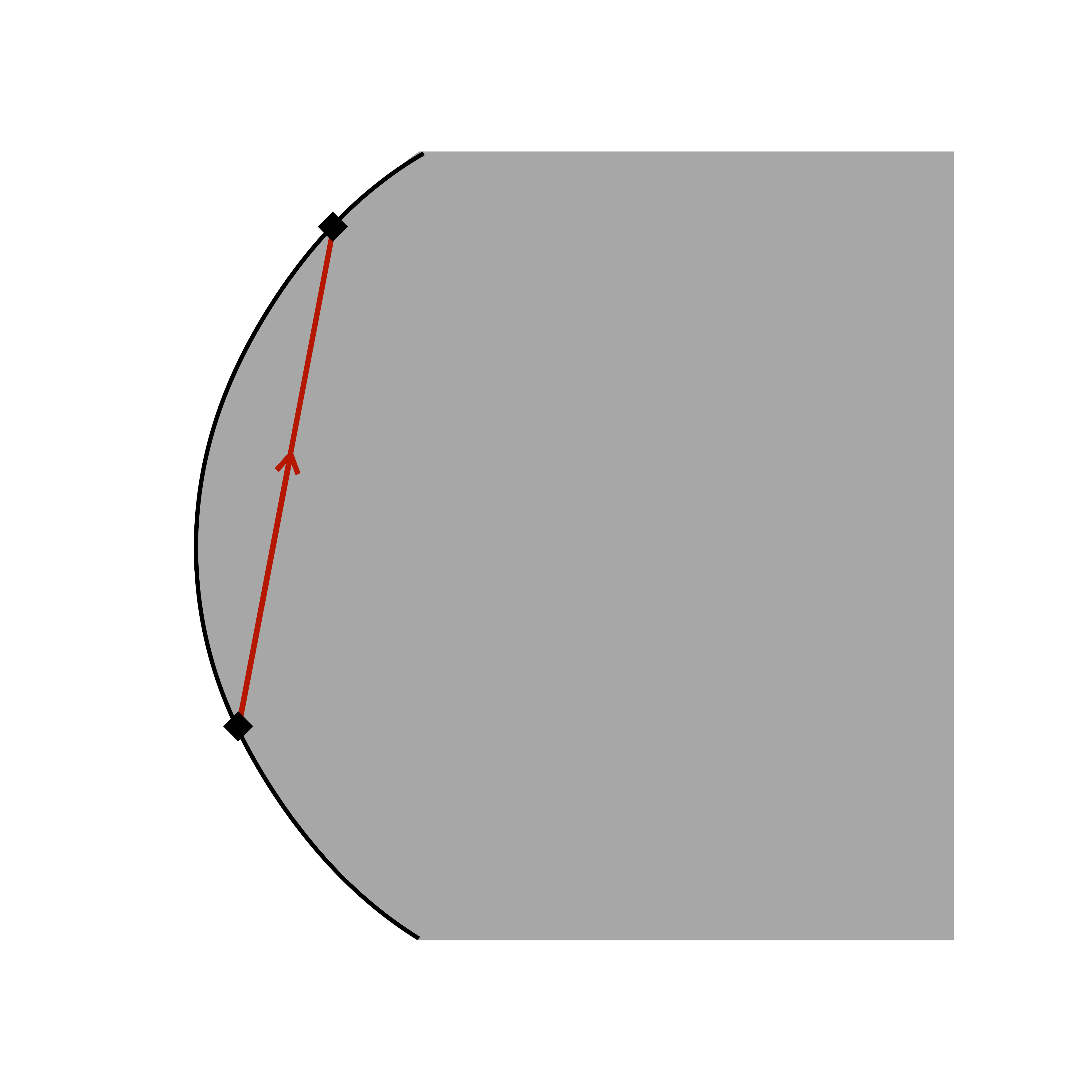}
    \includegraphics[width=5cm]{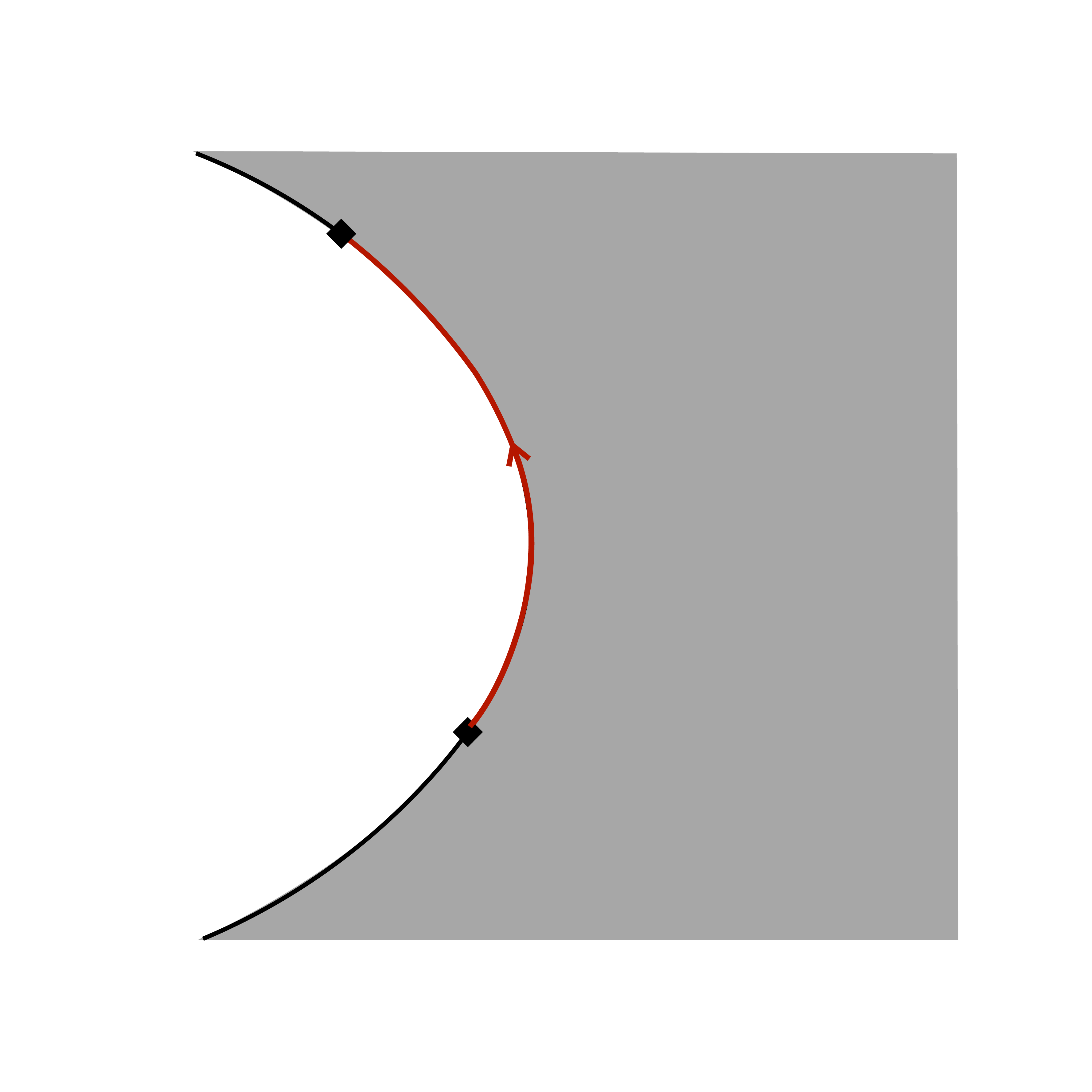}
    \caption{Time slices of (2+1)D static $\Sigma$ with different extrinsic curvatures on $\partial\Sigma$. The left, middle, right figure show $k_{\tilde{m}\tilde{n}}u^{\tilde{m}} u^{\tilde{n}} = 0$, $k_{\tilde{m}\tilde{n}}u^{\tilde{m}} u^{\tilde{n}} < 0$ and $k_{\tilde{m}\tilde{n}}u^{\tilde{m}} u^{\tilde{n}} > 0$, respectively. Consider sending a signal from $\partial\Sigma$ to another location on $\partial\Sigma$. The shortest paths are shown by red arrows. When $k_{\tilde{m}\tilde{n}}u^{\tilde{m}} u^{\tilde{n}} = 0$, the shortest path is a null geodesic on $\partial\Sigma$ which is also a null geodesic in $\Sigma$. When $k_{\tilde{m}\tilde{n}}u^{\tilde{m}} u^{\tilde{n}} < 0$, the shortest path is a null geodesic in the ambient $\Sigma\backslash\partial\Sigma$. When $k_{\tilde{m}\tilde{n}}u^{\tilde{m}} u^{\tilde{n}} > 0$, the shortest path is a null geodesic on $\partial\Sigma$ which is not a geodesic in $\Sigma$.}
    \label{fig:sigma_ex}
\end{figure}

In general, the causal structure of $\Sigma$ is determined by null geodesics in $\Sigma$ and those on $\partial \Sigma$. Figure \ref{fig:sigma_ex} shows one simple example of $\Sigma$ for each of $k_{\tilde{m}\tilde{n}}u^{\tilde{m}} u^{\tilde{n}} = 0$, $k_{\tilde{m}\tilde{n}}u^{\tilde{m}} u^{\tilde{n}} < 0$ and $k_{\tilde{m}\tilde{n}}u^{\tilde{m}} u^{\tilde{n}} > 0$.

In case 1, one can show that any two points on any null geodesic on $\partial\Sigma$ can be connected by a time-like or null geodesic in $\Sigma$. Therefore, the causal structure of $\Sigma$ is fully determined by null geodesics in $\Sigma$. As a result, one can simply apply the Gao-Wald theorem to show statement \ref{stat:adsbcft} as in section \ref{sec:proofadsbcft}.  

In case 2, however, there exist two points on $\partial\Sigma$ which can be connected by a null geodesic on $\partial\Sigma$ but not by any time-like or null geodesic in $\Sigma$. In this case, Gao-Wald theorem is not directly applicable and more inputs may be necessary to prove statement \ref{stat:adsbcft}. 

\paragraph{Towards Gravity Dual for Locally CFT with General Boundaries}~\par 
The AdS/BCFT construction, i.e. constructing a gravity dual for a BCFT defined on a given $\Sigma$ via finding an on-shell configuration of (\ref{eq:bulkaction}) satisfying the Neumann boundary condition (\ref{eq:Neumann}) on the end-of-the-world brane $Q$, is originally proposed for $\Sigma$ compatible with a BCFT \cite{Takayanagi11,FTT11}. A natural question is whether this construction can be extended to $\Sigma$ with general $\partial\Sigma$. 

The answer is no. It is pointed out in \cite{McKeown17} that, assuming $\Sigma$ is a portion of $\mathbb{R}^d$ ($d\geq3$) with one connected piece of boundary, the only configurations of $\partial\Sigma$ compatible with the above construction are $\mathbb{R}^{d-1}$ and $S^{d-1}$. The reason is because the Neumann boundary condition (\ref{eq:Neumann}) is too strong. 

Therefore, the Neumann boundary condition (\ref{eq:Neumann}) should be modified to be compatible with more general $\partial\Sigma$. This is reasonable since (\ref{eq:Neumann}) is originally proposed to maximally preserve the corresponding symmetries in the bulk, and there is no reason to expect it for less symmetric cases. There are two totally different recipes on how to change (\ref{eq:Neumann}) for it to be compatible with more general $\Sigma$. One is to introduce matter fields on the brane $Q$\cite{NTU12} by hand. Another one is to impose a mixed type of boundary conditions on $Q$ \cite{MCG17,CMG17}.

To check which one (either of the existing recipes or a new one) should be picked, we propose that the causality compatibility between $\Sigma$ and $\CM$ should be regarded as a principle to rule out inappropriate recipes. We will not go deep into it, but this is expected to be an interesting future direction.

\section{Causal Structure in the Intermediate Picture}\label{sec:causalityQ}

In this section, we discuss the causal structure in the intermediate picture for general configurations. Since the spacetime geometry associated to the intermediate picture is composed from the asymptotic boundary $\Sigma$ and the brane $Q$, we need to study the causal structure for three cases: sending a signal within $\Sigma$, within $Q$, and across $\Sigma$ and $Q$. The results for these three cases are given by statement \ref{stat:adsbcft}, \ref{stat:brane}, \ref{stat:across}, respectively.

Statement \ref{stat:adsbcft} has already been studied in section \ref{sec:proofadsbcft}. As for statement \ref{stat:across}, we have explicitly checked that the bulk introduces superluminal propagation in pure AdS case in section \ref{sec:geoex}. Besides, we have shown the existence of superluminal propagation without using the specific properties of the vacuum configuration. The only ingredient needed is the existence of the co-dimension 2 defect, which is naturally introduced by $\partial\Sigma$. Thus, presence of superluminal propagation in the intermediate picture, i.e. statement \ref{stat:across}, has already been shown for general configurations. 

Therefore, what we need to do is to show statement \ref{stat:brane} for general configurations. The analysis given here is parallel to that in \cite{Ishihara00}, where causality in brane universes is discussed. After that, we will summarize the causal structure in the intermediate picture of double holography, and discuss its relations to other holographic setups. 

\subsection{Showing Statement \ref{stat:brane} for General Configurations}\label{sec:proof3}

Here, we show that any null geodesic on $Q$ is also a null geodesic in $\CM$, under the Neumann boundary condition \eqref{eq:Neumann}. Then, it follows straightforwardly that\footnote{Mathematically, there is a small logical gap here. However, we will not bother us to worry about it. For readers who are skeptical of this point, see proposition 4.5.1 and proposition 4.5.10 in \cite{HE73}.} two spacetime points which are not causally connected on $Q$ are not causally connected in $\CM$ either (statement \ref{stat:brane}).
    
Consider an arbitrary curve on the brane $Q$ and use $u^a$ to denote its tangent vector. Then the covariant derivative on the brane $D_a$ and that in the bulk $\nabla_\mu$ can be related by the Gauss-Weingarten equation as 
        \begin{align}\label{eq:GWgen}
            u^\mu\nabla_\mu u^\nu = (u^a D_a u^c) {e^\nu}_{c} - K_{ab}u^a u^b n^\n        \end{align}
where $n^{\nu}$ is the normal vector to $Q$. Here, $K_{ab}$ is the extrinsic curvature of the brane. Taking the trace of \eqref{eq:Neumann} and substituting it back, we have 
        \begin{align}\label{eq:extrinsicbrane}
            K_{ab} = - \frac{1}{d-1} T h_{ab}.
        \end{align}
    
For a geodesic on the brane, we have
    \begin{align}
        u^a D_a u^b = 0.
    \end{align}
Moreover, for a null geodesic, $K_{ab}u^a u^b \propto h_{ab} u^a u^b = 0$ which follows from Eq. \eqref{eq:extrinsicbrane}, and hence 
    \begin{align}
        u^\mu\nabla_\mu u^\nu = 0 . 
    \end{align}
In other words, a null geodesic on the brane $Q$ is also a null geodesic in the bulk $\CM$. Note that the argument so far is extremely powerful since the Neumann boundary condition (\ref{eq:Neumann}) is almost the only requirement. 

\subsection{Causal Structure and Relations to Other Setups}\label{sec:vsothers}
So far, we have shown that all three statements \ref{stat:adsbcft} , \ref{stat:brane}, and \ref{stat:across} hold for generic configurations in double holography. Combining the three statements, to be compatible with the bulk picture, the effective theory in the intermediate picture should have the following property: 
\begin{itemize}
    \item Let $p$ and $q$ be two points which are space-like separated on $\Sigma\cup Q$. Then influencing $q$ by adding perturbations on $p$ is possible if and only if $p\in\Sigma ~ \wedge~q \in Q$ or $p\in Q ~ \wedge~q \in \Sigma$.
\end{itemize}
This is the causal structure of the intermediate picture in double holography. 

In the following, the relations and differences from causal structures in other holographic models are discussed.

\paragraph{v.s. the Gao-Wald theorem}~\par

Although statement \ref{stat:brane} and statement \ref{stat:adsbcft} are both important elements in the causal structure of the intermediate theory and look similar to each other, the mechanisms behind them are very different. 

As we have seen in the section \ref{sec:proof3}, the key ingredient which guarantees the causality within brane $Q$ and the bulk $\CM$ is the Neumann boundary condition (\ref{eq:Neumann}) imposed on $Q$. We would like to emphasize that neither restrictions on the bulk matter nor the on-shell condition of $\CM$ is required.

On the other hand, the Gao-Wald theorem which guarantees the no-shortcut statement in AdS/(B)CFT has a mechanism distinct from this. With Dirichlet boundary condition imposed on $\Sigma$, the most significant point of the Gao-Wald theorem is that it requires $\CM$ to be on-shell and the bulk matter to satisfy ANEC. 

\paragraph{v.s. the $T\bar{T}$-deformed CFT/cutoff AdS correspondence}~\par 
Pulling the boundary theory into the bulk in holographic duality causes a nonlocality in general. One of the most famous examples is the cutoff AdS. Consider a global AdS$_3$, make a finite radial cutoff and regard the cutoff surface as the manifold on which the boundary theory is defined. Then one can find that there exist two points which are space-like separated on the boundary manifold while causally connected in the bulk \cite{MMV16,LLST19,GMW20}. Therefore, the corresponding boundary theory should contain superluminal information propagations. It is proposed in \cite{MMV16} that the boundary theory corresponding to a cutoff AdS$_3$ is a $T\bar{T}$-deformed CFT \cite{Zamolodchikov04,SZ16,Cardy18} which indeed show a superluminal propagation pattern compatible with the bulk side. 

In this sense, it is not surprising that the intermediate theory in the double holography has a nonlocal causal structure since the boundary $\partial\CM = \Sigma \cup Q$ in this case is also pulled into the bulk. Instead, the truly special point is that no short cut can be made when sending a signal from $Q$ to $Q$, even if $Q$ is pulled into the bulk. This is what the Neumann boundary condition (\ref{eq:Neumann}) causes. Contrast to this, in cutoff AdS, Dirichlet boundary condition is imposed on the cutoff surface.

\paragraph{v.s. dS/flat brane-world holography}~\par
In this paper, we are focusing on the Karch-Randall brane $Q$, i.e. the asymptotically AdS brane intersecting the asymptotic boundary $\Sigma$ at a time-like corner. The tension for such a brane is restricted to $-(d-1)/L < T < (d-1)/L$. On the other hand, the analysis in section \ref{sec:proof3} has no restrictions on $T$. Therefore, the statement \ref{stat:brane} is also applicable to asymptotically dS brane realized by $|T| > (d-1)/L$, and asymptotically flat brane realized by $|T| = (d-1)/L$. Implications for the brane-world cosmology is discussed in \cite{Ishihara00}.

\paragraph{v.s. brane with matter localized on it}~\par

We would like to comment on what happens if one put matters on the brane $Q$ by hand. The presence of matters changes the Neumann boundary condition \eqref{eq:Neumann} as \footnote{Sign in front of energy-momentum tensor is different from \cite{Takayanagi11, FTT11}. This is because we are taking the direction of the normal vector to be opposite.}
\begin{align}\label{eq:NeumannwMat}
    K_{ab}-K h_{ab} - T h_{ab} + 8\pi G_N T^Q_{ab} = 0~.
\end{align}
Here, $T^Q_{ab}$ is the (localized) matter stress-energy tensor on the brane. Taking the trace of \eqref{eq:NeumannwMat}, we obtain
\begin{align}
    K = -\frac{d}{d-1} T - \frac{8 \pi G_N}{d-1} T^Q
\end{align}
with $T^Q = T^Q_{ab} h^{ab}$. Substituting it back to \eqref{eq:NeumannwMat}, we have
\begin{align}
    K_{ab} = -\frac{1}{d-1} T h_{ab} - 8\pi G_N \left(T^Q_{ab} - \frac{T^Q}{d-1}h_{ab}\right).
\end{align}
Now we impose null energy condition on the brane
\begin{align}
    T^{Q}_{ab} u^a u^b \ge 0~.
\end{align}
Here $u^a$ is an arbitrary null vector on the brane. Then we have 
\begin{align}
    K_{ab} u^a u^b = - 8\pi G_N T^Q_{ab}u^a u^b \le 0~,
\end{align}
which implies that brane is concave in null direction (see the middle panel of figure \ref{fig:sigma_ex} for example). Therefore, there exists a shortcut in the bulk. A physical interpretation of this consequence is that massive objects on the brane bends the brane and this creates shortcuts. This is discussed in \cite{Ishihara00} as the solution to the horizon problem in the context of brane cosmology.

\section{Commutators in the Intermediate Picture}\label{sec:commutator}

Until now, we have studied causal structures determined from the geometry of the background manifold. We have found that, the bulk causal structure is compatible with BCFT causality. On the other hand, a causality violation is expected in the intermediate theory $\CT^{\rm int}_{\Sigma\cup Q}$ for it to be compatible with the bulk causal structure. In this context, causality of a theory $\CT_{\CN}$ defined on manifold $\CN$ is a property that a signal cannot be sent between two space-like separated points in $\CT_{\CN}$. This notion of causality does not explain the fundamental mechanism underlying this property.

On the other hand, in the context of quantum many body systems (including QFTs, spin systems, etc.), there is a more fundamental notion of causality called microcausality. Microcausality of a theory $\CT_{\CN}$ is a property that (gauge invariant) operators commute outside the lightcone: 
\begin{align}
    [\mathcal{O}(p),\mathcal{O}(p')] = 0 ~~ \mbox{if} ~~  p~{\rm and}~p'~{\rm are~spacelike~separated}.
\end{align}
In this context, the previous notion of causality is often called macrocausality. Note that both macrocausaility and microcausality depends on the details of the theory. The difference is that microcausality says more about the details. 

In this section, we examine the commutator of the operators in the intermediate picture living on $\partial \CM = \Sigma \cup Q$. Here, we fix the bulk configuration $\CM$ to be the one considered in section \ref{sec:geoex}, i.e. a portion of pure AdS$_{d+1}$ with an end-of-the-world brane, as a concrete example. We calculate the vacuum expectation value of the commutator
\begin{align}\label{eq:comm}
    \braket{[\mathcal{O}(p),\mathcal{O}(p')]}^{\rm int}_{\partial \CM} = \braket{\mathcal{O}(p)\mathcal{O}(p')}^{\rm int}_{\partial \CM} - \braket{\mathcal{O}(p')\mathcal{O}(p)}^{\rm int}_{\partial \CM}~,
\end{align}
and see its behavior. In the following, we will use the $(z,t,\bm{x})$ coordinate introduced in \eqref{eq:Poincarecor3} and the $(\mu, t, y, \bm{\xi})$ coordinate introduced in \eqref{eq:metricFG} to describe a point in the bulk $\CM$. Accordingly, the $(t,\bm{x})$ coordinate and the $(t, y, \bm{\xi})$ are used to describe a point on $\Sigma$ or $Q$.

We would like to compute \eqref{eq:comm} using the holographic dictionary. In the usual AdS/CFT correspondence, the extrapolate dictionary \cite{HS11}
\begin{align}\label{eq:usualextrapolatedic}
	\braket{\mathcal{O}(t,\bm{x})\mathcal{O}(t', \bm{x}')}^{\rm CFT} \propto \lim_{z,z'\to 0} z^{-\Delta} z'^{-\Delta}\braket{\phi(z,t,\bm{x})\phi(z',t', \bm{x}')}^{\rm bulk}~,
\end{align}
is often used for calculation of the correlation functions. Here, $\mathcal{O}$ is a primary operator in the boundary CFT with scaling dimension $\Delta$, and $\phi$ is its corresponding field in the bulk. The left-hand side is the Wightman function of $\CO$ in the boundary CFT and the right-hand side is the corresponding Wightman function in the bulk. In the following, we consider $\phi$ to be a bulk scalar field with mass $m$ for simplicity. In this case, the scaling dimension is given by $\Delta = d/2 + \sqrt{d^2/4 + m^2}$. 

On the other hand, in our setup, the bulk $\CM$ is a portion of pure AdS, and its boundary $\partial\CM$ is composed by an end-of-the-world brane $Q$ placed at $\mu = \mu_*= \pi/2 - \theta_{*}$ plus a portion of the asymptotic boundary, instead of just a single asymptotic boundary as in the usual AdS/CFT. See figure \ref{fig:poincoor} for a sketch. A holographic dictionary applicable for our setup has already been proposed in \cite{Porrati01,Neuenfeld21}. For example, a correlation function between $p\in \Sigma$ and $p' \in Q$ can be computed by
\begin{align}\label{eq:extrapolatedicbrane}
	\braket{\mathcal{O}(p)\mathcal{O}(p')}^{\rm int}_{\partial \CM} \propto \lim_{z\to 0} z^{-\Delta} {\theta}'^{-\Delta}\braket{\phi(z,t,\bm{x})\phi(\mu',t',y',\bm{\xi}')}^{\rm bulk}_{\CM}|_{\theta' = \theta_*}~,
\end{align}
if we neglect the back reaction by the scalar field and look at the limit where the brane $Q$ is close to the asymptotic boundary ($\theta_* \ll 1$). Here, $\theta = \pi/2 - \mu$ (see equation \eqref{eq:metricFG2}). Of course, we can also consider the case in which both of the two points live on $\Sigma$ or $Q$. This dictionary, together with the usual one \eqref{eq:usualextrapolatedic}, implies that computing the bulk Wightman function is sufficient to obtain the correlation function in the intermediate picture. 

\begin{figure}
\centering
\includegraphics[keepaspectratio, scale=0.2]{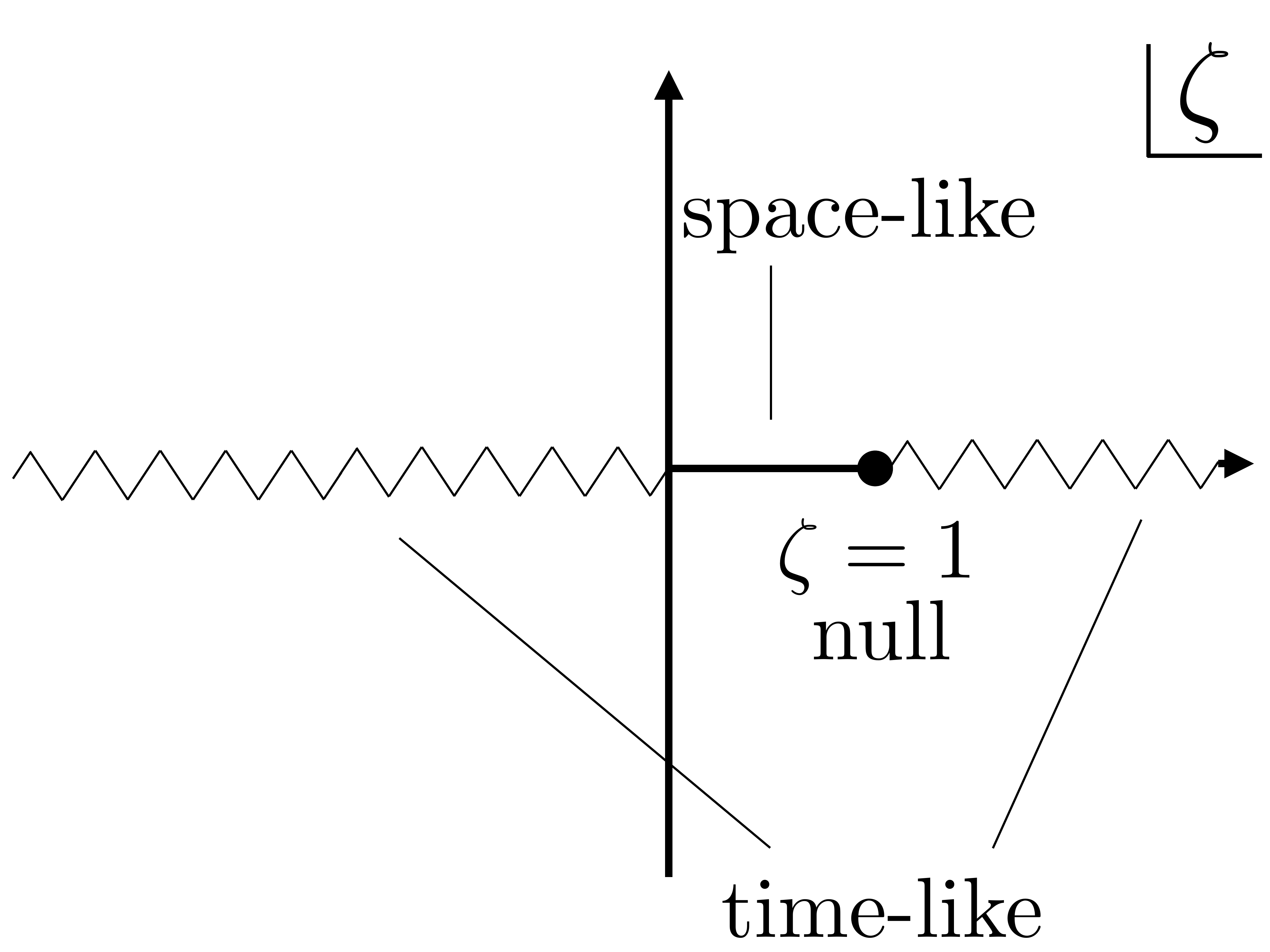}
\caption{The analytic structure of the bulk correlation function \eqref{eq:corrbulk}. There are two branch-cuts on the real axis with $\zeta > 1$ and $\zeta < 0$, respectively. The branch-cut with $\zeta > 1$ is introduced by the hypergeometric function $_2F_1$ and the branch-cut with $\zeta < 0$ is introduced by $\zeta^{\Delta}$. Using equation \eqref{eq:confinv}, we can show that region $\zeta>1~{\rm or}~\zeta <0$, $0<\zeta<1$, $\zeta = 1$ correspond to time-like, space-like, null separated two points in the bulk, respectively.}
\label{fig:analyticcorr}
\end{figure}

To calculate the bulk Wightman function, we introduce the conformal invariant
\begin{align}\label{eq:confinv}
	\zeta  = \frac{2}{\braket{X - X',X-X'} + 2} = \frac{2 z z'}{z^2 + z'^2 -(t-t')^2+ (\bm{x}-\bm{x}')^2}~.
\end{align}
Here, $\braket{\cdot,\cdot}$ is the inner product associated to the embedding space $\mathbb{R}^{2,d}$ and $X^A ({X'}^A)$ is the coordinate in $\mathbb{R}^{2,d}$ (see appendix \ref{sec:AdScoord} for details). Using the conformal invariant simplifies the derivation of correlation functions in the bulk and the final result is \cite{DF02}
\begin{align}\label{eq:corrbulk}
	\braket{\phi(z,t,\bm{x})\phi(z',t',\bm{x}')}^{\rm bulk}_{\CM} \propto \zeta^\Delta {}_2F_1(\Delta/2, 1/2 + \Delta/2, 1 + (\Delta - d/2);\zeta^2)~,
\end{align}
where $\Delta = d/2 + \sqrt{d^2/4 + m^2}$. Figure \ref{fig:analyticcorr} shows the analytic structure of \eqref{eq:corrbulk} on $\zeta$ plane. We let the branch-cut of $\zeta^{\Delta}$ run in the range $(-\infty,0)$ as in \cite{Fronsdal74}. This analytic structure is important for determining the microscopic causal structure in the bulk \cite{Fronsdal74}.

Let us calculate the commutator for three cases: $p, p' \in \Sigma; p,p'\in Q$; and $p\in \Sigma, p'\in Q$. We introduce the following quantity for later use,
\begin{align}
    \Delta\beta(p,p') \equiv \left\{ 
    \begin{array}{ll}
        -(t-t')^2+ (\bm{x}-\bm{x}')^2 & (p, p' \in \Sigma) \\
        -(t-t')^2 + (\bm{\xi}-\bm{\xi}')^2 + (y-y')^2 & (p, p' \in Q) \\
        -(t-t')^2 + (\bm{\xi}-\bm{\xi}')^2 + (y+y')^2 & (p\in \Sigma, p'\in Q)
    \end{array}
    \right.
    .
\end{align}
With such a $\Delta \beta$, we have 
\begin{align}
    & \Delta\beta > 0 \Longleftrightarrow ~p~{\rm and}~p'~{\rm are~space-like~separated~on }~\partial\CM. \nonumber\\
    & \Delta\beta = 0 \Longleftrightarrow ~p~{\rm and}~p'~{\rm are~null~separated~on }~\partial\CM.\\
    & \Delta\beta < 0 \Longleftrightarrow ~p~{\rm and}~p'~{\rm are~time-like~separated~on }~\partial\CM.\nonumber
\end{align}
where $\partial\CM = \Sigma \cup Q$.

The case $p, p' \in \Sigma$ is the same as the usual AdS/CFT and can be calculated straightforwardly. From the definition of $\zeta$ \eqref{eq:confinv}, we have $\zeta \to 0$ near the asymptotic boundary region $z,z' \sim 0$. Now the asymptotic behavior of the hypergeometric function tells us that the behavior of the bulk correlator around the asymptotic boundary is like
\begin{align}\label{eq:confinvgreen}
	\braket{\phi(z,t,\bm{x})\phi(z',t',\bm{x}')}^{\rm bulk}_{\CM} \to \zeta^{\Delta} = \left(\frac{2 z z'}{z^2 + z'^2 -(t-t')^2+ (\bm{x}-\bm{x}')^2 + i\epsilon}\right)^{\Delta}~.
\end{align}
Here, we put $i\epsilon$ ($\epsilon \ll 1$) in this way to correctly obtain the Wightman function\footnote{See, for example, the appendix of \cite{SvR09} for details.}. Then the dictionary \eqref{eq:usualextrapolatedic} leads to 
\begin{align}
	\braket{\mathcal{O}(p)\mathcal{O}(p')}^{\rm int}_{p,p'\in \Sigma}  &\propto \left(\frac{1}{ -(t-t')^2+ (\bm{x}-\bm{x}')^2 + i \epsilon}\right)^{\Delta}~.
\end{align}

Using the above expression of Wightman function, we obtain the commutator for $p, p' \in \Sigma$ as
\begin{align}\label{eq:commSigma}
    \braket{[\mathcal{O}(p),\mathcal{O}(p')]}^{\rm int}_{p,p'\in \Sigma} \propto \left(\frac{1}{ -(t-t')^2+ (\bm{x}-\bm{x}')^2 + i \epsilon}\right)^{\Delta} - \left(\frac{1}{ -(t-t')^2+ (\bm{x}-\bm{x}')^2 - i \epsilon}\right)^{\Delta}~.
\end{align}
We use the fact that $\braket{\mathcal{O}(p')\mathcal{O}(p)}$ is given by the complex conjugate of $\braket{\mathcal{O}(p)\mathcal{O}(p')}$.
Since the branch-cut of $\zeta^{\Delta}$ runs on the negative half of the real axis, \eqref{eq:commSigma} vanishes for
\begin{align}
    \Delta \beta = -(t-t')^2+ (\bm{x}-\bm{x}')^2 > 0~.
\end{align}
In other words, the commutator vanishes for space-like separated points in $\Sigma$. Therefore, microscopic causality is preserved within $\Sigma$.

Now we move on to the case $p,p'\in Q$. In this case, it is more convenient to use the $(\mu,t,y,\bm{\xi})$ coordinate. When the two bulk points are sitting at $\mu = \mu_*= \pi/2 - \theta_{*}$, the conformal invariant $\zeta$ turns out to be
\begin{align}\label{eq:confinvQ}
    \zeta = \frac{2 \cos^2\left(\frac{\pi}{2} - \theta_*\right) yy'}{\Delta \beta + 2 \cos^2\left(\frac{\pi}{2} - \theta_*\right) yy'}~,
\end{align}
where
\begin{align}
    \Delta \beta = -(t-t')^2 + (\bm{\xi}-\bm{\xi}')^2 + (y-y')^2~.
\end{align}
Then the dictionary \eqref{eq:extrapolatedicbrane} gives the Wightman function on the brane as
\begin{align}\label{eq:WightmanQ}
	\braket{\mathcal{O}(p)\mathcal{O}(p')}^{\rm int}_{p,p'\in Q} 
	\propto &\left(\frac{2 yy'}{\Delta \beta +i\epsilon + 2 yy' \sin^2 \theta_* }\right)^\Delta \nonumber \\
	&~~~~\times {}_2F_1\left(\Delta/2, 1/2 + \Delta/2, 1 + (\Delta - d/2);\left(\frac{\Delta \beta + i \epsilon}{2 y y' \sin^2\theta_*} + 1\right)^{-2}\right) \nonumber \\
	\equiv &~G^{Q}_{+}(t,y,\bm{\xi},t',y',\bm{\xi}')
\end{align}
Substituting to equation \eqref{eq:comm}, we obtain
\begin{align}\label{eq:commQ}
    \braket{[\mathcal{O}(p),\mathcal{O}(p')]}^{\rm int}_{x,x'\in Q} \propto G^{Q}_+(t,y,\bm{\xi},t',y',\bm{\xi}') - \left[G^{Q}_{+}(t,y,\bm{\xi},t',y',\bm{\xi}')\right]^{*}~.
\end{align}

The behavior of this function is determined by the analytic structure of \eqref{eq:WightmanQ}, which is the same as that of the bulk Wightman function (see figure \ref{fig:analyticcorr}). Using \eqref{eq:confinvQ}, relation between the value of $\zeta$ and the causal relation of two points $p,p'$ on the brane (time-like for $\Delta \beta <0$, space-like for $\Delta \beta > 0$, and null for $\Delta \beta = 0$) is
\begin{align}
    \Delta \beta < 0 &\to \zeta >1, \zeta < 0~,\\
    \Delta \beta = 0 &\to \zeta = 1~,\\
    \Delta \beta > 0 &\to 0 < \zeta <1~.
\end{align}
Combining with figure \ref{fig:analyticcorr}, we observe that the commutator vanishes for $\Delta \beta > 0$, i.e. two space-like separated points on the brane. Thus microscopic causality is preserved within $Q$.

Finally, let us consider the commutator for $p\in \Sigma,p'\in Q$. In this case, the conformal invariant $\zeta$ can be written as 
\begin{align}
    \zeta = \frac{2 z y' \cos\left(\frac{\pi}{2} - \theta_*\right)}{z^2 + \Delta \beta - 2 y y' (1 - \sin\left(\frac{\pi}{2} - \theta_*\right))}.
\end{align}
where
\begin{align}
    \Delta \beta = -(t-t')^2 + (\bm{\xi}-\bm{\xi}')^2 + (y+y')^2~.
\end{align}
Substituting this to \eqref{eq:confinvgreen} and using the dictionary \eqref{eq:extrapolatedicbrane} by taking $z \to 0$, we obtain
\begin{align}
	\braket{\mathcal{O}(p)\mathcal{O}(p')}^{\rm int}_{p\in \Sigma,p'\in Q} &\propto \left(\frac{1}{\Delta \beta - 2 y y' (1 - \cos\theta_*) + i \epsilon}\right)^{\Delta}~.
\end{align}
Note that the contribution from the hypergeometric function disappears thanks to the $z \to 0$ limit, unlike the $p,p'\in Q$ case. Since $\braket{\mathcal{O}(p')\mathcal{O}(p)}$ is the complex conjugate of $\braket{\mathcal{O}(p)\mathcal{O}(p')}$, commutator \eqref{eq:comm} is given by
\begin{align}
\begin{aligned}
	\braket{[\mathcal{O}(p),\mathcal{O}(p')]}^{\rm int}_{p\in \Sigma,p'\in Q} \propto & \left(\frac{1}{\Delta\beta - 2 y y' (1-\cos\theta_*) + i\epsilon}\right)^{\Delta}\\
	&- \left(\frac{1}{\Delta \beta - 2 y y' (1-\cos\theta_*) - i\epsilon}\right)^{\Delta}~.
\end{aligned}
\end{align}

Recall that we take the branch-cut of $\zeta^{\Delta}$ to run in the range $(-\infty,0)$. Thus, the commutator vanishes in the region
\begin{align}\label{eq:microviolation}
	\Delta \beta - 2 y y' (1-\cos\theta_*) > 0~,
\end{align}
which is strictly smaller than the space-like region indicated by the geometry of $\partial \CM$
\begin{align}
	\Delta \beta > 0~.
\end{align}
Therefore, the commutator does not necessarily vanish even when the two points are space-like separated on $\partial\CM = \Sigma \cup Q$. This clearly shows a breakdown of the microscopic causality and an existence of some nonlocal interaction\footnote{It is known that any local relativistic theory satisfies microcausality. Since we expect that the intermediate theory $\CT^{\rm int}_{\Sigma\cup Q}$ is relativistic, it must be nonlocal.} between $Q$ and $\Sigma$ in the intermediate picture.

Comparing \eqref{eq:microviolation} and \eqref{eq:causalbulk}, we can see that they are in exactly the same form, and it is straightforward to see that 
\begin{align}
    &\braket{[\mathcal{O}(p),\mathcal{O}(p')]}^{\rm int}{\rm ~ vanishes}~ \nonumber\\
    \Longleftrightarrow &~p~{\rm and}~p'~{\rm cannot~communicate~through~the~bulk}~\CM.
\end{align}
This fact implies that the origin of the microscopic causality violation (or emergence of nonlocal interaction between $Q$ and $\Sigma$) is the causal structure of the bulk $\CM$. 

We will use ``violation of causality" and ``nonlocality" interchangeably when talking about the effective theory in the intermediate picture, $\CT^{\rm int}_{\Sigma\cup Q}$.

\section{Nonlocality and Subregions in Double Holography}\label{sec:EWRC}
So far, we have seen that the effective theory in the intermediate picture should have a violation of causality and hence nonlocality to be compatible with the causal structure in the bulk picture. 

In this section, we are going to discuss physical consequences and features of the nonlocality observed in the intermediate picture. Starting from a discussion on the notion of domain of dependence when nonlocality is involved, we will review typical subregions in the usual AdS/CFT correspondence, such as causal wedges and entanglement wedges. After that, we will examine the relation between some typical subregions in double holography. In particular, we will see that a straightforward analog of subregion/subregion duality in the AdS/CFT correspondence does not serve as a subregion/subregion duality in the intermediate/bulk correspondence. 

In the end of this section, we will point out that the nonlocality appearing in the intermediate picture is sensitive at IR (or long range). We will compare it with other known nonlocality in quantum gravity, and give an intuitive explanation about how it can arise in the gravitational path intergral language.  

\subsection{Breakdown of Domain of Dependence}\label{sec:DoD}
The nonlocal nature appearing in the intermediate picture prompts us to reconsider the notion of {\it domain of dependence}. To proceed, let us split the usual notion of domain of dependence into two. In the following, we use $\mathcal{N}$ to denote a $d$-dimensional spacetime and suppose there is a field theory $\CT_{\CN}$ living on it. 

For a space-like subregion $A\subset\CN$, consider the set consisting of all points $p\in \CN$ with the property that every causal curve through $p$ intersects $A$. Let us call it the {\it geometrical domain of dependence} of $A$ and denote it as $\CD_G(A)$. 

For the same subregion $A$, consider the set consisting of all points $p\in\CN$ with the property that every local observable at $p$ can be determined from the initial condition on $A$. Let us call it the {\it effective domain of dependence} of $A$ and denote it as $\CD_E(A)$. 

Note that $\CD_G(A)$ only depends the geometry of $\CN$, while $\CD_E(A)$ also depends on the effective theory $\CT_{\CN}$ living on it. If $\CT_{\CN}$ is a local relativistic theory, $\CD_G(A)$ and $\CD_E(A)$ degenerate to the conventional domain of dependence. In general, however, $\CD_G(A) \neq \CD_E(A)$.

\paragraph{$\CD_G(A) \neq \CD_E(A)$ in the intermediate picture}~\par
The mismatch between the geometrical domain of dependence and the effective domain of dependence also occurs in the intermediate picture of double holography. Let us see it using the same configuration as in section \ref{sec:geoex} and section \ref{sec:commutator}. See figure \ref{fig:vacuumslice}, \ref{fig:shortpath} and \ref{fig:poincoor} for sketches. In this case, the manifold we consider is $\Sigma \cup Q$ and the effective theory on it $\CT^{\rm int}_{\Sigma \cup Q}$ should be compatible with the causal structure in the bulk picture. Here, the index ``int" stands for the intermediate picture. 

For simplicity, we focus on $d=2$ and use the $(\rho,t, y)$ coordinate introduced in \eqref{eq:corrhoty} to parameterize the bulk $\CM$. Accordingly, $\partial\CM = \Sigma \cup Q$ can be parameterized by $(t, y)$. In the following, we will use $y>0$ to parameterize $\Sigma$, and $y<0$ to parameterize $Q$. This is the same as what we did when computing \eqref{eq:tbrane} and \eqref{eq:causalbulk}. 

Let us take the subregion $A$ as a finite interval of $\Sigma$ given by $0< a\leq y \leq b$ on time slice $t=0$. Obviously, the geometrical domain of dependence $\CD^{\rm int}_G(A)$ is the region surrounded by $t=\pm(y-a)$ and $t=\pm(y-b)$. 

Let us then move on to determine the effective domain of dependence $\CD^{\rm int}_E(A)$. Note that while the geometrical domain of dependence $\CD^{\rm int}_G(A)$ is determined by geometrical inputs from the boundary manifold $\Sigma \cup Q$, the effective domain of dependence $\CD^{\rm int}_E(A)$ should be determined by geometrical inputs from the bulk manifold $\CM$, since we expect the effective theory in the intermediate picture is equivalent to a local theory in the bulk picture. For any point $q \in (\Sigma\cup Q)\backslash A$, one can always find a point $p \in Q$ such that $p$ can communicate with\footnote{Here, if $p$ can send a signal to $q$ or receive a signal from $q$, then we say that $p$ can communicate with $q$, and vice versa.} $q$ through the bulk $\mathcal{M}$ but not with any point on $A$. Let us explicitly give such a $p$ for different types of $q$ for the the vanishing tension $T=0$ case.\footnote{The discussion can be straightforwardly extended to general tension cases.} The coordinates of $p$ and $q$ are denoted as $(t_p,y_p)$ and $(t_q,y_q)$, respectively. It is sufficient to consider $t_q\geq0$ without loss of generality. 
\begin{itemize}
    \item For $0<t_q\leq y_q$, $(t_p,y_p) = (t_q-y_q^2/t_q,t_q-y_q^2/t_q)$ can send a signal to $q$ but cannot communicate with any point on $A$.
    \item For $|y_q|\leq t_q$, $(t_p,y_p) = (0,0)$ can send a signal to $q$ but cannot communicate with any point on $A$.
    \item For $y_q < 0<t_q\leq |y_q|$ or $t_q=0$, $(t_p,y_p) = (t_q,y_q)$ can send a signal to $q$ but cannot communicate with any point on $A$.
\end{itemize}
In other words, any point not living on $A$ cannot be determined only from $A$. Therefore, $\CD^{\rm int}_E(A) = A$. 

As a result, for the finite interval considered above,  $\CD^{\rm int}_G(A) \supset \CD^{\rm int}_E(A) = A$. Note that $\CD^{\rm int}_E(A)$ is codimension-1 while $\CD^{\rm int}_G(A)$ is codimension-0 with respect to $\Sigma \cup Q$. The breakdown of the notion of conventional domain of dependence in the intermediate picture prompts us to reconsider naive discussions assuming $\CD^{\rm int}_G(A) = \CD^{\rm int}_E(A)$ including state matching with other pictures, causal wedge reconstruction and entanglement wedge reconstruction. These topics cannot be totally split from each other, but we would like to start by comparing $A$ in the intermediate picture, $A$ in the BCFT picture and the corresponding region in the bulk picture. 

\subsection{Subregions and States Associated to \texorpdfstring{$A$}{A}}

In this subsection, we focus on a single interval $A$ on $\Sigma\backslash\partial\Sigma$ and consider the physics associated to it in the three pictures of double holography. 

To proceed, let us distinguish two notions. Consider a manifold $\CN$ and a theory $\CT_{\CN}$ living on it. A subregion $\CD\subseteq\CN$ is a geometrical object, while the information contained in $\CD$ should be determined by also taking the theory living on it, $\CT_{\CN}$, into account. Let us use $(\CD, \CT_{\CN})$ to denote the information contained in $\CD$. Although $\CD$ is often used to refer to $(\CD, \CT_{\CN})$, we would like to make a distinction here to avoid confusion. 

\paragraph{Subregions and states in the BCFT picture}~\par
With this in mind, let us consider the domain of dependence of $A$ in the BCFT picture. First of all, the BCFT picture has a BCFT $\CT^{\rm BCFT}_\Sigma$ living on $\Sigma$. Since $\CT^{\rm BCFT}_\Sigma$ is local and relativistic, $\CD^{\rm BCFT}_G(A)$ and $\CD^{\rm BCFT}_E(A)$ coincide. Therefore, we just use $\CD^{\rm BCFT}(A)$ to denote it and call it the domain of dependence. Almost by definition, the reduced density matrix $\rho^{\rm BCFT}_A$ contains the same information as $\left(\CD^{\rm BCFT}(A), \CT^{\rm BCFT}_{\Sigma}\right)$. Let us denote this as 
\begin{align}
    \rho^{\rm BCFT}_A \simeq \left(\CD^{\rm BCFT}(A), \CT^{\rm BCFT}_{\Sigma}\right)
\end{align}
with a slight abuse of notation. 

\paragraph{Subregions in the bulk picture}~\par
\begin{figure}[H]
    \centering
    \includegraphics[width=16.5cm]{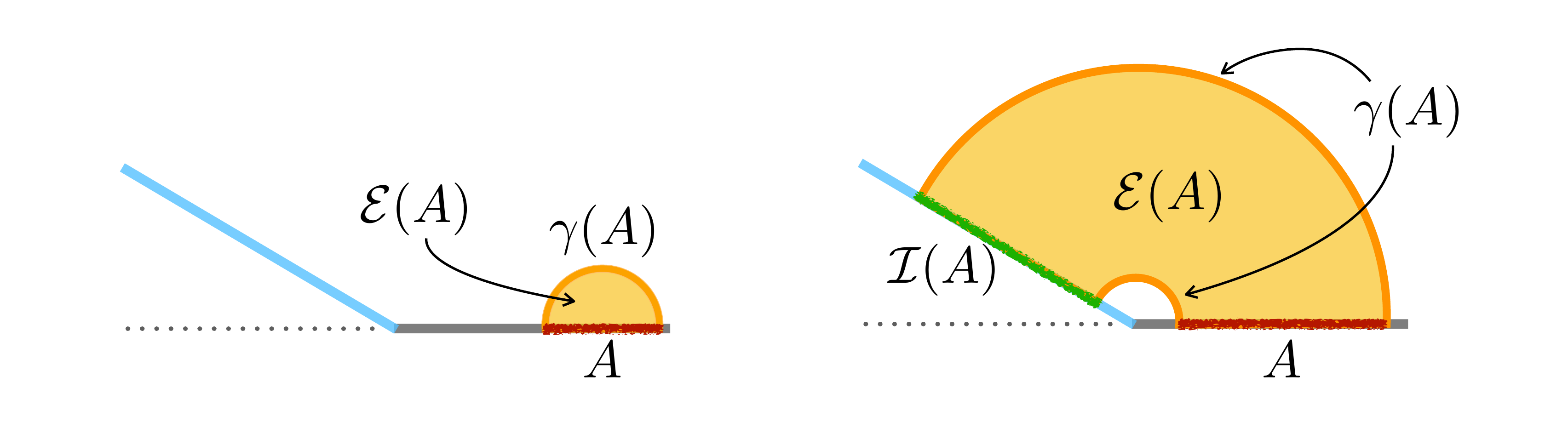}
    \caption{Two examples of entanglement wedges for a single interval $A$ (shown in red). For simplicity, here we show the projection on a time slice in AdS$_3$. The asymptotic boundary $\Sigma$ is shown in grey, and the end-of-the-world brane $Q$ is shown in blue. The HRT surface $\gamma(A)$ is shown in orange, and the entanglement wedge $\CE(A)$ is shaded in yellow. In the left panel, $\gamma(A)$ does not touch the brane $Q$. In the right panel, $\gamma(A)$ touches the brane $Q$ and hence an island $\CI(A)$ (shown in green) arises.}
    \label{fig:HRTEW}
\end{figure}

Let us then go to the bulk picture and consider subregions associated to $A$. First of all, the {\it causal wedge} of $A$, $\CC(A)$, is defined as the intersection of the causal future and the causal past of $\CD^{\rm BCFT}(A)$ in the bulk $\CM$:
\begin{align}
    \CC(A) = J^+\left(\CD^{\rm BCFT}(A)\right) \cap J^-\left(\CD^{\rm BCFT}(A)\right). 
\end{align}
Second, the {\it entanglement wedge} \cite{CKJFvR12,Wall12} of $A$, $\CE(A)$, is defined as the bulk geometrical domain of dependence of a space-like bulk region surrounded by $A$ and its Hubeny-Rangamani-Takayanagi (HRT) surface \cite{HRT07,RT06a,RT06b} $\gamma(A)$. Here, the HRT surface of $A$, $\gamma(A)$, is a space-like bulk surface which satisfies the following conditions:
\begin{itemize}
    \item It is codimension-2 with respect to the bulk manifold $\CM$. 
    \item It shares boundaries with $A$, i.e. $\partial \gamma(A) = \partial A$. 
    \item It is homologous to $A$, i.e. one can bring $A$ to $\gamma(A)$ by a continuous deformation in the bulk. In particular, the end-of-the-world brane $Q$ is regarded as trivial. 
    \item It is an extremal surface. 
    \item If there are multiple candidates satisfying these conditions, $\gamma(A)$ is given by the one with the minimal area. 
\end{itemize}
Due to the treatment of the end-of-the-world brane $Q$, there is a chance that $\gamma(A)$ touches $Q$ and $\CE(A)$ intersects $Q$, as shown in the right panel of figure \ref{fig:HRTEW}. In this case, there exists a space-like region $\CI(A)\subset Q$ such that $\CD^{\rm int}_G\left(\CI(A)\right)$ coincides $\CE(A)\cap Q$. This is guaranteed by statement \ref{stat:brane}. Such an $\CI(A)$ is called the {\it island} of $A$ \cite{AMMZ19,RSJvRWW19,CMNRS20}. 

It is known that the causal wedge is contained in the entanglement wedge \cite{CKJFvR12,Wall12,AKLL16}, i.e.
\begin{align}
    \CC(A) \subseteq \CE(A), 
\end{align}
Therefore, information in the causal wedge can be recovered from the information in the entanglement wedge. Let us denote this as 
\begin{align}
    \left(\CC(A), \CT^{\rm bulk}_\CM\right) \preceq \left(\CE(A), \CT^{\rm bulk}_\CM\right).
\end{align}

\paragraph{Subregions and states in the intermediate picture}~\par
Returning to the intermediate picture, geometrically
\begin{align}
    \CD^{\rm int}_G(A) \supset \CD^{\rm int}_E(A) = A. 
\end{align}
Therefore, one can recover $\left(\CD^{\rm int}_E(A), \CT^{\rm int}_{\Sigma\cup Q}\right)$ from $\left(\CD^{\rm int}_G(A), \CT^{\rm int}_{\Sigma\cup Q}\right)$. On the other hand, by definition of $\CD^{\rm int}_E(A)$, $\left(\CD^{\rm int}_G(A), \CT^{\rm int}_{\Sigma\cup Q}\right)$ cannot be recovered from $\left(\CD^{\rm int}_E(A), \CT^{\rm int}_{\Sigma\cup Q}\right)$. Let us denote this relation as 
\begin{align}\label{eq:rhoint}
    \rho^{\rm int}_A \simeq \left(\CD^{\rm int}_E(A), \CT^{\rm int}_{\Sigma\cup Q}\right) \prec \left(\CD^{\rm int}_G(A), \CT^{\rm int}_{\Sigma\cup Q}\right),
\end{align}
where the ``$\simeq$" follows from the definition of $\CD^{\rm int}_E$.

Now we have introduced several subregions associated to $A$ in the three pictures. Let us see how they correspond to each other. 

\paragraph{BCFT picture v.s. Bulk picture: Entanglement wedge reconstruction}~\par
Subregions in the BCFT picture and the bulk picture are related by the entanglement wedge reconstruction, i.e.
\begin{align}\label{eq:bcftbulk}
    \rho^{\rm BCFT}_A
    \simeq
    (\CD^{\rm BCFT}(A), \CT^{\rm BCFT}_{\Sigma})
    \simeq 
    \left(\CE(A), \CT^{\rm bulk}_\CM\right)
    \simeq 
    \rho^{\rm bulk}_{\CE(A)}. 
\end{align}
See \cite{DHW16,CHPSSW17,CPS19} for proofs and methods for entanglement wedge reconstruction. 

\paragraph{Intermediate picture v.s. Bulk picture: An assumption}~\par 
Since the correspondence between the intermediate picture and the bulk picture is not an ordinary AdS/CFT duality, we do not have a standard way to specify the correspondence between subregions in the two pictures.\footnote{One may assume a naive version of subregion/subregion duality by considering a straightforward analog of entanglement wedge in the intermediate/bulk picture. However, we will see in the next subsection that such a naive construction does not hold as a subregion/subregion duality.} However, we would like to make the following assumption and argue that it holds. 
\begin{assumption}\label{assum:subregion}
    The entanglement wedge $\CE(A)$ in the bulk picture is equivalent to its intersection with the asymptotic boundary and the brane, $\CD^{\rm int}_G\left(A\right) \cup \CD^{\rm int}_G\left(\CI(A)\right)$, in the intermediate picture, i.e.
    \begin{align}\label{eq:intbulk}
        \left(\CE(A), \CT^{\rm bulk}_\CM \right) 
        \simeq
        \left(\CD^{\rm int}_G\left(A\right) \cup \CD^{\rm int}_G\left(\CI(A)\right), \CT^{\rm int}_{\Sigma\cup Q} \right). 
    \end{align}
    Note that $\CI(A)$ can be an empty set. 
\end{assumption}
In other words, this assumption argues that a doubly holographic structure holds also for entanglement wedges. Since an entanglement wedge has already provided a correspondence between a gravitational theory in the bulk and a field theory on the asymptotic boundary, the assumption above is just as justified as many other doubly holographic models. In this way, we argue that this assumption holds.

\paragraph{Intermediate picture v.s. BCFT picture}~\par 
Let us then compare the intermediate picture and the BCFT picture. Geometrically, 
\begin{align}
    \CD^{\rm BCFT}(A) = \CD^{\rm int}_G(A) \supset \CD^{\rm int}_E(A) = A. 
\end{align}
The next question is whether $\left(\CD^{\rm BCFT}(A), \CT^{\rm BCFT}_{\Sigma}\right)$ and $\left(\CD^{\rm int}_G(A), \CT^{\rm int}_{\Sigma\cup Q}\right)$ contain the same information. Combining (\ref{eq:bcftbulk}) and (\ref{eq:intbulk}), we have 
\begin{align}
    \rho^{\rm BCFT}_A \simeq
    (\CD^{\rm BCFT}(A), \CT^{\rm BCFT}_{\Sigma})
    &\simeq 
    \left(\CE(A), \CT^{\rm bulk}_\CM\right) \nonumber\\
    &\simeq 
    \left(\CD^{\rm int}_G\left(A\right) \cup \CD^{\rm int}_G\left(\CI(A)\right), \CT^{\rm int}_{\Sigma\cup Q} \right) \nonumber\\
    &\succeq
    \left(\CD^{\rm int}_G(A), \CT^{\rm int}_{\Sigma\cup Q}\right).
\end{align}
Moreover, combining this with (\ref{eq:rhoint}), we have 
\begin{align}\label{eq:rhobcftint}
    \rho^{\rm BCFT}_A 
    \succ 
    \rho^{\rm int}_A.
\end{align}
(\ref{eq:rhobcftint}) is an important consequence. It implies that, even for the same subregion $A$ on $\Sigma$, the associated state in the BCFT picture and that in the intermediate picture are different from each other. This consequence is consistent with the arguments that $\rho^{\rm BCFT}_A$ is a fine-grained state while $\rho^{\rm int}_A$ corresponds to a coarse-grained state\cite{BW20}. Note that \eqref{eq:rhobcftint} holds even when $A$ does not have an island, i.e. $\CI(A) = \emptyset$.

\subsection{A Tentative Subregion Duality and its Breakdown}\label{sec:tenwedge}

In the previous subsection, we argue that a subregion duality between the intermediate picture and the bulk picture, assumption \ref{assum:subregion}, is expected to hold. However, this argument does not cover all the intermediate subregions. For example, for a subregion $A\subset\Sigma$ with a non-vanishing island $\CI(A)\subset Q$, assumption \ref{assum:subregion} tells us that the bulk dual for the intermediate subregion $\CD^{\rm int}_G\left(A \cup \CI(A)\right)$ is the entanglement wedge $\CE(A)$, but nothing about what the bulk dual of the intermediate subregion $A$ or that of $\CD^{\rm int}_G\left(A\right)$ is.\footnote{Note again that the intermediate subregion and the boundary subregion should be distinguished even when they have the same geometry. The bulk dual of the boundary subregion $A$ is $\CE(A)$.} 

In the following, we will give a straightforward analogy of the entanglement wedge in the intermediate/bulk correspondence and test if it is qualified as a dual bulk subregion to an intermediate subregion. 

To clarify the statement, let us introduce some new concepts and terminologies. 

\begin{figure}[H]
    \centering
    \includegraphics[width=16.5cm]{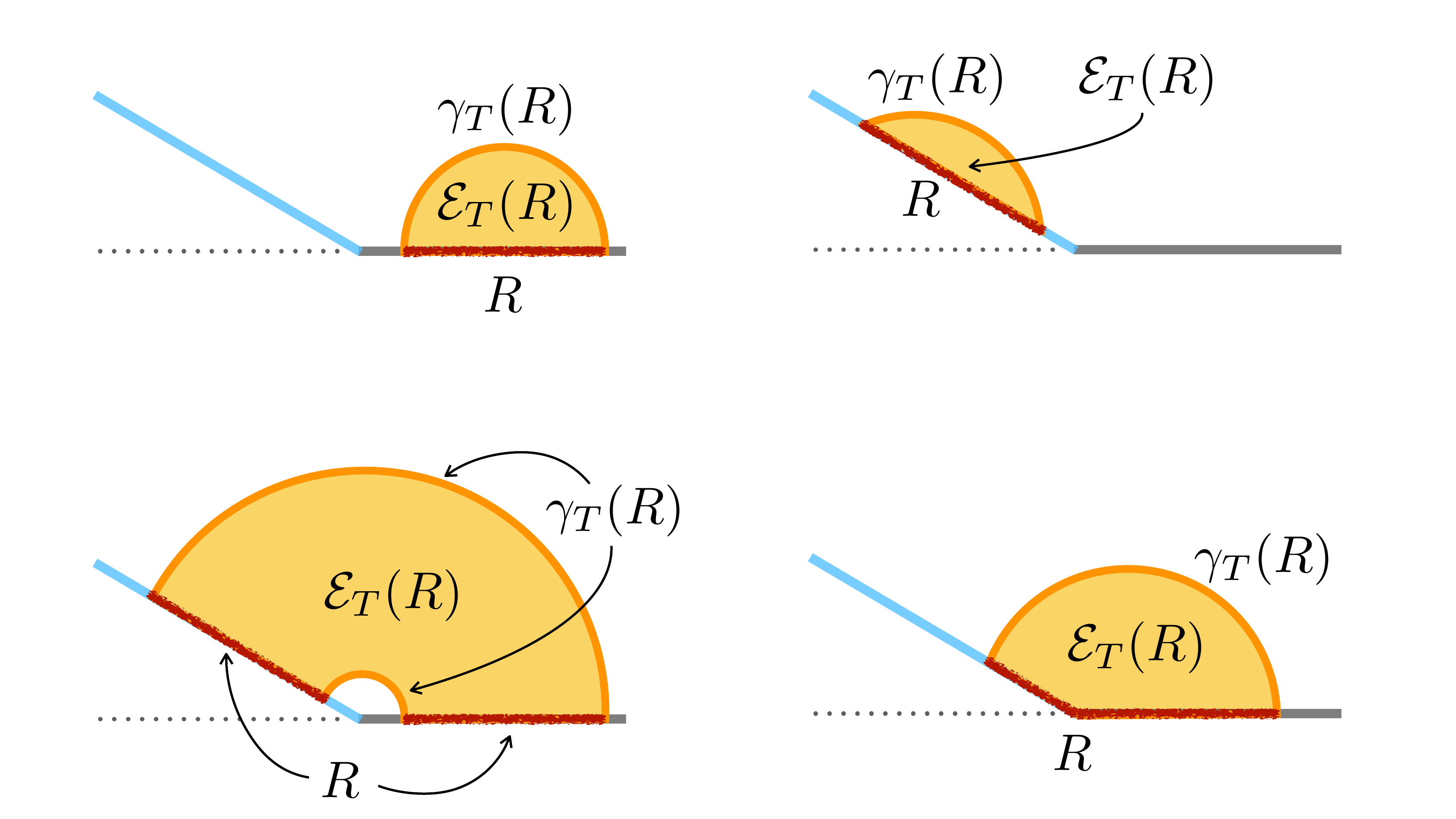}
    \caption{Four examples of tentative entanglement wedges for an intermediate subregion $R$ (shown in red). For simplicity, here we show the projection on a time slice in AdS$_3$. The asymptotic boundary $\Sigma$ is shown in grey, and the end-of-the-world brane $Q$ is shown in blue. The tentative HRT surface $\gamma_T(A)$ is shown in orange, and the tentative entanglement wedge $\CE_T(A)$ is shaded in yellow.}
    \label{fig:THRTEW}
\end{figure}

For a space-like intermediate subregion $R\subset \Sigma \cup Q = (\partial\CM)$ which is codimendion-1 with respect to $\Sigma \cup Q$, the {\it tentative HRT surface} $\gamma_T(R)$ is the minimal extremal surface which is codimension-2 with respect to $\CM$, ending on $\partial R$, and homologous to $R$. The {\it tentative entanglement wedge} $\CE_T(R)$ is the bulk domain of dependence of a space-like region surrounded by $R$ and $\gamma_T(R)$. See figure \ref{fig:THRTEW} for some examples.  

In other words, tentative HRT surfaces and tentative entanglement wedges are given by applying the definitions of HRT surfaces and entanglement wedge directly to the intermediate picture instead of the (B)CFT picture. We use ``tentative" to imply that we still need to investigate whether they satisfy important properties owned by conventional HRT surfaces and entanglement wedges. 

Note that a main difference between a tentative HRT surface and a HRT surface is the treatment of brane $Q$. In the definition of tentative HRT surfaces, $Q$ is treated as a part of the ``boundary". Therefore, $\gamma(R)$ cannot touch $Q$ unless $R$ has an overlap with $Q$. The difference can be clearly observed by comparing figure \ref{fig:HRTEW} and figure \ref{fig:THRTEW}. 

\begin{figure}[H]
    \centering
    \includegraphics[width=10cm]{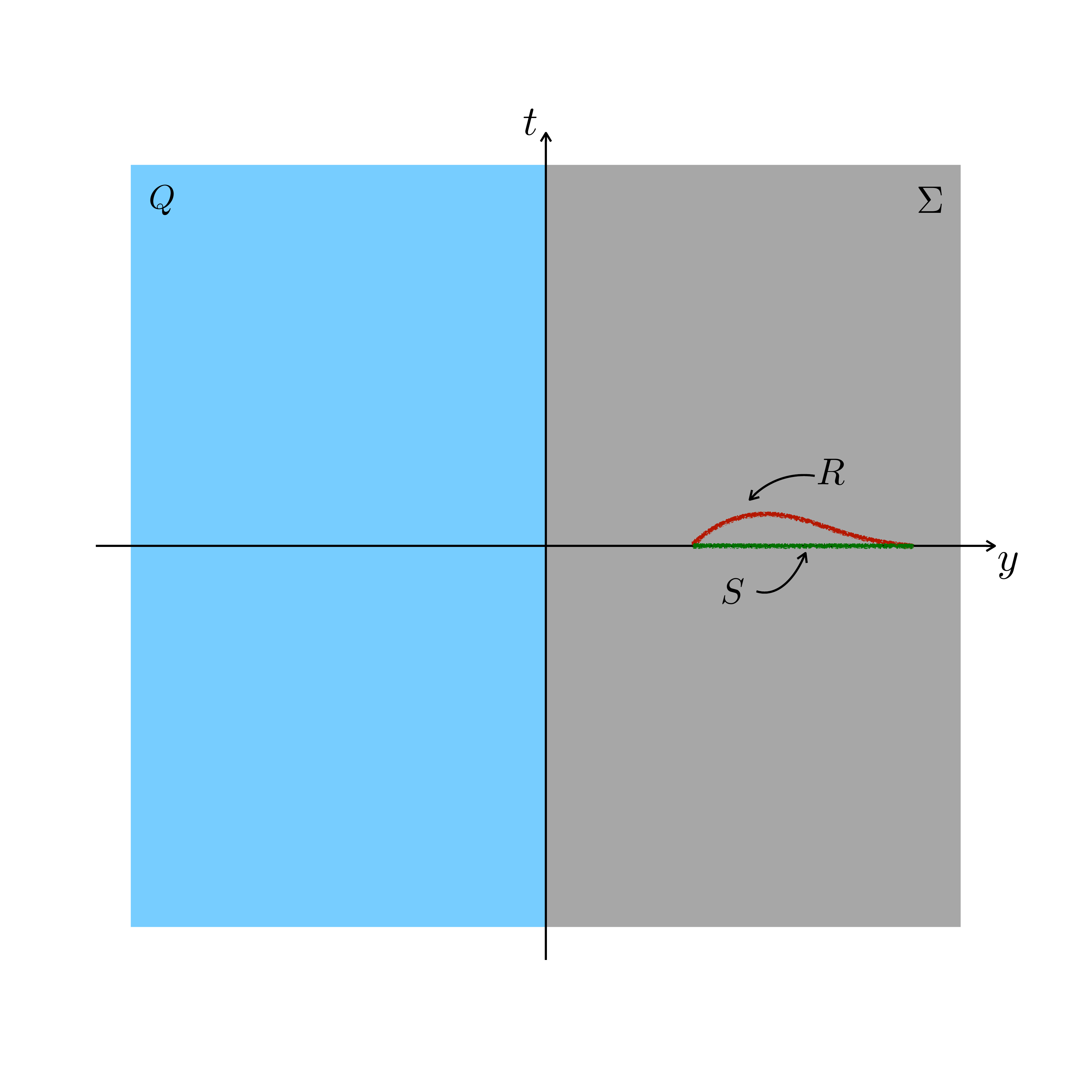}
    \caption{An example in which $\CE_T(S) = \CE_T(R)$ while $\CD^{\rm int}_E(S)\neq \CD^{\rm int}_E(R)$. The figure shows the intermediate picture in the vacuum configuration with $d=2$. Taking $S$ (green) and $R$ (red) as spatial intervals on $\Sigma$ with $\partial S = \partial R$ but $S\neq R$, we have $\CE_T(S) = \CE_T(R)$ from the definition of tentative entanglement wedges, while $\CD_E(S) = S \neq R = \CD_E(R)$ from the discussion in section \ref{sec:DoD}.}
    \label{fig:SandR}
\end{figure}

Since in the bulk/BCFT correspondence, $(\CE(A), \CT^{\rm bulk}_\CM) \simeq (\CD^{\rm BCFT}(A), \CT^{\rm BCFT}_\Sigma) \simeq \rho^{\rm BCFT}_A$ works as a subregion duality for any $A\subset\Sigma$, there are two natural guesses one may make at first glance: 
\begin{itemize}
    \item $(\CE_T(R), \CT^{\rm bulk}_\CM)$ and $(\CD^{\rm int}_E(R), \CT^{\rm int}_{\Sigma\cup Q}) \simeq \rho^{\rm int}_A$ may be equivalent for any intermediate subregion $R\subset\Sigma\cup Q$, 
    \item $(\CE_T(R), \CT^{\rm bulk}_\CM)$ and $(\CD^{\rm int}_G(R), \CT^{\rm int}_{\Sigma\cup Q})$ may be equivalent for any intermediate subregion $R\subset\Sigma\cup Q$. 
\end{itemize}
However, we can easily see that the first guess fails by considering $S, R\subset\Sigma$ such that $\partial S = \partial R$ and hence $\CE_T(S) = \CE_T(R)$ while $\CD^{\rm int}_E(S)\neq \CD^{\rm int}_E(R)$. See figure \ref{fig:SandR} for such an example. 

On the other hand, we have even already seen a support to the second guess: assumption \ref{assum:subregion} tells us this should hold at least for $R = A \cup \CI(A)$. In this case, $\CE_T\left(A \cup \CI(A)\right) = \CE(A)$. 

However, in spite of all the ``justifications", the second guess does not works for arbitrary $R\subset\Sigma\cup Q$ either. To see this, let us firstly review some important properties of conventional entanglement wedges which protect the subregion duality in the usual AdS/CFT correspondence. Then, we will show that these properties are not satisfied by tentative entanglement wedges. 

\paragraph{$\CC \subseteq \CE$ and nesting for entanglement wedges}~\par
$\CC \subseteq \CE$ \cite{HHLM14,Wall12,EW14} and the entanglement wedge nesting \cite{CKJFvR12,Wall12} are known as properties that entanglement wedges should satisfy for the entanglement wedge reconstruction to work in the ordinary AdS/CFT or AdS/BCFT correspondence. 

$\CC \subseteq \CE$ states that the causal wedge $\CC(A)$ is completely contained in the entanglement wedge $\CE(A)$ for any $A\subset \Sigma$. Since $\CC(A)$ is a bulk region which can causally communicate with $\CD^{\rm (B)CFT}(A)$, $\CE(A)$ should at least contain $\CC(A)$ to be dual to $\CD^{\rm (B)CFT}(A)$ \cite{HHLM14,Wall12,EW14}. 

{\it Entanglement wedge nesting} states that, for any $B\subset A \subset \Sigma$, $\CE(B) \subset \CE(A)$. If entanglement wedge nesting was not satisfied, the subregion duality for $A$
\begin{align}
    (\CE(A), \CT^{\rm bulk}_\CM) \simeq (\CD^{\rm BCFT}(A), \CT^{\rm (B)CFT}_\Sigma)
\end{align}
and the subregion duality for $B$ 
\begin{align}
    (\CE(B), \CT^{\rm bulk}_\CM) \simeq (\CD^{\rm (B)CFT}(B), \CT^{\rm (B)CFT}_\Sigma)
\end{align}
would not be able to hold at the same time \cite{CKJFvR12,Wall12}. 

$\CC \subseteq \CE$ and entanglement wedge nesting are not independent from each other. It is discussed in \cite{AKLL16} that entanglement wedge nesting implies $\CC \subseteq \CE$ implies locality of the boundary theory $\CT^{\rm (B)CFT}_\Sigma$, i.e.
\begin{align}\label{eq:nesting1}
    {\rm entanglement~wedge~nesting} \Longrightarrow \CC \subseteq \CE \Longrightarrow {\rm locality~of~}\CT^{\rm (B)CFT}_\Sigma. 
\end{align}

\paragraph{Breakdown of $\CC \subseteq \CE_T$ and nesting for tentative entanglement wedges}~\par
Let us then go back to the intermediate/bulk correspondence. Accordingly, we can define $\CC \subseteq \CE$ and nesting for tentative entanglement wedges, by simply replacing $A\subset\Sigma$ with $R\subset \Sigma\cup Q$ in the previous discussions. Let us call the two corresponding statements 
$\CC \subseteq \CE_T$ and tentative entanglement wedge nesting, respectively. 

\begin{figure}[H]
    \centering
    \includegraphics[width=7.5cm]{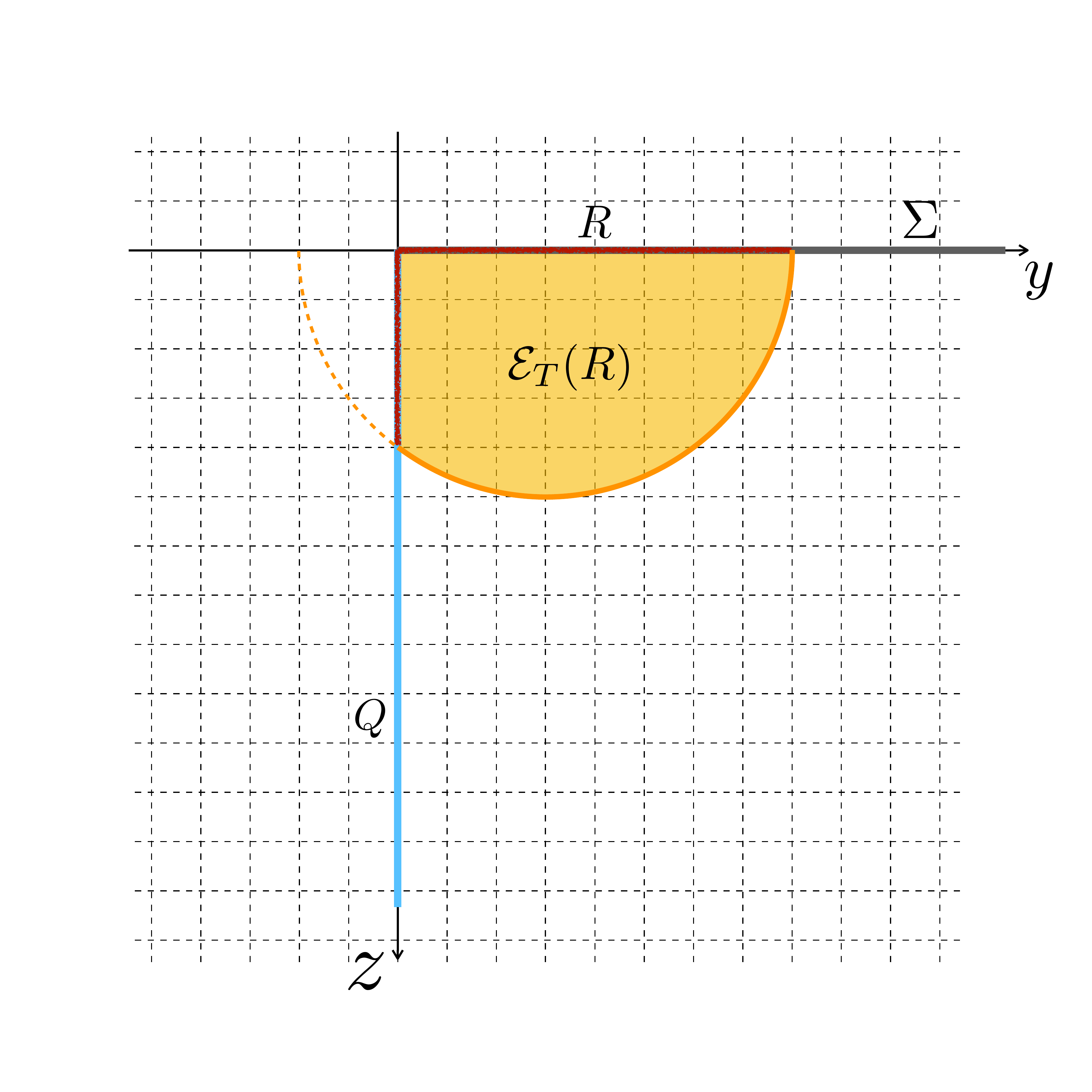}
    \includegraphics[width=7.5cm]{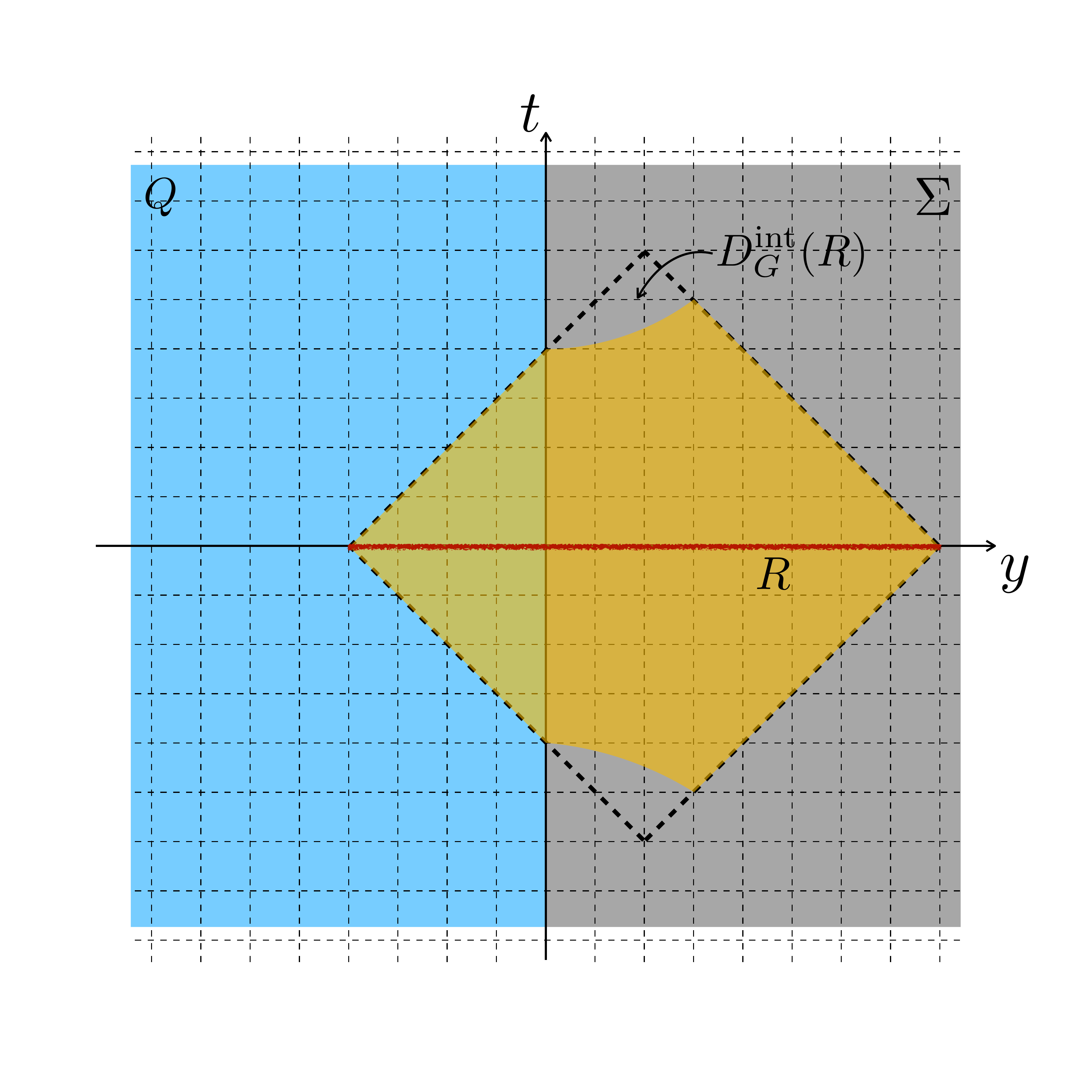}
    \caption{A simple example showing the breakdown of $\CC \subseteq \CE_T$. Consider a $d=2$ vacuum setup with a vanishing tension $T=0$. The left figure shows the $t=0$ slice, and the right figure shows $\Sigma\cup Q$. In this case, $Q$ is perpendicular to $\Sigma$, and $z$-coordinate coincides the minus direction of $y$-coordinate. Taking subregion $R$ (red) as $-4\leq y \leq 8$ and $t=0$. The tentative entanglement wedge $\CE_T(R)$ is shown in yellow. The geometrical domain of dependence $\CD_G^{\rm int}(R)$ is the region surrounded by the bold dotted lines in the right picture and goes outside of $\CE_T(R)$. Therefore, $\CC(R) \supset \CD_G^{\rm int}(R)$ also goes outside of $\CE_T(R)$.} 
    \label{fig:TEWbreakdown}
\end{figure}

\begin{figure}[H]
    \centering
    \includegraphics[width=7.5cm]{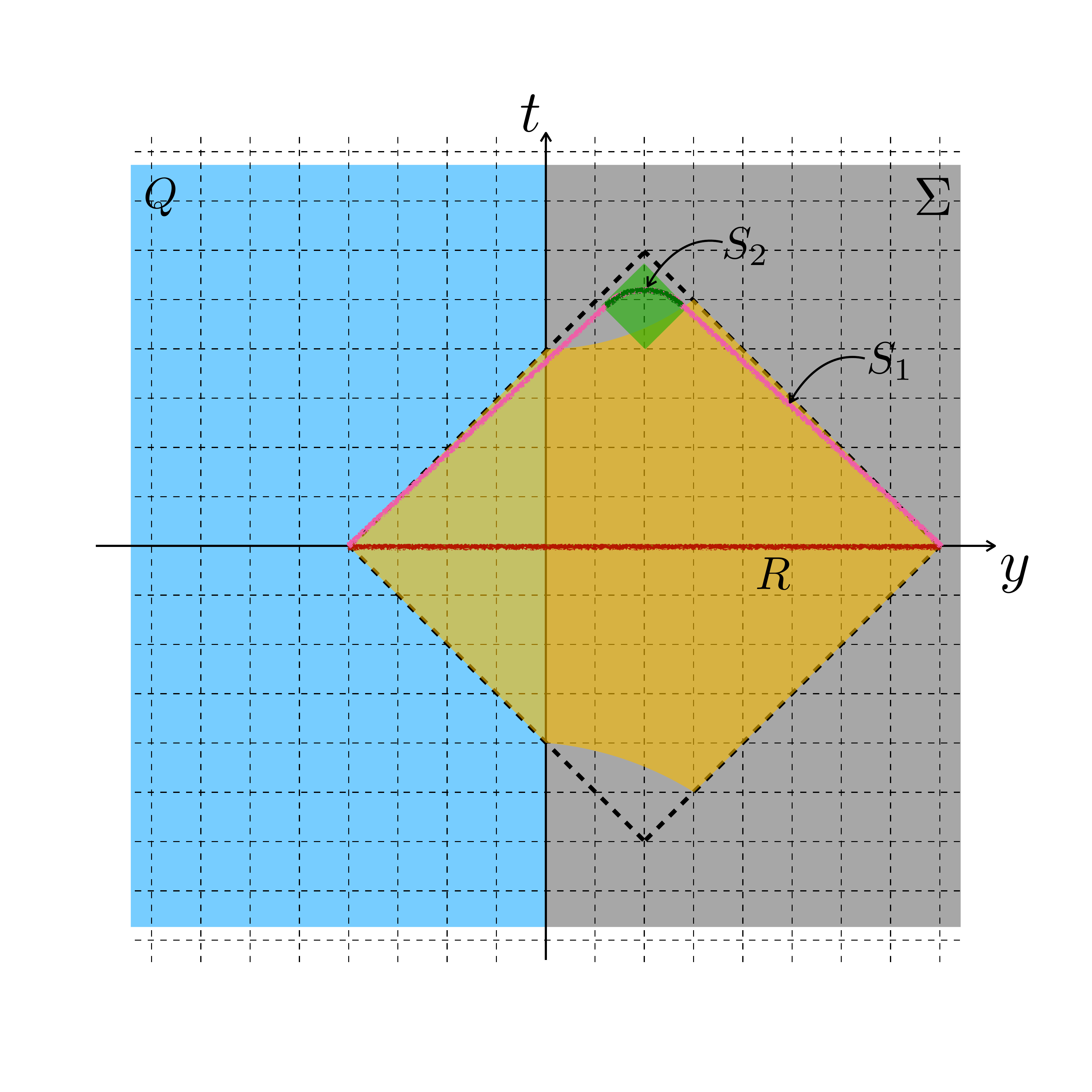}
    \caption{A simple example showing the breakdown of tentative entanglement wedge nesting. Consider the identical setup with figure \ref{fig:TEWbreakdown} and take a subregion $S_1$ (shown in pink) such that $\partial S_1 = \partial R$. By definition, $\CE_T(S_1) = \CE_T(R)$ and its intersection with $\Sigma \cup Q$ is shown in yellow. Consider $S_2\subset S_1$ shown in dark green. Its entanglement wedge $\CE_T(S_2)$ intersects $\Sigma \cup Q$ at the region shown in light green. Clearly, $\CE_T(S_2) \not\subseteq \CE_T(S_1)$ and hence tentative entanglement nesting breaks down.} 
    \label{fig:nestingbreak}
\end{figure}

Most straightforwardly, we have already known that $\CT^{\rm int}_{\Sigma\cup Q}$, the boundary theory in this case, is nonlocal. Therefore, by simply applying (\ref{eq:nesting1}), we have 
\begin{align}
    {\rm nonlocality~of~}\CT^{\rm int}_{\Sigma\cup Q} &\Longrightarrow {\rm breakdown~of~} \CC \subseteq \CE_T \nonumber\\
    &\Longrightarrow
    {\rm breakdown~of~tentative~entanglement~wedge~nesting}.
\end{align}
We can also confirm the breakdown of $\CC \subseteq \CE_T$ and tentative entanglement wedge nesting directly. See figure \ref{fig:TEWbreakdown} and \ref{fig:nestingbreak} for a simple example. As a result, $\left(\CE_T(R),\CT^{\rm bulk}_{\CM}\right)$ is not equivalent to $\left(\CD^{\rm int}_G(R),\CT^{\rm int}_{\Sigma\cup Q}\right)$ in general. 

As a summary of this subsection, for an arbitrary intermediate subregion $R$, its tentative entanglement wedge $\left(\CE_T(R),\CT^{\rm bulk}_{\CM}\right)$ is dual to neither its effective domain of dependence $\left(\CD^{\rm int}_E(R),\CT^{\rm int}_{\Sigma\cup Q}\right)$ nor its geometrical domain of dependence $\left(\CD^{\rm int}_G(R),\CT^{\rm int}_{\Sigma\cup Q}\right)$. More should be investigated to answer the question whether there is a bulk subregion dual to $\left(\CD^{\rm int}_E(R),\CT^{\rm int}_{\Sigma\cup Q}\right)$ or $\left(\CD^{\rm int}_G(R),\CT^{\rm int}_{\Sigma\cup Q}\right)$ for an arbitrary intermediate subregion $R$.\footnote{Of course, any intermediate subregion must have a dual in the bulk theory. The nontrivial point is whether this dual has a semiclassical description similar to an entanglement wedge.} We would like to leave this as a future problem. 

\subsection{IR-sensitive Nonlocality and Quantum Gravity}
Until now, we have discussed the causal structure in double holography, pointed out that the effective theory in the intermediate picture $\CT^{\rm int}_{Q\cup \Sigma}$ should allow (effectively) superluminal information propagation when sending a signal from $Q$ to $\Sigma$, and thus contain a nonlocality, in order to be compatible with the bulk picture. This was verified from both the geometrical causal structure (i.e. geodesics) and the microscopic causal structure (i.e. commutators). 

As a result, the conventional notion of domain of dependence breaks down, and no longer serves as a region associated to a density matrix of a spatial subregion. Accordingly, conventional subregion duality also breaks down. 

Although our explicit examples are mostly given in the vacuum background for simplicity, these results hold for any on-shell bulk configurations. Double holography was originally proposed in the study of Hawking radiation. In that context, the intermediate picture, where a black hole lives on $Q$ and can communicate with the heat bath $\Sigma$, can be straightforwardly regarded as a setup of black hole evaporation. The benefits of double holography is that this ``black hole communicating with heat bath" setup in the intermediate picture is related to well-defined BCFT picture and classical bulk picture. In this context, the intermediate picture can be regarded as an effective theory of quantum gravity defined on a classical geometrical background $Q\cup \Sigma$. This fact implies that, when defined on a classical geometrical background, the effective theory of quantum gravity should have a nonlocal effect, at least for the intermediate picture in double holography. 

\begin{figure}[H]
    \centering
    \includegraphics[width=16cm]{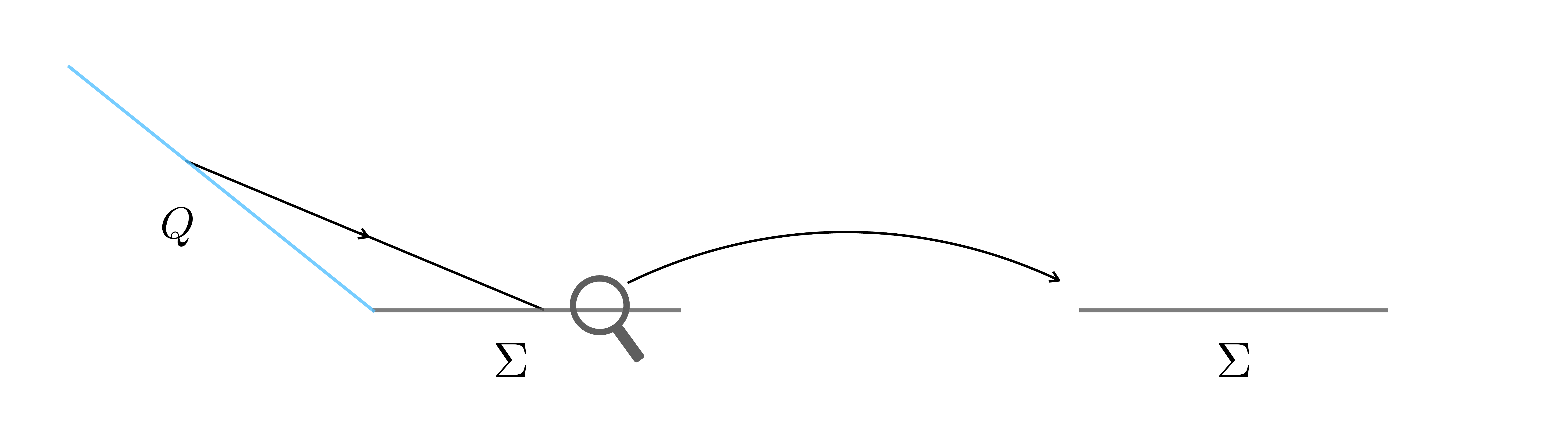}
    \caption{The IR-sensitive nonlocality in the intermediate picture. The left figure shows a time slice of the intermediate picture where the arrow shows a superluminal propagation from $Q$ to $\Sigma$ induced from a shortcut in the bulk. When zooming in to UV, one can only see $\Sigma$, and no nonlocality will be observed within $\Sigma$.}
    \label{fig:IRsens}
\end{figure}

Let us emphasize one key feature of this nonlocal effect in $\CT^{\rm int}_{Q\cup \Sigma}$. The fact that information propagation within $Q$ or $\Sigma$ cannot be superluminal implies that the nonlocal effect can be ignored if we zoom in to the UV. On the other hand, if we zoom out to the IR, the nonlocal effect becomes significant. For the latter feature, we say that the intermediate picture is {\it IR-sensitive}. See figure \ref{fig:IRsens}. There is, however, a special region in which the nonlocal effect is significant at any length scale. This region is $\partial\Sigma = \partial Q$.

\begin{figure}[H]
    \centering
    \includegraphics[width=16cm]{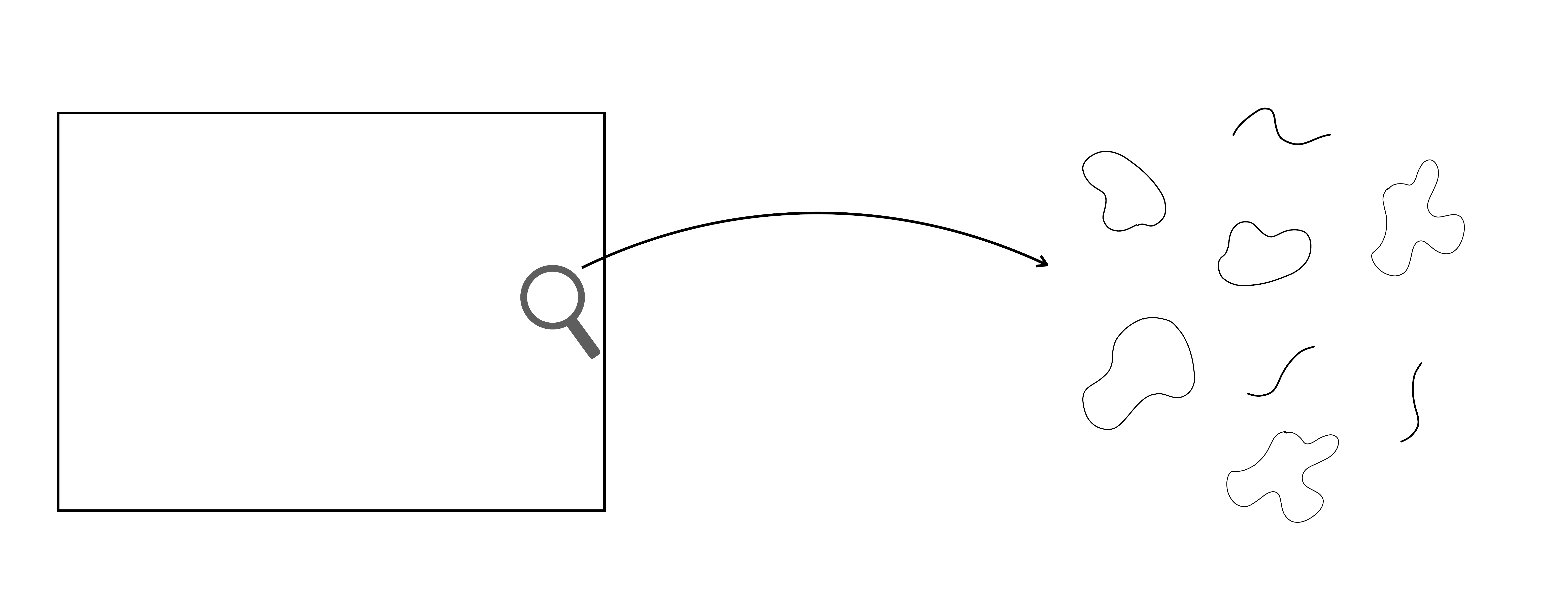}
    \caption{The UV-sensitive nonlocality known in the string theory. When zooming in a field theory (left) which has a string origin to UV, nonlocality caused by the extension of strings becomes significant.}
    \label{fig:UVsens}
\end{figure}

We would like to understand this IR-sensitive nonlocality as a nature of quantum gravity. However, before proceeding, we would like to note that this behavior is very different from the nonlocality known in string theory \cite{LPSTU95,Giddings06}. Intuitively, the extending structure of strings causes a non-vanishing commutator even outside of the light cone. This effect is significant at the string scale, and is expected to be invisible at large length scale in which strings look like point particles. Accordingly, we can say that this is a {\it UV-sensitive} nonlocal effect. See figure \ref{fig:UVsens}. Note that the IR-sensitive nonlocality found in the intermediate picture and the UV-sensitive nonlocality known in string theory do not contradict with each other, but they are expected to be caused by different mechanisms. 

Let us go back and consider what kind of effect in quantum gravity can give rise to an IR-sensitive nonlocality. A gravitational path integral\cite{Coleman88,GS88,MM20} can be written as: 
\begin{align}
    Z_{\rm QG} = \sum_{\substack{\rm topology\\\rm of~\CN}} \int \CD g \int \CD \phi ~ \exp\left(iI\left[(\CN,g),\phi\right]\right) 
\end{align}
where $\phi$ denotes matter fields and $I\left[(\CN,g),\phi\right]$ denotes a local gravitational action:
\begin{align}
    I\left[(\CN,g),\phi\right] = \frac{1}{16\pi G_N} \int_{\CN} \sqrt{-g} ~(R-2\Lambda) + \int_{\CN} \sqrt{-g} ~\CL[\phi] + ({\rm boundary~terms}) . 
\end{align}
At weak gravity limit $G_N \rightarrow 0$, it is straightforward to expect that the effective theory is a local QFT on a classical spacetime $(\CN_0, g_0)$ determined from the on-shell condition. In this case, the partition function can be written as:
\begin{align}
    Z_{\rm QG} \sim \int \CD \phi ~ \exp\left(iI\left[(\CN_0,g_0),\phi\right]\right).
\end{align}
On the other hand, a full quantum gravitational picture should include corrections from other geometries. In particular, it is natural to expect that geometries containing spacetime wormholes (and hence with higher topology), though suppressed at $\CO(e^{-1/G_N})$ order, should be also included as quantum corrections. Accordingly, scattering amplitude on such (possibly off-shell) geometry should also be counted unless there are strong evidences showing that it vanishes. As a result, if we use an effective theory defined on the dominant geometry to include corrections coming from higher topology, this effective theory is expected to be nonlocal and include superluminal phenomena. We conjecture that the mechanism depicted above can be thought as a fundamental origin of the nonlocality appearing in the intermediate picture. See (\ref{eq:wormhole}) for example. 
\begin{align}\label{eq:wormhole}
    &\imineq{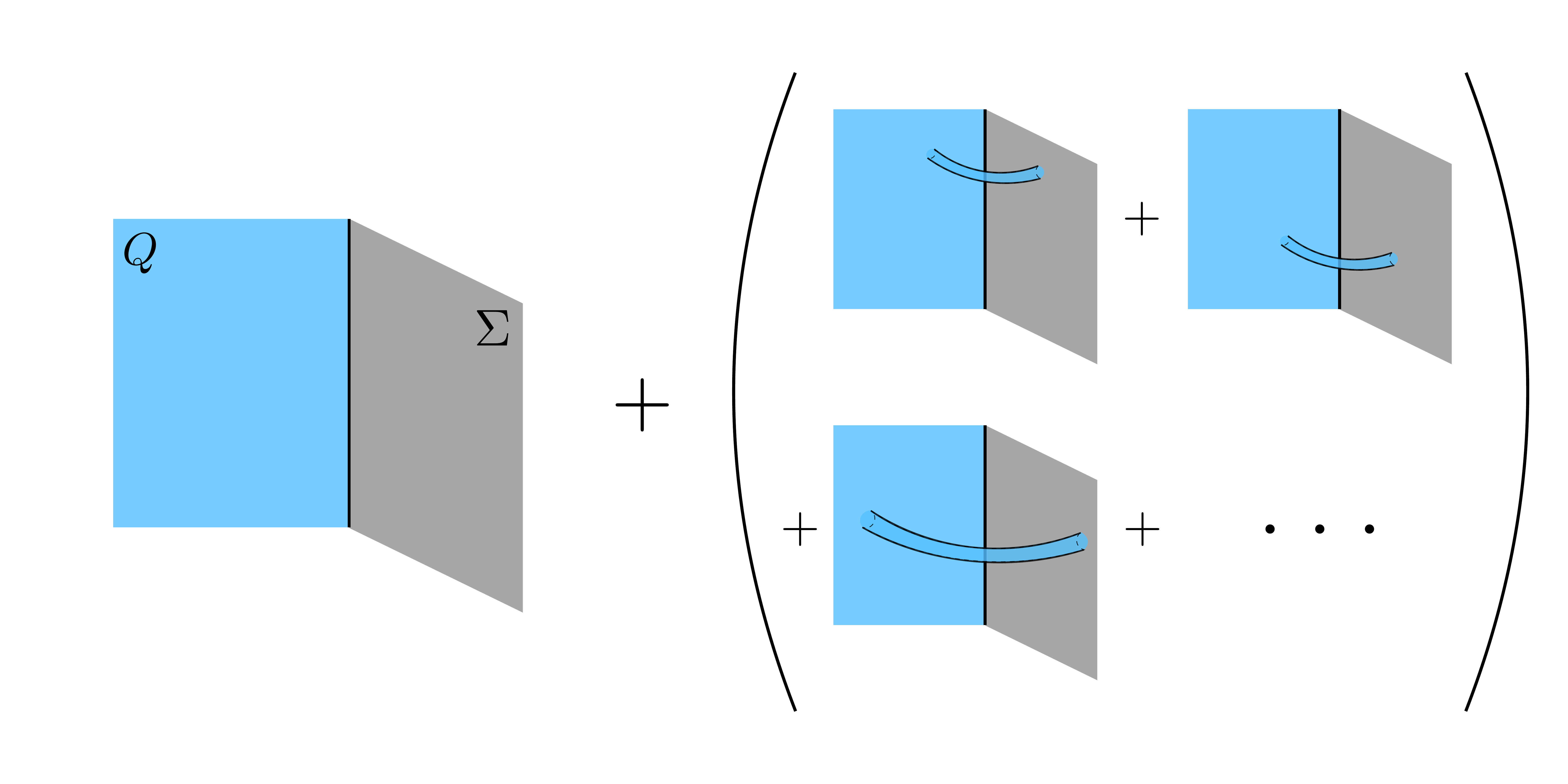}{27} \nonumber\\
    \simeq & \imineq{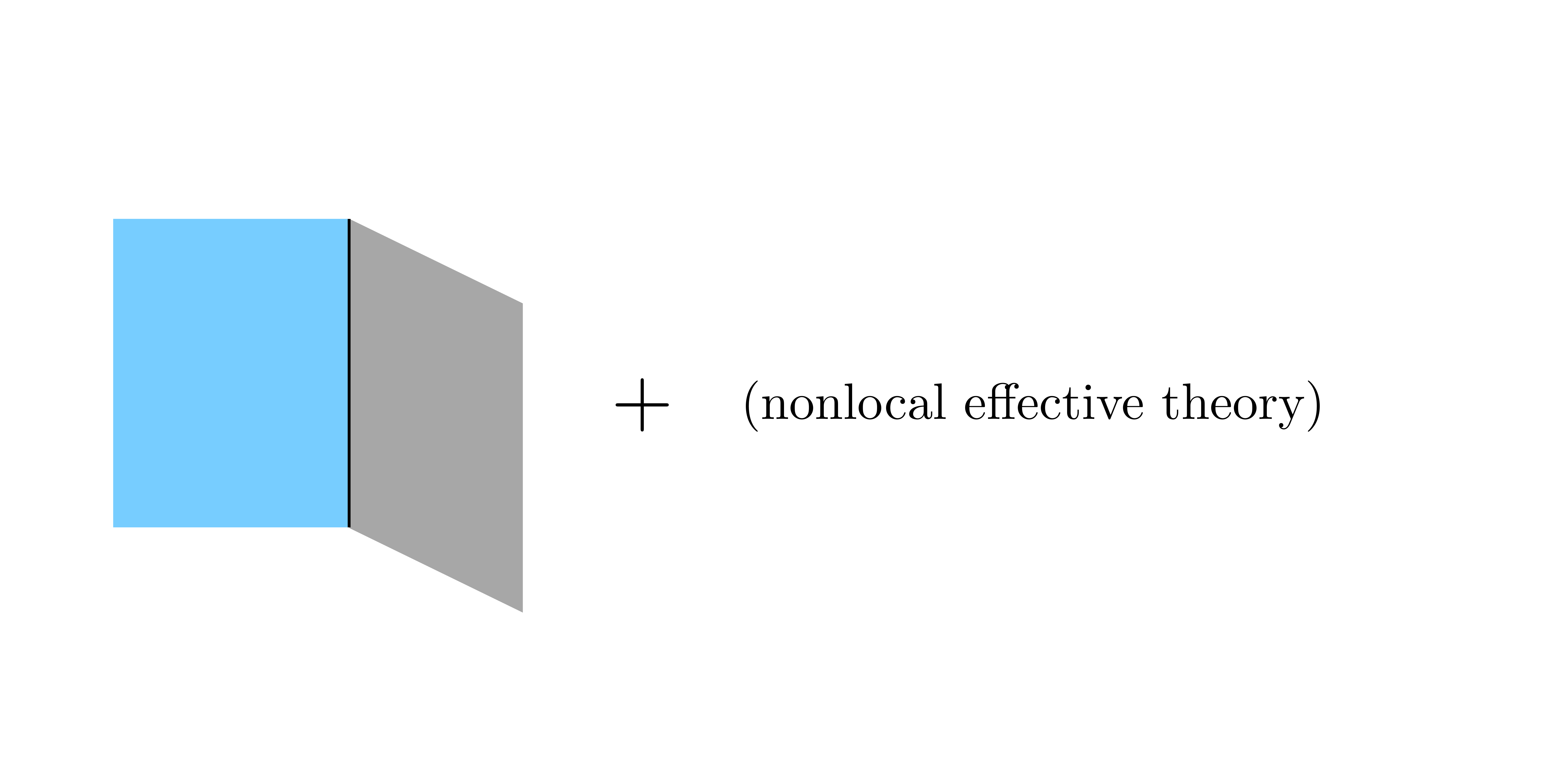}{27} \nonumber\\
    \simeq & \imineq{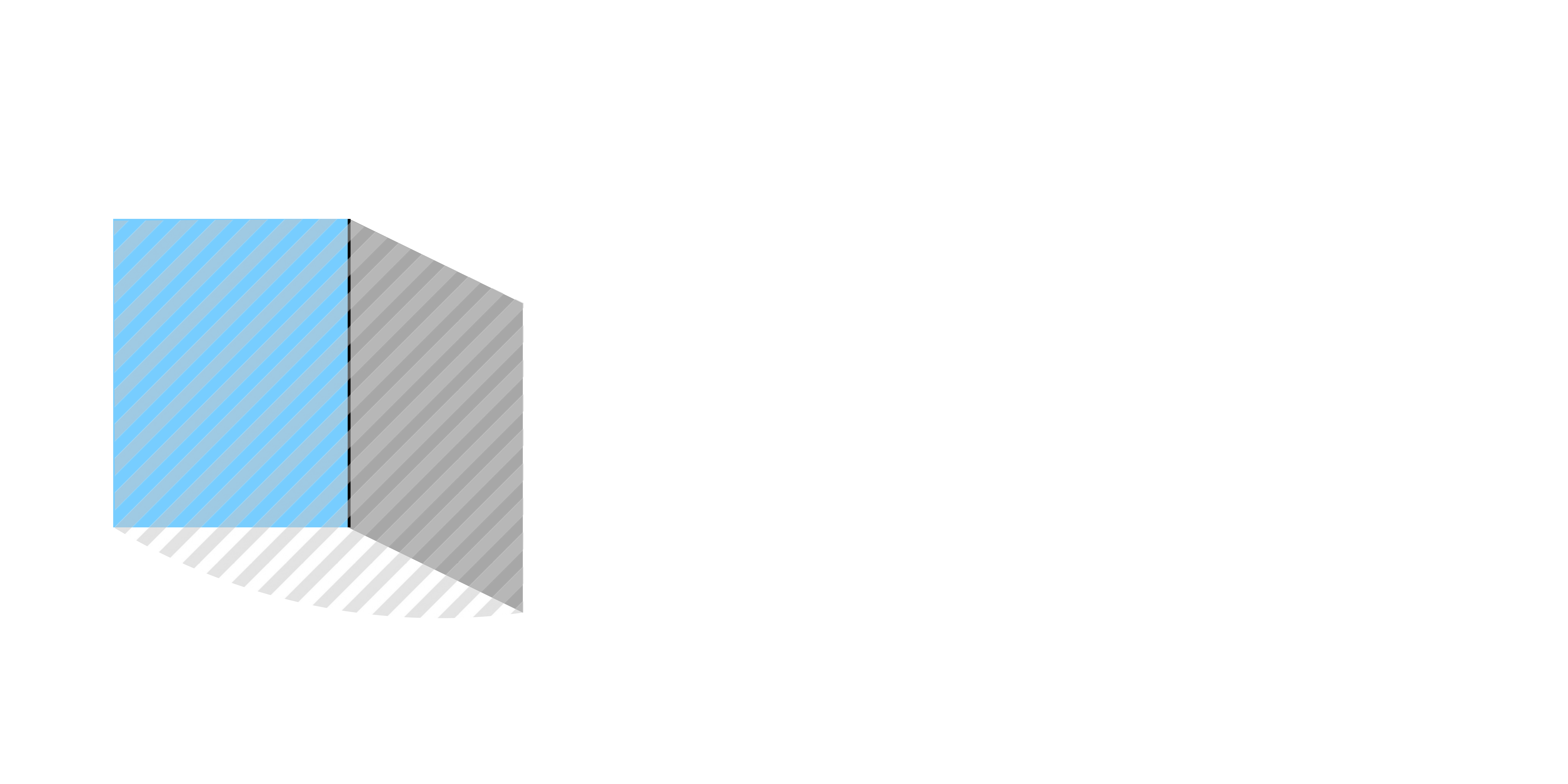}{27}
\end{align}
In this equation, we show time slices of the intermediate picture. We can see that the nonlocal effects in the effective theory arises from an ``ensemble" of non-dominant geometries with spacetime wormholes. This ``ensemble" also gives an effective extra-dimension. It is an intriguing future direction to test this idea by directly and carefully studying gravitational path integrals.

\section{Conclusions and Discussions} \label{sec:conclu}

In this paper, we focused on the causal structures in double holography and studied the relations between the three equivalent pictures. Especially, we have found that the bulk causal structure is compatible with causality in the BCFT picture, while it requires a violation of causality in the intermediate picture. We kept spacetime dimensions and configurations as general as possible. Therefore, all the doubly holographic models considered recently, e.g.\cite{AMS19,RSJvRWW19,CMNRS20,BW20,GK20,AKTW20,GKPRRRS20,AKSTW20,Neuenfeld21,Neuenfeld21-2,AKSTW21,Uhlemann21} are covered by this paper. 

Besides a specific case study for the vacuum configuration, we have extended the original Gao-Wald theorem to generic AdS/BCFT setups, and found that the bulk causal structures are compatible with causality in the BCFT picture. We have also discussed the possibility of using the compatibility of causality as an input to restrict constructions of bulk duals for general locally CFT with boundaries. 

On the other hand, we have found that being compatible with the bulk structure requires a violation of causality in the intermediate picture. More specifically, superluminal propagations exist if and only if we consider the communication between the gravitational region $Q$ and the non-gravitational region $\Sigma$. We have also confirmed this point in the microscopic causal structure by computing the commutators for a light primary field in the vacuum configuration. The breakdown of both macrocausality and microcausality indicates that there should be a nonlocal interaction in the intermediate picture, when considering it as an effective theory defined on $Q\cup\Sigma$. 

We discussed the physical consequences of the emergent nonlocality found in the intermediate effective theory. We have shown that the existence of nonlocality results in the breakdown of the conventional notion of domain of dependence. Based on this, the correspondences between subregions in the three pictures were reconsidered. Especially, we have shown that a subregion in the BCFT picture and the same geometrical subregion in the intermediate picture contain different information. More specifically, the latter one can be reconstructed from the former one, while the inverse is impossible, as shown in \eqref{eq:rhobcftint}. Another intriguing consequence is that, a straightforward analog of entanglement wedges to the intermediate/bulk correspondence, which we call tentative entanglement wedges, do not serve as duals of corresponding intermediate subregions.

We have also commented that the nonlocality found in this paper is significant at IR, i.e. at long range. This feature is similar to that of a spacetime wormhole. Therefore, a possible quantum gravitational origin of this sort of nonlocality is contributions coming from wormhole configurations in the gravitational path integral.

There are many possible future directions of our work. Here, we list some of them and give brief comments. 

Although we have concluded that the effective theory in the intermediate picture should include a nonlocal interaction, the effective action is not given. In double holography, the bulk allows shortcuts and works like a traversable wormhole. Therefore, it is natural to expect that the nonlocal term is given by a bilocal deformation such as double trace deformation \cite{GJW16,MSY17}, at least at the leading order, though more efforts are needed to examine the explicit form. 

In this paper, we have discussed causal issues in double holography by considering sending a signal from one point to another point on on-shell bulk geometries. Recently, causal puzzles and their resolutions associated with codimension-1 subregions \cite{BG20} and off-shell geometries \cite{HHTW21} are discussed in the AdS/CFT correspondence. It would be interesting to study similar aspects in the framework of double holography.  

Another promised direction is to derive the nonlocality in more general gravity-coupled-to-heat-bath setups without using holography. Inspired by the relation between the two methods for deriving the island formula: one by double holography \cite{AMMZ19} and the other by gravitational path integral \cite{AHMST19,PSSY19}, it is natural to expect that the nonlocality comes from the wormhole configurations, as also commented in the main text. However, we need a quantitative analysis to justify this intuition and make it more clear. Also, it would be interesting to understand its relation with early discussions on spacetime wormholes \cite{Coleman88,GS88,MM20}.

Last but not least, the implication of the IR-sensitive nonlocality in the context of black hole physics, especially the information loss problem, is also interesting. Although we did not focus on black holes, the results obtained in this paper are general enough to apply to these configurations. There have been many discussions on whether long-range nonlocality should be involved in a black hole evaporation process \cite{Strings21-1,Strings21-2}. Results in this paper indicate that long-range nonlocality indeed exists in doubly holographic models, but the role it plays is not clear so far. It would also be important to readdress the firewall problem \cite{AMPS12,Neuenfeld21-2} in double holography, with the IR-sensitive nonlocality involved.

\section*{Acknowledgements}
We are grateful to Ibrahim Akal, Bowen Chen, Yuya Kusuki, Tatsuma Nishioka, Tadashi Takayanagi, Kotaro Tamaoka, Takahiro Tanaka, Tomonori Ugajin and Zi-zhi Wang for useful discussions. We would like to thank Hao Geng, Andreas Karch and Federico Piazza very
much for valuable comments on this paper. ZW is supported by the ANRI Fellowship and Grant-in-Aid for JSPS Fellows No. 20J23116.

\appendix

\section{\texorpdfstring{AdS$_{d+1}$}{AdS} spacetime}\label{app}

In this appendix, we summarize coordinates and geodesics for AdS spacetime considered in this paper.

\subsection{Coordinates}\label{sec:AdScoord}
Anti-de Sitter (AdS) spacetime is a spacetime with a constant negative curvature in the Lorentz signature. This is given by the hypersurface
\begin{align}\label{eq:AdSdef}
	\braket{X,X} \equiv -(X^0)^2 - (X^{d+1})^2 + \bm{X}^2 = -L^2
\end{align}
in the $\mathbb{R}^{2,d}$, with its inner product given by $\braket{\cdot,\cdot}$. Here, coordinate in $\mathbb{R}^{2,d}$ is defined by $(X^0,X^{d+1},\bm{X})$ and its metric is given by
\begin{align}
	ds^2 = - (dX^0)^2 - (dX^{d+1})^2 + d\bm{X}^2~.
\end{align}
The quantity $L$ is AdS radius, which is set to $L=1$ in the following. There are several convenient coordinates in AdS spacetime used in this paper. One is the global coordinate $(\tau,r,\bm{\Omega})$, defined by
\begin{align}
	X^0 &= \sqrt{1 +r^2} \cos \tau~, & X^{d+1} &= \sqrt{1+r^2} \sin \tau ~, & \bm{X} &= r \bm{\Omega} ~.
\end{align}
Here, $\bm{\Omega}$ is the point on $S^{d}$. Then the metric is given by
\begin{align}
	ds^2 = - (1+r^2) d\tau^2 + \frac{dr^2}{1+r^2} + r^2 d\bm{\Omega}^2.
\end{align}
Note that we should take covering space to avoid the closed time like curve. 

The coordinate we are most interested in is the Poincar\'e coordinate. This is defined by
\begin{align}
\begin{aligned}
	X^0 &= \frac{z}{2} \left(1 + \frac{1 - t^2 + \bm{x}^2}{z^2}\right)~, & X^{d+1} &= \frac{t}{z} ~,\\
	X^d &= \frac{z}{2} \left(1 + \frac{-1- t^2 + \bm{x}^2}{z^2}\right)~,& X^i &= \frac{x^i}{z}~.
\end{aligned}
\end{align}
where $\bm{x} = (x^1,\dots,x^{d-1})$. The metric is
\begin{align}\label{eq:metricPoin}
	ds^2 = \frac{dz^2 - dt^2 + d\bm{x}^2}{z^2}
\end{align}
In this coordinate, asymptotic boundary of the AdS spacetime is given by $z = 0$.

\subsection{Geodesics}\label{sec:AdSgeodesic}

Geodesics in the AdS spacetime can be easily obtained by using the fact that they are embedded in $\mathbb{R}^{2,d}$. Suppose that $\mathcal{M}$ with $\dim\mathcal{M} = d+1$ (in our case AdS$_{d+1}$ spacetime) is embedded in the $\mathcal{N}$ with $\dim\mathcal{N} = d+2$ (in our case $\mathbb{R}^{2,d}$ spacetime). Then, Gauss-Weingarten equation relates the covariant derivative of $\mathcal{N}$ (which is denoted by $D$) and the covariant derivative of the $\mathcal{M}$ (which is denoted $\nabla$) by
\begin{align}\label{eq:GWapp}
	u^A D_{A} v^B = (u^\mu \nabla_\mu v^\nu) {e^B}_{\nu} -K_{\mu\nu} u^\mu v^\nu n^B~.
\end{align}
Here, ${e^A}_\mu$ is the basis of $T(\mathcal{M})$, $K_{\mu\nu}$ is the extrinsic curvature of $\mathcal{M}$ and $n^A$ is the normal vector to $\mathcal{M}$. Therefore, an equation for the affine parametrized geodesic on $\mathcal{M}$ can be written in terms of covariant derivative on $\mathcal{N}$ as 
\begin{align}
	u^A D_A u^B = - \left(g_{\mu\nu}u^\mu u^\nu\right) n^B~.
\end{align}
where $g_{\mu\nu}$ is the metric of AdS spacetime. Here, we used the fact that  the extrinsic curvature of AdS spacetime embedded in $\mathbb{R}^{2,d}$ is given by \footnote{We can easily obtain this result from Gauss equation
\begin{align}
	R_{ABCD}~{e^A}_\mu {e^B}_\nu {e^C}_\rho {e^D}_\sigma = R_{\mu\nu\rho\sigma} - (K_{\mu\sigma}K_{\nu\rho} - K_{\mu\rho}K_{\nu\sigma})~.
\end{align}
Of course, Riemann curvature $R_{ABCD}$ of $\mathbb{R}^{2,d}$ is zero, and we know that the curvature of the AdS spacetime is given by
\begin{align}
	R_{\mu\nu\rho\sigma} = - (g_{\mu\rho} g_{\nu\sigma} - g_{\mu\sigma} g_{\nu\rho})~.
\end{align}
Hence the extrinsic curvature of AdS$_{d+1}$ is given by
\begin{align}
	K_{\mu\nu} = g_{\mu\nu}~.
\end{align}
}
\begin{align}\label{eq:3}
	K_{\mu\nu} = g_{\mu\nu}~.
\end{align}

Let us write a geodesic as $\gamma(\lambda) = (X^0(\lambda),X^{d+1}(\lambda), \bm{X}(\lambda))$. Its tangent vector $u$ is given by
\begin{align}
	u(\lambda) \equiv \frac{d X}{d\lambda}~.
\end{align}
Then, the  geodesic equation of AdS spacetime is given by
\begin{align}\label{eq:A18}
	\frac{d^2 X(\lambda)}{d \lambda^2} = - (g_{\mu\nu} u^{\mu}(\lambda) u^\nu(\lambda)) X(\lambda)~.
\end{align}
Note that we chose $n = X$ in \eqref{eq:GWapp}. This is possible, because if we regard a point on the AdS as a vector in $\mathbb{R}^{2,d}$, they are always orthogonal to the hypersurface \eqref{eq:AdSdef}. This can be seen by taking derivative of the definition of the AdS spacetime \eqref{eq:AdSdef}. This gives
\begin{align}\label{eq:(u,x)}
	\frac{d}{d \lambda}\braket{X(\lambda), X(\lambda)} = 2\braket{u,X} = 0~,
\end{align}
where $\braket{\cdot,\cdot}$ is the usual flat metric on $\mathbb{R}^{2,d}$.

Now the geodesic equation \eqref{eq:A18} can be easily solved. For time-like geodesics ($g_{\mu\nu}u^\mu u^\nu = -1$),
\begin{align}
	X(\lambda) = Y \cosh \lambda + Z \sinh\lambda~,
\end{align}
for space-like geodesics ($g_{\mu\nu}u^\mu u^\nu = 1$),
\begin{align}
	X(\lambda) = Y \cos \lambda + Z \sin\lambda~,
\end{align}
and for null geodesics ($g_{\mu\nu}u^\mu u^\nu = 0$)
\begin{align}\label{eq:A22}
	X(\lambda) = Y  + Z \lambda~.
\end{align}
Here, $Y$ and $Z$ denotes the initial condition. From equation \eqref{eq:(u,x)}, they must satisfy
\begin{align}\label{eq:A23}
	\braket{Y,Z} = 0~.
\end{align}
Also, tangent vector $Z$ must satisfy
\begin{align}\label{eq:A24}
	\braket{Z,Z} =
	\begin{cases}
		-1 & (\mbox{ for time-like case})~,\\
		+1 & (\mbox{ for space-like case})~,\\
		0 & (\mbox{ for null case})~.
	\end{cases}
\end{align}

\newpage
\bibliographystyle{jhep}
\bibliography{BraneCausality.bib}

\end{document}